\documentclass[prb,aps,twocolumn,floatfix,amsmath,amssymb,superscriptaddress]{revtex4-1}
\usepackage{comment,enumerate}
\usepackage{graphicx}
\usepackage{tabularx}
\usepackage{times}
\usepackage[usenames,dvipsnames]{color}
\usepackage{amsmath}
\usepackage{dcolumn}
\usepackage{latexsym,amsmath,amssymb,bm,euscript}
\usepackage[
	colorlinks=true,
	urlcolor=BlueViolet,
    citecolor=Maroon,
	plainpages=false,
 	pdfpagelabels,
 	bookmarksnumbered
 	]{hyperref}

\newcommand{\sgn}{\textrm{sgn}}

\renewcommand{\(}{\left(}
\renewcommand{\)}{\right)}

\newcommand{\Hc}{{\mathrm{H.c.}}}

\definecolor{gray}{rgb}{0.4,.4,0.4}
\definecolor{purple}{rgb}{0.6,.0,0.6}
\definecolor{darkgreen}{rgb}{0.0,.6,0.}
\newcommand{\vect}[1]{\boldsymbol{#1}}
\def\beq{\begin{equation}}
\def\eeq{\end{equation}}

\newcommand{\hiddensubsection}[1]{
\begin{center}
\small
\textbf{#1}
\normalsize
\end{center}}
\newcommand{\hiddensubsubsection}[1]{
\begin{center}
\small
\textit{#1}
\normalsize
\end{center}}

\begin{document}
\title{Symmetry and duality in bosonization of two-dimensional Dirac fermions}

\author{David F. Mross}
\affiliation{Department of Condensed Matter Physics, Weizmann Institute of Science, Rehovot, 76100, Israel}

\author{Jason Alicea}
\affiliation{Department of Physics and Institute for Quantum Information and Matter, California Institute of Technology, Pasadena, CA 91125, USA}
\affiliation{Walter Burke Institute for Theoretical Physics, California Institute of Technology, Pasadena, CA 91125, USA}

\author{Olexei I. Motrunich}
\affiliation{Department of Physics and Institute for Quantum Information and Matter, California Institute of Technology, Pasadena, CA 91125, USA}
\affiliation{Walter Burke Institute for Theoretical Physics, California Institute of Technology, Pasadena, CA 91125, USA}

\begin{abstract}
Recent work on a family of boson-fermion mappings has emphasized the interplay of symmetry and duality: Phases related by a particle-vortex duality of bosons (fermions) are related by time-reversal symmetry in their fermionic (bosonic) formulation. 
We present exact mappings for a number of concrete models that make this property explicit on the operator level. 
We illustrate the approach with one- and two-dimensional quantum Ising models, and then similarly explore the duality web of complex bosons and Dirac fermions in $(2+1)$ dimensions.
\end{abstract}
\maketitle

\tableofcontents
\section{Introduction}\label{sec.introduction}

Mapping models of spins or bosons to fermions has a long history in condensed-matter physics.
In $(1+1)$-dimensional systems such mappings are based on the Jordan-Wigner transformation, which introduces non-local string operators that ensure change of statistics between objects at arbitrary spatial separation.
Despite this non-locality, many local $(1+1)$-dimensional spin or boson models map exactly onto local fermion models. For example, the one-dimensional (1D) transverse-field Ising model maps to a chain of free Majorana fermions that become massless at the phase transition.\cite{SchultzMattisLieb1964, ZuberItzykson1977, Shankar_Acta} Chern-Simons flux attachment\cite{FradkinCS} generalizes this technique to two-dimensional (2D) systems; here a ``statistical'' gauge field  fulfills the same role as the Jordan-Wigner string in 1D.

An illuminating application of the latter approach is the description of electronic fractional quantum Hall states as superfluids of Chern-Simons bosons.\cite{GirvinCB, ZhangCB, ReadCB, LeeAnyonSuperconductivity}
Another important early application directly related to the topics here is the study of phase transitions involving topological states.\cite{WeiFisherWu_1993, Wen_2000,Maissam14}

Dualities provide alternative reformulations complementary to those obtained by statistical transmutation.  Classic examples include Kramers-Wannier duality for Ising spins and particle-vortex duality for bosons.\cite{DasguptaHalperin, MatthewDungHai}  More recently, a fermionic counterpart has been discovered that maps free 2D Dirac fermions to dual Dirac fermions coupled to a gauge field.\cite{Son, WangSenthil2015, MetlitskiVishwanath2015, metlitskiduality, diracduality}
In all these cases the dual quasiparticles are highly non-local objects in terms of the original microscopic degrees of freedom, but exhibit the same statistics.

The presence of symmetries can yield interesting consequences for systems that are amenable to both duality and statistical transmutation.  Due to the non-local relation between various representations, symmetries that act locally in one set of variables can act highly nontrivially in another.  In 1D, for instance, translation symmetry in a free Majorana chain implements duality for the Ising model that arises upon transmuting back to spins. Two important recent works by Seiberg, Senthil, Wang and Witten\cite{SeibergSenthilWangWitten} and by Karch and Tong\cite{KarchTong} have extended this symmetry-duality correspondence to 2D systems. These groups established that phases for a free 2D Dirac fermion that are related by time-reversal symmetry are related by particle-vortex duality when expressed in terms of bosons coupled to a Chern-Simons field. Similarly, time-reversal symmetry for microscopic bosons corresponds to particle-vortex duality for Dirac fermions with Chern-Simons coupling.  

This paper aims to elevate the symmetry-duality interplay from the level of quantum phases to explicit properties of operators describing physical degrees of freedom in various representations. To illustrate the basic principles in a simplified setting, we first review the correspondence noted above between duality for the transverse-field Ising chain and translation symmetry in the Majorana-fermion representation, and then generalize this correspondence to a class of 2D spin Hamiltonians.
The main body of the paper then analyzes $(2+1)$-dimensional models of bosons, vortices, Dirac fermions, and dual Dirac fermions that can be explicitly mapped between one another via dualities and an analogue of Chern-Simons flux attachment.
We specifically formulate these models as coupled-wire arrays, which makes operator-based mappings possible.  

\begin{figure}[ht]
\includegraphics[width=\columnwidth]{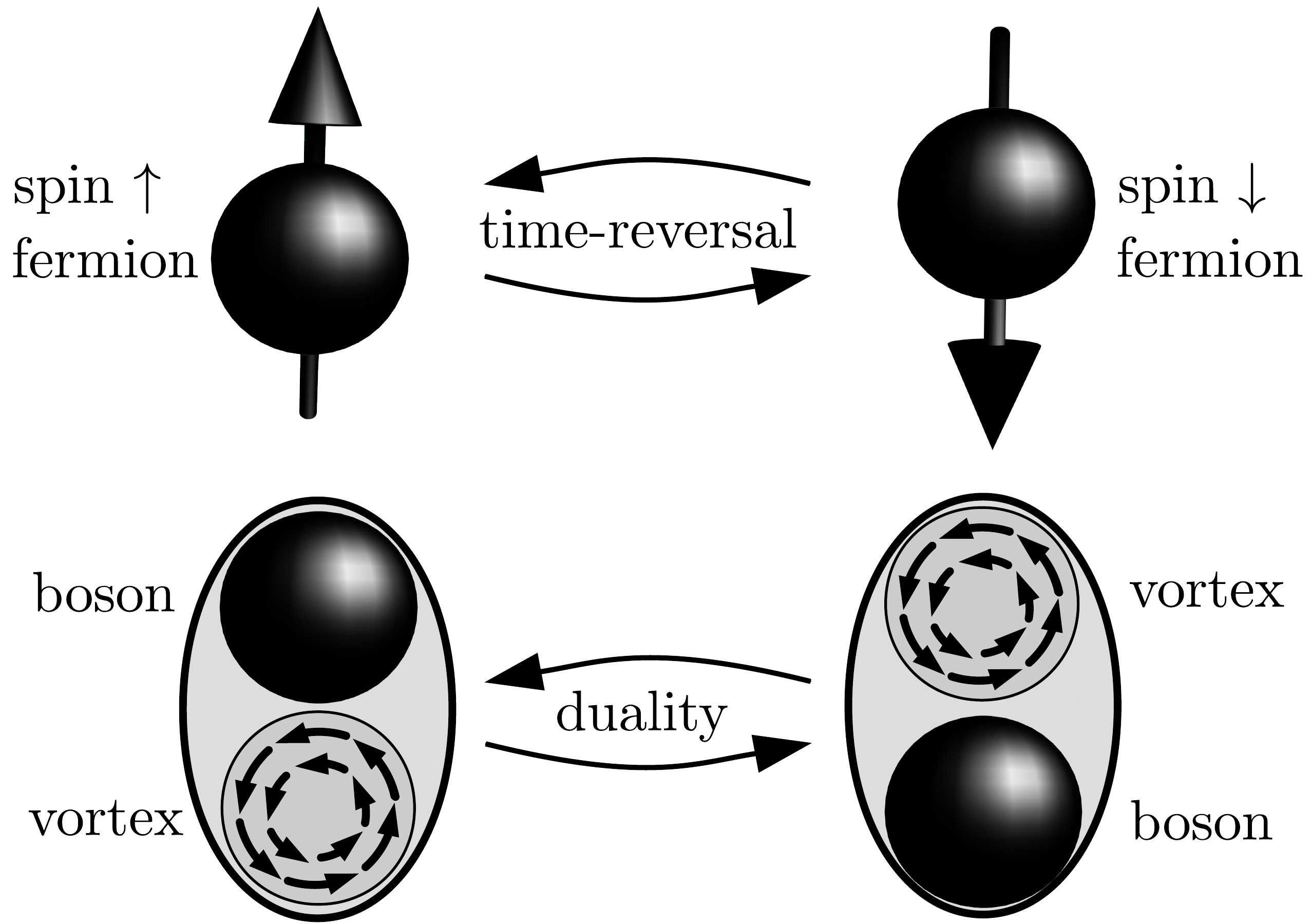}
\caption{In our formulation, the fermion spin is interpreted as the relative orientation of boson and vortex.
Under fermionic time-reversal the spin is flipped, corresponding to replacing bosons $\leftrightarrow$ vortices, i.e., duality.
This schematic picture will be made precise in the course of this paper.}
\label{fig:intfig}
\end{figure}

The coupled-wire approach paints an intuitive physical picture for the underlying transformations as well as the connection between symmetry and duality.  Representing Dirac fermions by bosons requires not only statistical transmutation; one must also augment the latter with an internal degree of freedom that encodes spin. Fermionic statistics can be achieved by forming bound states of bosons and vortices.  Importantly, in our discrete wire setups, vortices live on the dual lattice and are thus naturally displaced from the bosons---similar to the ``dipole picture'' of composite fermions\cite{Pasquier98,Read1998,ShankarMurthy97,DHLee98,Stern99,MurthyShankarRMP} which was recently revisited in Ref.~\onlinecite{WangSenthilReview}.
The Dirac-fermion spin correlates with the relative orientation of the boson-vortex bound state: for spin up the vortex sits just below the boson, while for spin down the orientation is reversed.  
Fermionic time reversal swaps up and down spins, and correspondingly swaps bosons and vortices, i.e., implements bosonic duality as sketched in Fig.~\ref{fig:intfig}.  
Dual Dirac fermions meanwhile arise simply by attaching the opposite vorticity to each boson. (As a result, the dual fermions also have opposite chirality in the wire formulation and opposite sign of their velocity in the continuum 2D description compared to the ``direct'' fermions.)
Bosonic time-reversal symmetry reverses vorticity and thus likewise implements duality for the Dirac fermions.

We flesh out the above picture in the framework of several explicit sets of wire models. The first set of models is schematically described by
\small
\begin{align}
&i \vect{j}_\text{boson} \cdot \vect{A}  &\leftrightarrow & && i \vect{j}_\text{Dirac} \cdot \vect{c} - i \frac{c dc}{8\pi} + i \frac{A dc}{2\pi} - i \frac{A dA}{4\pi} \nonumber \\
&\ \ \ \ \ \ \ \ \ \ \ \  \updownarrow& &   &&  \ \ \ \ \ \ \ \ \ \ \ \ \updownarrow \label{duality1} \\
&i \vect{j}_\text{vortex} \cdot \tilde{\vect{a}} + i \frac{A d\tilde{a}}{2\pi}  &\leftrightarrow &  && i \vect{j}_\text{dual Dirac} \cdot \tilde{\vect{c}} + i \frac{\tilde{c} d\tilde{c}}{8\pi} + i \frac{A d\tilde{c}}{2\pi} + i \frac{A dA}{4\pi} ~. \nonumber
\end{align}
\normalsize
Here $\vect{j}$ denotes space-time currents in a given representation indicated by the subscript, $\vect{A}$ is the external vector potential, and all other variables are dynamical gauge fields.
Vertical and horizontal arrows connect theories related by duality and statistical transmutation, respectively.
The left side sketches the familiar duality between bosons and vortices coupled to a gauge field that mediates long-range vortex interactions.
As indicated on the right, these systems in turn map to self-dual (in a sense that will be made precise later on) Dirac fermions coupled to a level-1/2 Chern-Simons gauge field. 
We will show explicitly that time-reversal symmetry for the bosons imposes exact self-duality for the fermions.  

We can similarly summarize the second set of models by
\small
\begin{align}
&i \vect{j}_\text{Dirac} \cdot \vect{A}   & \leftrightarrow &   && i \vect{j}_\text{boson} \cdot \vect{c} + i \frac{c dc}{4\pi} - i \frac{A dc}{2\pi} + i\frac{A dA}{8\pi} \nonumber \\
&\ \ \ \ \ \ \ \ \ \ \ \  \updownarrow& &   &&  \ \ \ \ \ \ \ \ \ \ \ \ \updownarrow\label{duality2} \\
&i \vect{j}_\text{dual Dirac} \cdot \tilde{\vect{a}} + i \frac{A d\tilde{a}}{4\pi}    &\leftrightarrow &  
&&i \vect{j}_\text{vortex} \cdot \tilde{\vect{c}} - i \frac{\tilde{c} d\tilde{c}}{4\pi} + i \frac{A d\tilde{c}}{2\pi} - i \frac{A dA}{8\pi} ~. \nonumber 
\end{align} 
\normalsize
The left side now represents the recently discovered duality between free Dirac fermions and dual Dirac fermions coupled to a gauge field.\cite{Son, metlitskiduality, WangSenthil2015, diracduality}
These theories map to self-dual Chern-Simons bosons, with time-reversal in the fermionic representation implementing duality in the bosonic representation.
Note that on the level of continuum theories, the models in Eqs.~\eqref{duality1} and \eqref{duality2} can be related by the conventional flux attachment technique,
where the conventional fermion features two Dirac nodes, one of which is very massive and ``integrated out''.
However, the action of symmetries on the statistically transmuted variables becomes obscured during this process, yet remains precise in our wire scheme.

We explore a third set of models that exhibit time-reversal symmetry both in the bosonic and fermionic representations---thus prohibiting Chern-Simons terms for any of the dynamical gauge fields.  We propose that these wire models yield the relations
\begin{eqnarray}
i \vect{j}_\text{boson} \cdot \vect{a} + {\cal L}[a] \  \ \ \ &\leftrightarrow &\ \ \ i \vect{j}_\text{Dirac} \cdot \vect{c} + \frac{1}{2} {\cal L}[c] \label{duality3} \nonumber \\
\updownarrow  \ \ \ \ \ \ \ \ \ \ \ \ \ \ \ \ \ \ & &  \ \ \ \ \ \ \ \ \ \updownarrow\label{eqn.superduality} \\
i \vect{j}_\text{vortex} \cdot \tilde{\vect{a}} + {\cal L}[\tilde{a}] \ \ \ \  &\leftrightarrow &\ \ \ 
i \vect{j}_\text{dual Dirac} \cdot \tilde{\vect{c}} + \frac{1}{2} {\cal L}[\tilde{c}] ~, \nonumber
\end{eqnarray}
where in momentum space
\begin{align}
{\cal L}[a] = \frac{1}{4\pi |\vect k|} |\vect{k} \times \vect{a}|^2 ~, \nonumber
\end{align}
and for brevity we suppressed the external gauge field $\vect{A}$ (which can be introduced as in the previous theories).
It is known that bosons with such kind of marginally long-range interactions mediated by ${\cal L}[a]$ can be exactly self-dual.\cite{FradkinKivelson,Geraedts2012_rangedloops}
Furthermore, this interaction does not break time-reversal symmetry, and consequently the fermionized description is also both time-reversal symmetric and self-dual.

\begin{figure}
\includegraphics[width=.9\columnwidth]{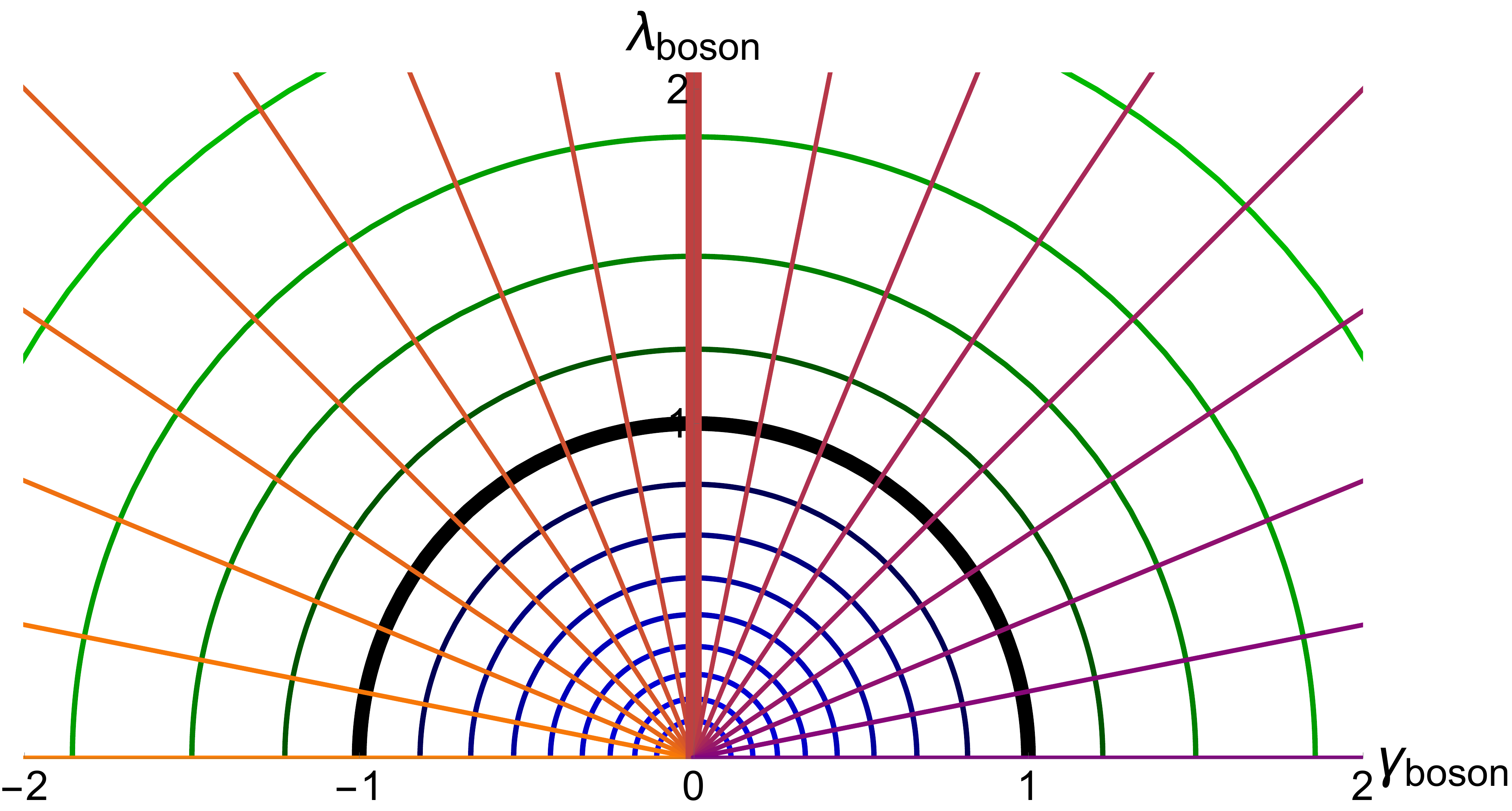}
\includegraphics[width=.9\columnwidth]{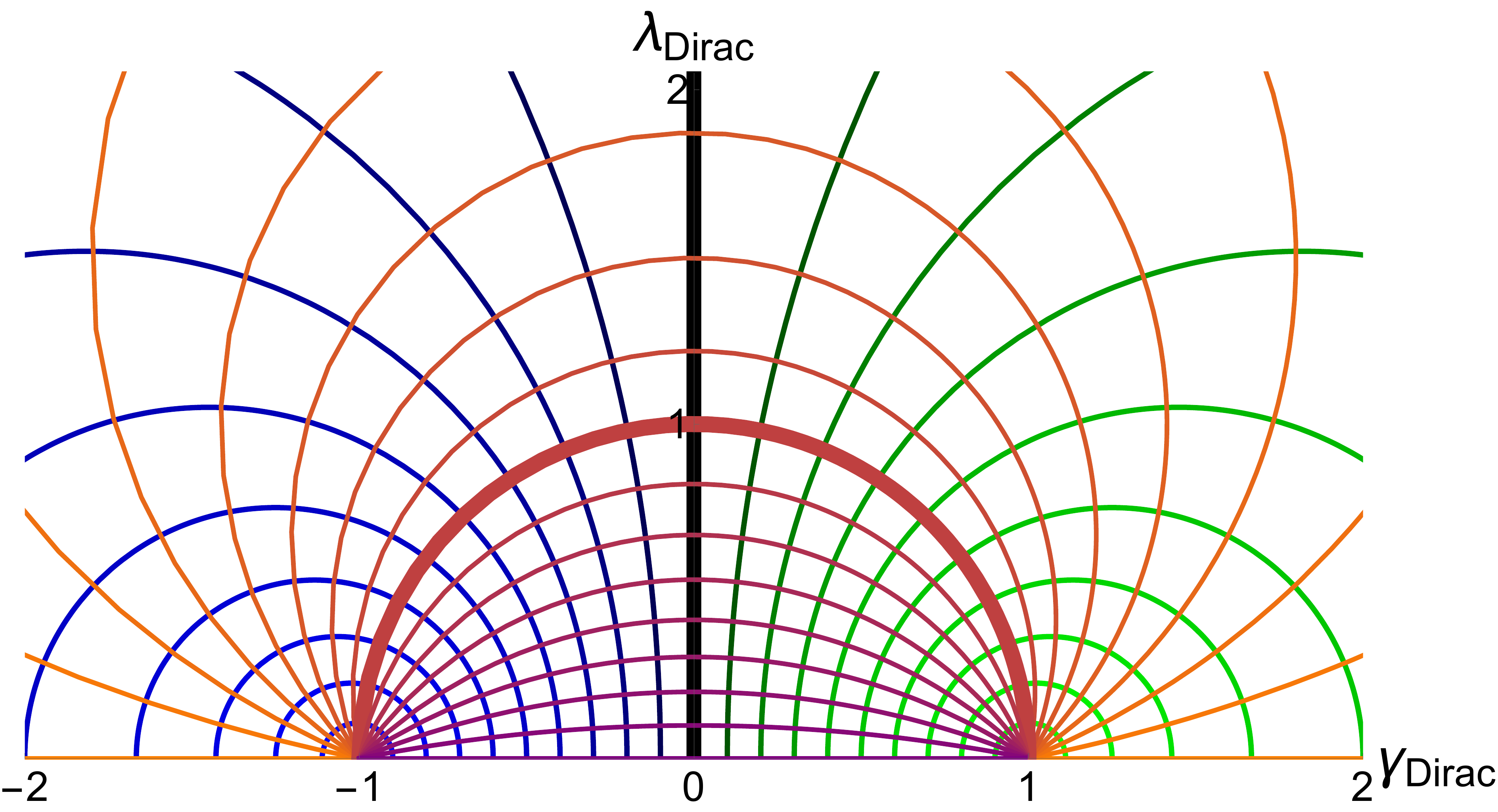}
\caption{Eq.~\eqref{eqn.cfmap} describes a conformal map from the complex half-plane onto itself. 
The upper figure depicts (orthogonal) lines of fixed radius or angle.
The lower figure shows the same lines in the transformed coordinate system. 
A number of important special cases are: 
(i) Purely imaginary $z_\text{boson}$, corresponding to time-reversal invariant boson systems, map onto the unit circle $|z_\text{Dirac}| = 1$ for fermions, i.e., self-dual models.
The limits $z_\text{boson} \rightarrow i \infty$ and $z_\text{boson} = i 0^+$ correspond to $z_\text{Dirac} = +1$ and $z_\text{Dirac} = -1$, i.e., the fermion models with half-integer Chern-Simons described in Eq.~\eqref{duality1}.
(ii) Self-dual boson models, $|z_\text{boson}| = 1$, map onto purely imaginary $z_\text{Dirac} = i \tan[\text{arg}(z_\text{boson})/2]$, i.e., time-reversal invariant fermions.
(iii) The special point $z_\text{boson} = z_\text{Dirac} = i$ is invariant under the conformal map and corresponds to a model that is simultaneously self-dual and time-reversal invariant for both bosons and fermions.}
\label{fig.cfmap}
\end{figure}

\begin{figure*}
\centering
\includegraphics[width=.475\textwidth]{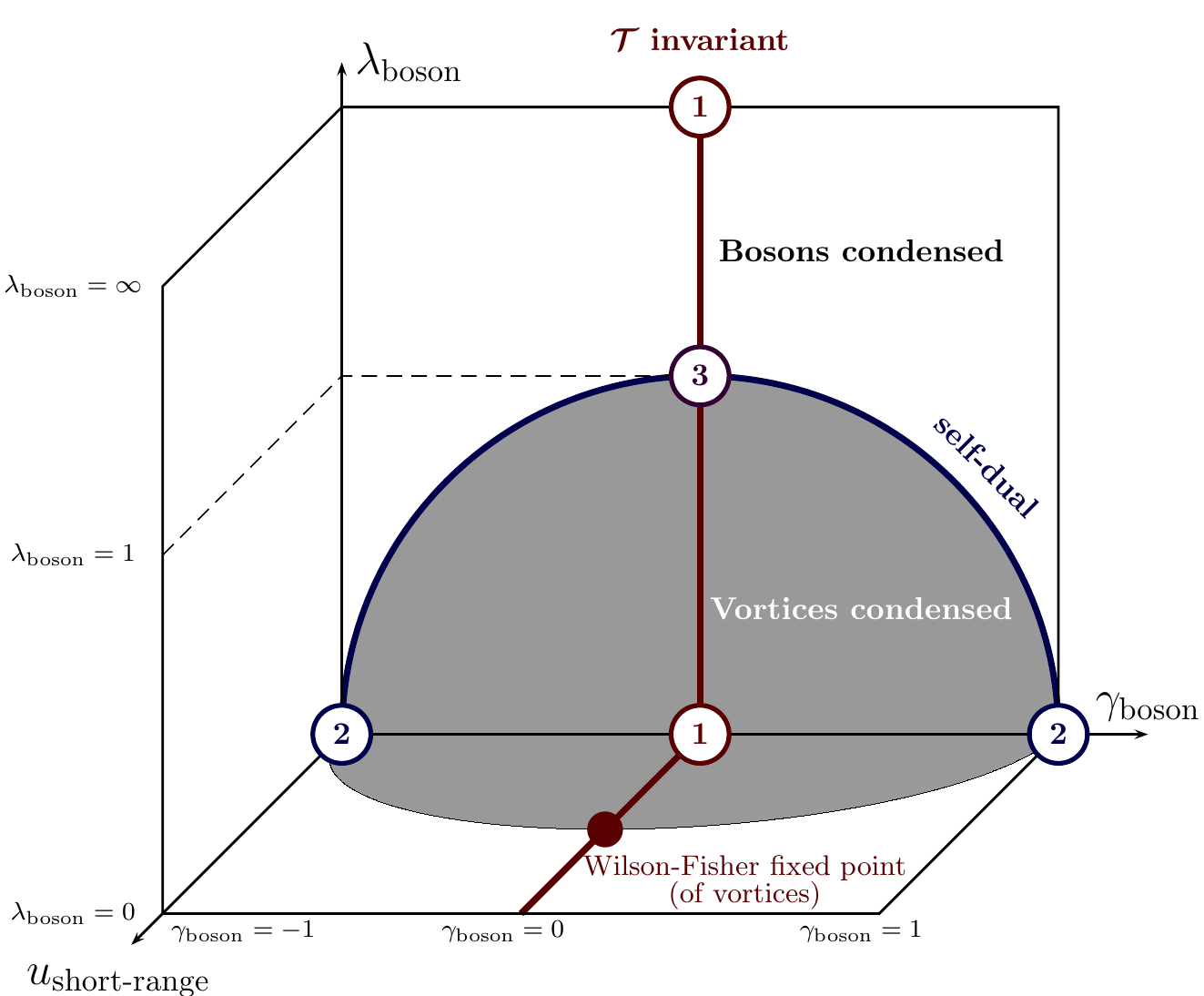}
\includegraphics[width=.475\textwidth]{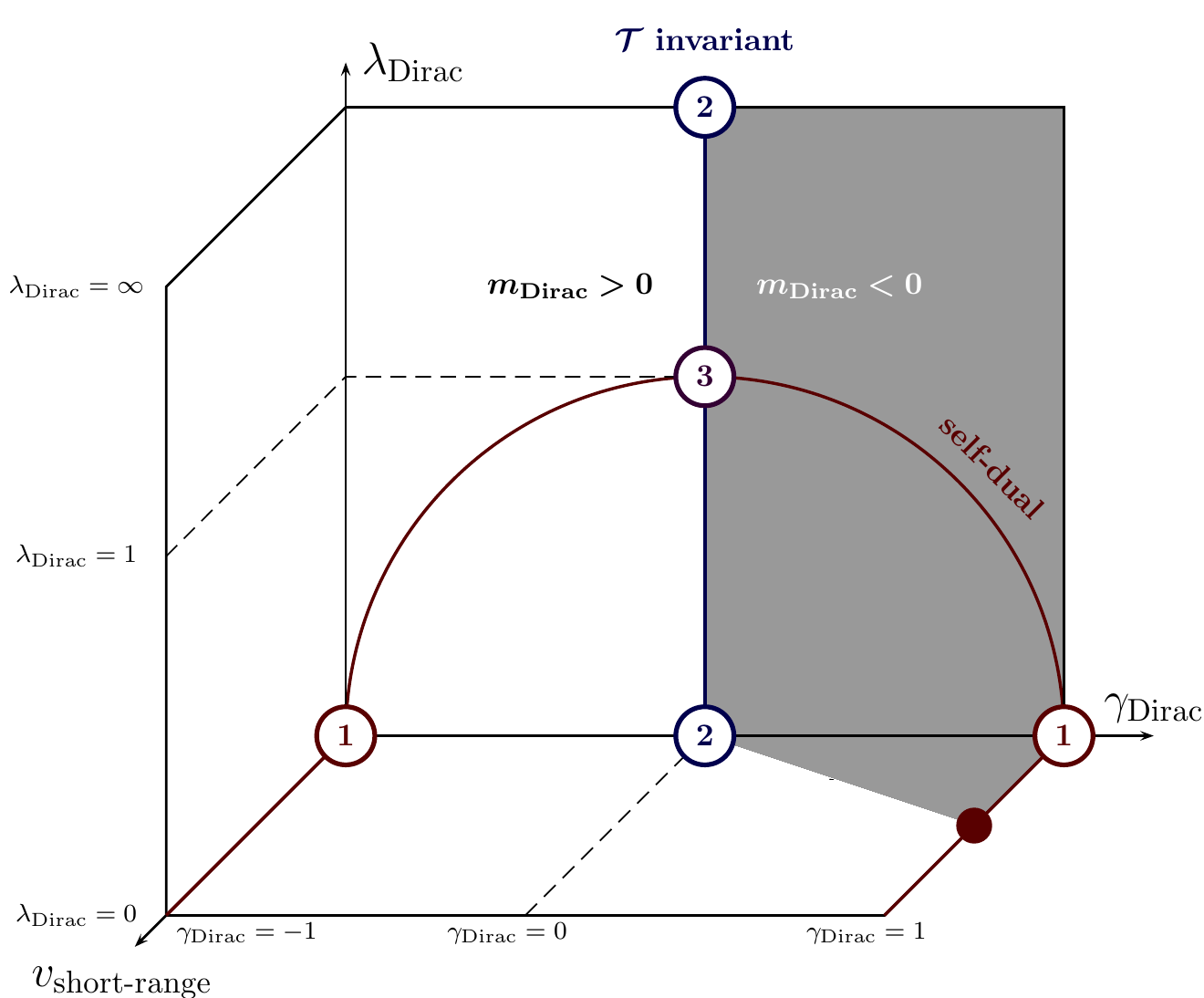}
\caption{Left: Phase diagram of bosons with Chern-Simons term ($\gamma_\text{boson}$), marginally long-range interactions ($\lambda_\text{boson}$), and additional non-universal short-range interactions ($u_\text{short-range}$). Time-reversal symmetry is present at $\gamma_\text{boson}=0$. When additionally $\lambda_\text{boson}=0$, the bosons interact via a gauge field that obeys Maxwell dynamics. This corresponds to vortices with purely short-range interactions which may either condense or become gapped. The transition between the two phases is governed by the Wilson-Fisher fixed point. In contrast, $\gamma_\text{boson}=0$ and $\lambda_\text{boson}=\infty$ formally corresponds to bosons with purely short-range interactions. These two limits,  are related by the standard boson-vortex duality of Eq.~\eqref{duality1}. Finite, non-zero $\lambda_\text{boson}$ interpolates between the two with self-duality realized at $\lambda_\text{boson}=1$, corresponding to Eq.~\eqref{duality3}. 
This point describes a quantum phase transition between the same phases as the Wilson-Fisher fixed point (either bosons or vortices condense), but is of a different universality class because of the marginally-long-range interactions (strictly speaking, long-distance properties of the two phases also change qualitatively for finite non-zero $\lambda_\text{bos}$). 
For non-zero $\gamma_\text{bosons}$, this point extends into a self-dual line of phase transitions at $\lambda_\text{bosons}^2 + \gamma_\text{bosons}^2 = 1$; this includes the case of bosons with purely statistical interactions described by Eq.~\eqref{duality2}. In the figure, we assumed that this line of self-dual phase transitions lies in the $u_\text{short-range} = 0$ plane, which is something we can realize in explicit wire models. Right: The same models can be transcribed into Dirac fermions with statistical ($\gamma_\text{Dirac}$), marginally long-range ($\lambda_\text{Dirac}$), and additional short-range interactions ($v_\text{short-range}$).
The self-dual line of bosons maps onto time-reversal invariant fermions $\gamma_\text{Dirac}=0$ which are critical.
It separates gapped phases with opposite sign of the Dirac mass $m_\text{Dirac}$ which corresponds to either bosons or vortices condensing.
The $\gamma_\text{Dirac} = 0$ line includes the special cases of $N=1$ QED$_3$ ($\lambda_\text{Dirac} = 0$), Dirac fermions with short range interactions ($\lambda_\text{Dirac} = \infty$) and self-dual Dirac fermions ($\lambda_\text{Dirac} = 1$).
As for the case of bosons, the self-dual point extends to a self-dual line for $\gamma_\text{Dirac} \neq 0$.
It includes the case of Dirac fermions with purely statistical interactions, $\lambda_\text{Dirac} = 0$ and $\gamma_\text{Dirac} = 1$, which is self-dual and for which short range interactions $v_\text{short-range}$ can drive a phase transition in the Wilson-Fisher universality class.
}
\label{SummaryFig}
\end{figure*}

We will show that all of the equivalences encapsulated by Eqs.~\eqref{duality1} through \eqref{duality3} may be viewed as special cases of generalized mappings connecting theories
\begin{align*}
&i \vect{j}_\text{boson} \cdot \vect{a} + \lambda_\text{boson} {\cal L}[a] - i \gamma_\text{boson} \frac{a da}{4\pi} ~, \\
&i \vect{j}_\text{vortex} \cdot \tilde{\vect{a}} + \lambda_\text{vortex} {\cal L}[\tilde a] - i \gamma_\text{vortex} \frac{\tilde{a} d\tilde{a}}{4\pi} ~, \\
&i \vect{j}_\text{Dirac} \cdot \vect{c} + \frac{\lambda_\text{Dirac}}{2} {\cal L}[c] - i \gamma_\text{Dirac} \frac{c dc}{8\pi} ~, \\
&i \vect{j}_\text{dual Dirac} \cdot \tilde{\vect{c}} + \frac{\lambda_\text{dual Dirac}}{2} {\cal L}[\tilde {c}] - i \gamma_\text{dual Dirac} \frac{\tilde{c} d\tilde{c}}{8\pi} ~.
\end{align*}
These theories are equivalent for a specific relationship between the parameters $\lambda_j,\gamma_j$ that can be expressed succinctly by adopting the notation $z_j =  \gamma_j + i \lambda_j$. From duality one finds the relations $z_\text{vortex} = - z_\text{boson}^{-1}$ and $z_\text{dual Dirac} = - z_\text{Dirac}^{-1}$.  This ``modular structure'' has been previously analyzed for the case of bosons in Ref.~\onlinecite{FradkinKivelson}.
Moreover, we will show that the fermionic and bosonic theories are related via
\begin{align}
z_\text{Dirac} = \frac{z_\text{boson} - 1}{z_\text{boson} + 1} ~. \label{eqn.cfmap}
\end{align}
Equation~\eqref{eqn.cfmap} describes a conformal map illustrated in Fig.~\ref{fig.cfmap}.  Figure~\ref{SummaryFig} presents an overview of the various field theories related by this general duality and of the corresponding phases; for a discussion see the caption.  

We organize the remainder of the paper as follows.
Section~\ref{sec.ising} discusses the interplay between symmetry and duality for 1D and 2D quantum Ising models.
In Sec.~\ref{sec.bosons} we review boson-vortex duality from the viewpoint of coupled wires, while in Sec.~\ref{sec.fermions} we similarly discuss duality for Dirac fermions.
Section~\ref{sec.bosonization} relates the boson and fermion formulations through an analogue of flux attachment, which is where the symmetry-duality correspondence becomes manifest.
The ``mother'' equivalences summarized above are derived in Sec.~\ref{sec.modular}. Section~\ref{GappedPhases} briefly discusses applications of our methods to gapped phases, and Sec.~\ref{sec.furtherapps} pursues extensions to systems hosting two Dirac fermion flavors.
A brief conclusion and outlook appears in Sec.~\ref{sec.conclusions}.
Supplementary details can be found in several Appendices.

\section{Symmetry-duality relation in Ising models}
\label{sec.ising}

\subsection{Transverse-field Ising chain}
\label{sec.1dising}

As an instructive warm-up exercise, we examine the 1D transverse-field Ising model,
\begin{align}
H_\sigma = - J \sum_r \sigma_r^z \sigma_{r+1}^z - h \sum_r \sigma_r^x ~,
\label{Hsigma}
\end{align}
with $r$ integers that label sites.  This system provides a simple (and well-known) example where symmetry in one representation corresponds to duality in another.  Moreover, the changes of variables that link these representations loosely parallel those that we exploit in later sections for 2D systems.  

First we define dual variables on half-integer sites via
\begin{subequations}
\begin{align}
&\tau_{r+1/2}^z = \prod_{r' < r+1/2} \sigma_{r'}^x ~,\\
&\tau_{r+1/2}^x = \sigma_r^z \sigma_{r+1}^z ~.
\end{align}
\end{subequations}
Under this duality transformation the Hamiltonian becomes
\begin{align}
H_\tau = - J \sum_r \tau_{r+1/2}^x - h \sum_r \tau_{r-1/2}^z \tau_{r+1/2}^z ~,
\label{Htau}
\end{align}
which is self-dual at the critical point $J = h$.  Note that we have assumed an infinite chain above so that boundary terms can be ignored (see below, however).  From the viewpoint of the original spins, one can view $\sigma^z$ as an order parameter and $\tau^z$ as a ``disorder parameter'': $\tau^z$ creates a domain-wall defect in the original language and thus condenses in the disordered phase with $h>J$.  

\begin{figure}[ht]
\includegraphics[width=\columnwidth]{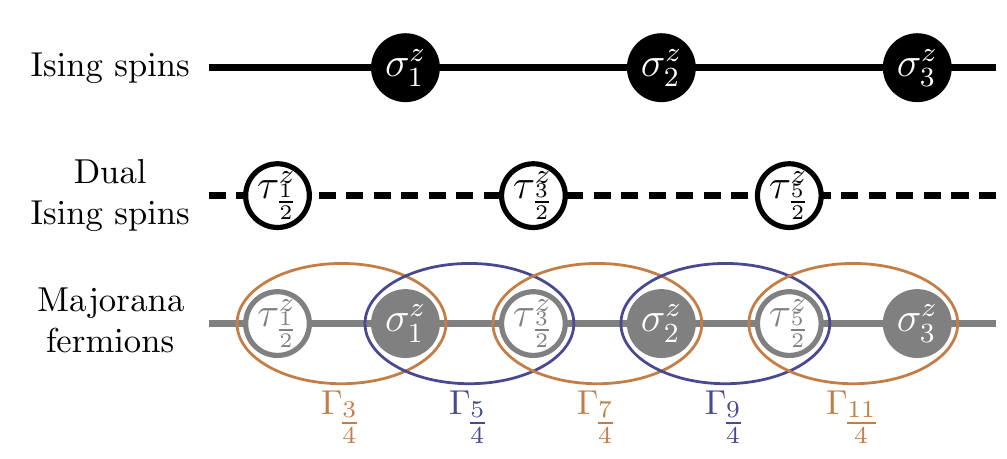}
\caption{Different sets of variables for representing the transverse-field Ising model.
Microscopic Ising spins reside on sites labeled by integers. 
Dual Ising spins reside on the dual lattice sites denoted by half-integers.
Majorana fermions are obtained by combining direct Ising spins with dual Ising spins and are thus most naturally associated with sites labeled by quarter-integers.}
\label{fig:ising}
\end{figure}

By combining order and disorder operators one can alternatively describe the model in terms of Majorana fermions $\Gamma$,\cite{SchultzMattisLieb1964, ZuberItzykson1977, Shankar_Acta}
\begin{subequations}
\begin{align}
&\Gamma(r-1/4) = \left(\prod_{r'<r} \sigma_{r'}^x \right) \sigma_r^z = \tau_{r-1/2}^z \sigma_r^z ~, \\
&\Gamma(r+1/4) = -\left(\prod_{r'<r} \sigma_{r'}^x \right) \sigma_r^y = i \sigma_r^z \tau_{r+1/2}^z ~.
\end{align}
\end{subequations}
In particular, the two inequivalent fermions defined above arise by appending a $\tau^z$ operator to the left or right of an original spin $\sigma^z$.  These Majorana operators thus naturally reside on ``quarter-integer'' (i.e., odd integers divided by four) sites $r \pm 1/4$ located midway between direct and dual lattice sites.  (See Fig.~\ref{fig:ising} for a summary of the representations.)  It is often convenient to instead enumerate fermion sites by integers $j$ with 
\begin{equation}
  \gamma_j \equiv \Gamma(j/2+1/4).
\end{equation}
Using $\gamma_j$ Majorana operators, the Hamiltonian takes the form
\begin{align}
H_\gamma = \sum_j \frac{1}{2} \left[J+h + (-1)^j (J-h)\right] \(i \gamma_j \gamma_{j+1} \) ~.
\label{Hgamma}
\end{align}
In this representation, $J$ and $h$ favor competing dimerization patterns for the Majorana fermions.

At the self-dual point of the Ising model, $J=h$, the two dimerization terms compete to a draw, and the Majorana chain is therefore gapless.  Here the chain preserves a formal unitary symmetry
\begin{equation}
T: \gamma_j \to \gamma_{j+1} ~, \qquad i \to i ~,
\label{TIsing}
\end{equation}
as well as an anti-unitary symmetry
\begin{equation}
T': \gamma_j \to (-1)^j \gamma_{j+1} ~, \qquad i \to -i ~.
\label{T'Ising}
\end{equation}
The latter can be viewed as $T$ composed with time reversal $\mathcal{K}$ in the Ising chain that acts as simple complex conjugation in the $\sigma^z$ basis; in the Majorana representation we have
$\mathcal{K}: \gamma_j \to (-1)^{j+1} \gamma_j, i \to -i$.
Requiring either $T$ or $T'$ protects gaplessness of the Majorana chain, while $\mathcal{K}$ by itself of course does not.  We stress, however, that for any strictly 1D fermionic system both $T$ and $T'$ are anomalous in the sense that neither commutes with the total-fermion-parity operator $P = \prod_j (i\gamma_{2j-1} \gamma_{2j})$ (which in Ising language translates to $P = \prod_r \sigma^x_r$).  These  symmetries can nevertheless arise microscopically at the edge of a weak 2D topological superconductor composed of an array of 1D Kitaev chains; for example, $T$ would then correspond to a simple  translation by one wire that preserves the total electron parity.  

Deducing the action of $T$ and $T'$ on the spin variables requires some care, specifically regarding the origin of strings that appear under duality and in the definition of the fermions.
A careful treatment (see Appendix~\ref{app.ising} for details) yields 
\begin{align}
T \text{ or } T': \quad &\sigma_r^x ~\to~ \tau_{r+1/2}^x ~, \nonumber \\
&\sigma_r^z ~\to~ i \sigma_0^z \tau_{r+1/2}^z ~. 
\label{IsingT}
\end{align}
Thus, the anomalous fermionic symmetries $T$ and $T'$ implement duality for the original spin variables, modulo the additional factor $i \sigma_0^z$ that arises because of the non-local strings involved in our definitions of the dual operators. Note that $\sigma_0^z$ anticommutes with all $\tau_{r+1/2}^{z}$ operators, which ensures, e.g., that $\sigma_r^z \sigma_{r+1}^z \to \tau_{r+1/2}^z \tau_{r+3/2}^z$ as appropriate for duality.

\subsection{2D generalization}
\label{sec.2dising}

The symmetry-duality correspondence for the 1D quantum Ising chain can be generalized to higher-dimensional models.  As an illustration we will examine the following 2D square-lattice Hamiltonian,
\begin{align}
  H_\sigma =&-J_{z} \sum_{\bf r} \sigma^z_{\bf r} \sigma^z_{\bf r + \hat{x}}-h\sum_{\bf r} \sigma^x_{\bf r} \label{Ising2D}\\
& -K \sum_{\bf r} \sigma^z_{\bf r}\sigma^z_{\bf r + \hat{x}}\sigma^z_{\bf r + \hat{y}}\sigma^z_{\bf r + \hat{x} + \hat{y}}-J_{x} \sum_{\bf r} \sigma^x_{\bf r} \sigma^x_{\bf r + \hat{y}} ~.\nonumber
\end{align}  
Here ${\bf r} = (x,y)$ labels square-lattice sites delineated by integers $x,y$.  Importantly, the above Hamiltonian commutes with $\prod_{{\bf r} \in {\rm row}}\sigma^x_{\bf r}$, where the product runs over all ${\bf r}$ in \emph{any} row of the 2D lattice. 

Equation~\eqref{Ising2D} admits a number of noteworthy special cases: {\it (i)} For $K=J_{x}=0$, $H_\sigma$ describes an array of decoupled transverse-field Ising chains, which we discussed in the previous subsection. {\it (ii)} The $J_{x}=J_{z}=0$ limit  was studied by Xu and Moore in Ref.~\onlinecite{XuMoore} and shown to be self-dual when $h = K$. {\it (iii)} Finally, at $h = K =0$, $H_{\sigma}$ reduces to the quantum compass model (see Ref.~\onlinecite{RMPcompass} for a recent review). Reference~\onlinecite{ChenChen2007} formulated this limit of the model in terms of Jordan-Wigner fermions.  Moreover,  special cases {\it (ii)} and {\it (iii)} were shown to be related by a duality mapping in Ref.~\onlinecite{NussinovFradkin}.     

We adopt a straightforward extension of 1D Ising duality and define dual operators as \begin{subequations}
\begin{align}
 &\tau^z_{{\bf r + \hat{x}}/2} = \prod_{{\bf r'} < {\bf r +  \hat{x}}/2} \sigma^x_{\bf r'} ~, \\
&\tau^x_{{\bf r + \hat{x}}/2} = \sigma^z_{\bf r}\sigma^z_{\bf r + \hat{x}} ~.
\end{align}
\end{subequations}
In the first equation the string of $\sigma^x$'s begins at the bottom-left site, and runs rightward through each row until the termination at $\sigma^x_{\bf r}$ in ``typewriter'' fashion.  
The corresponding dual Hamiltonian reads
\begin{align}
H_\tau = &-J_{z}\sum_{\bf r} \tau^x_{{\bf r + \hat{x}}/2} -h \sum_{\bf r}
\tau^z_{{\bf r - \hat{x}}/2} 
\tau^z_{{\bf r + \hat{x}}/2}\nonumber\\& -K \sum_{\bf r} \tau^x_{{\bf r + \hat{x}}/2} \tau^x_{{\bf r + \hat{x}}/2 + {\bf \hat{y}}}
 \\
&-J_{x}\sum_{\bf r} \tau^z_{{\bf r - \hat{x}}/2}\tau^z_{{\bf r + \hat{x}}/2}\tau^z_{{\bf r - \hat{x}}/2 + {\bf \hat y}} \tau^z_{{\bf r + \hat{x}}/2 + {\bf \hat{y}}} ~ \nonumber,
\end{align}  
and, similar to the 1D Ising model, is self-dual when $h = J_{z}$ and $K=J_{x}$. (Appendix~\ref{app.2dising} discusses the relationship between self-duality of $H_\sigma$ and  self-duality of the Xu-Moore model from Ref.~\onlinecite{XuMoore}.)  Note that the spin conservation for each row ensures that the Hamiltonian remains local in terms of $\tau^{x,z}$ variables.  Pairwise exchanges between different rows, e.g., $\sigma^z_{\bf r} \sigma^z_{\bf r + \hat{y}}$, spoil these conserved quantities and would yield non-local terms in the dual Hamiltonian.

Combining order and disorder operators once again allows the model to be recast in terms of Majorana fermions:
\begin{subequations}
\begin{align}
&\Gamma({\bf r}-{\bf \hat x}/4) = \tau_{{\bf r - \hat{x}}/2}^z \sigma_{\bf r}^z ~, \label{Gamma2Da}\\
&\Gamma({\bf r}+{\bf \hat x}/4) = i \sigma_{\bf r}^z \tau_{{\bf r + \hat{x}}/2}^z ~.
\label{Gamma2Db}
\end{align}
\end{subequations}
In terms of relabeled Majorana operators
\begin{equation}
  \gamma_{j,y} \equiv \Gamma(j/2+1/4,y)
\end{equation}
we obtain the fermionized Hamiltonian
\small
\begin{align}
  &H_\gamma = \sum_{j,y} \frac{1}{2} \left[J_z+h + (-1)^j (J_z-h)\right] \(i \gamma_{j,y} \gamma_{j+1,y} \)\label{2DIsingMajorana}\\
  &+\sum_{j,y} \frac{1}{2}[K + J_x + (-1)^j(K-J_x)]\gamma_{j,y} \gamma_{j+1,y}\gamma_{j,y+1} \gamma_{j+1,y+1}~\nonumber.
\end{align}
\normalsize
At $h = J_z$ and $K = J_x$, corresponding to the self-dual point in the spin representation, $H_\gamma$ preserves the symmetries $T, T'$ defined in Eqs.~\eqref{TIsing} and \eqref{T'Ising} (with subscripts $y$ appended trivially). If one views the fermionic Hamiltonian as describing an array of strictly 1D Kitaev chains coupled through $J_x, K$, then both $T$ and $T'$ are anomalous in the same sense as in the previous subsection.  Both symmetries can, nevertheless, arise microscopically as simple, global-parity-preserving translations at the surface of a 3D weak topological superconductor. One can readily check that these anomalous symmetries implement duality for the 2D Ising variables, up to boundary operators---providing a higher-dimensional generalization of the well-known 1D result reviewed earlier. 

We remark as an aside that the fermionic Hamiltonian also preserves analogues of $T,T'$ that implement translations by one Majorana site along $y$ instead of $x$.  These symmetries are \emph{not} anomalous (in the wire-array realization noted above) and persist even away from the $h = J_z$, $K = J_x$ limit.  In general one can  construct strictly 2D Majorana systems that preserve simple translations along one---and only one---direction of the lattice.  

Critical properties of the model \eqref{Ising2D} near self-duality have, to our knowledge, not been explored. (The quantum compass model arising when $K=h=0$ is known to exhibit a first-order phase transition \cite{DorierBeccaMila}, but here we are interested in a different critical point that requires nonzero $K=J_x$.) The fermionic formulation, Eq.~\eqref{2DIsingMajorana}, suggests the possibility of interesting 2D phases analogous to the ``Exciton Bose Liquid'' \cite{ArunLeonMatthew,Sachdev2002,Balents2005,Xu2007,Motrunich2007,Tay2011} that may arise in boson models with ring exchange but no direct hopping. There, interactions spontaneously generate coherence between different rows/columns, yielding a 2D phase with emergent gapless fields (see Ref.~\onlinecite{Tay2011} for this viewpoint). Likewise, $H_\gamma$ for $h = J_z=0$ exclusively features ring-exchange terms that may give rise to new 2D phases.  (At non-zero $h = J_z$, Majorana fermions can only hop in the $x$-direction, though propagation along $y$ may be generated spontaneously by the ring-exchange interaction.) Simulations of the spin model \eqref{Ising2D} could provide direct evidence of such an exotic phase. The fermionic representation opens a window for complementary field-theoretic studies, and may be particularly enlightening in the limit of weakly coupled critical Ising chains.  The latter perspective moreover suggests a natural interpretation of the putative critical theory in terms of gapless Majorana fermions, likely coupled to an emergent gauge field.

\section{Particle-vortex duality for bosons}\label{sec.bosons}
We turn now to particle-vortex duality for bosons.  
This section first reviews the familiar Mott transition of bosons in $2+1$ dimensions from a viewpoint that facilitates an explicit mapping to fermions (see Sec.~\ref{sec.bosonization}).
In particular, we formulate the theory as an array of quantum wires hosting charge-$e$ bosons that can enter various 2D phases depending on the strength of interactions and inter-wire couplings.
We then introduce a non-local mapping from bosons to vortices and review two dualities in this framework: 
{\it (i)} Duality between short-range-interacting bosons and vortices coupled to a gauge field, and {\it (ii)} duality between bosons coupled to a Chern-Simons gauge field with coefficient $\pm 1$ and vortices coupled to a Chern-Simons gauge field with coefficient $\mp 1$.

\subsection{Coupled-wire formulation}

\label{sec:bosonwires}
Consider a 2D array of quantum wires enumerated by integers $y$, each hosting bosons $\Phi_y \sim e^{i \varphi_y}$ with density $\rho_y = \partial_x \theta_y/\pi$.
The $\varphi, \theta$ fields obey the commutator
\begin{align}
[\partial_x \theta_y(x), \varphi_{y'}(x')] = i \pi \delta_{y,y'} \delta(x-x') ~, \label{eqn.canonicalcommutator}
\end{align}
which implies that $\Phi_y^\dagger$ creates a  boson with unit charge as desired while $e^{2i \theta_y}$ creates a $2\pi$ phase slip in wire $y$.
We write the Euclidean action as
\begin{equation}
S = \int_{x, \tau} \sum_y \left[ \frac{i}{\pi} \partial_x \theta_y \, \partial_\tau \varphi_y + \mathcal{L}_\text{ LL} + \mathcal{L}_{\text{phase-slip}} + \mathcal{L}_\text{ hop} \right] ~.
\label{boson_action}
\end{equation}
Here
\begin{align}
{\cal L}_\text{LL} = \frac{v}{2\pi} (\partial_x \varphi_y)^2 + \frac{u}{2\pi} (\partial_x \theta_y)^2 ~
\label{LLL}
\end{align}
describes independent Luttinger liquids of bosons in each wire, with short-range density-density interactions encoded by the $u$ coupling. Inter-wire boson hoppings generate 
\begin{align}
{\cal L}_\text{hop} = -g_1 \cos(\varphi_{y+1} - \varphi_y) ~.
\label{Lhop}
\end{align}
Finally, when the boson density is commensurate with the underlying lattice, considering specifically integer density per site, each wire contains an additional term
\begin{align}
{\cal L}_\text{phase-slip} = -g_2 \cos(2 \theta_y) ~.
\label{Lumklapp}
\end{align}

Nonzero $g_1, g_2$ generically destabilize the decoupled boson Luttinger liquids.
When $g_1$ is relevant and flows to strong coupling, bosons can hop coherently between the wires and form a superfluid.
Conversely, when $g_2$ is relevant and flows to strong coupling, $2\pi$ phase slips proliferate and drive a transition to a Mott-insulator phase.

\begin{figure}
\includegraphics[width=\columnwidth]{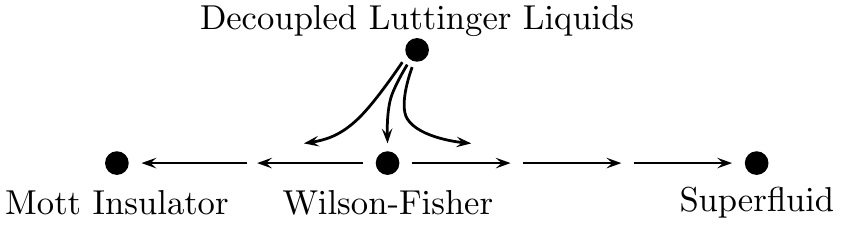}
\caption{Schematic RG flow of the bosonic wire model defined in Eq.~\eqref{boson_action}.}
\label{fig:boson}
\end{figure}

The boson density-density interaction $u$ in ${\cal L}_\text{LL}$ determines which of these competing couplings dominates.  
At weak coupling (small $g_1, g_2$), the renormalization group (RG) flow equations are
\begin{subequations}
\begin{align}
\frac{d g_1}{d \ell} &= \(2 - \frac{1}{2} \sqrt{\frac{u}{v}} \) g_1 ~,\\
\frac{d g_2}{d \ell} &= \(2 - \sqrt{\frac{v}{u}} \) g_2 ~, 
\end{align}
\end{subequations}
with $\ell$ a logarithmic rescaling factor.  
For strong repulsion, $u \gg v$, $g_1$ rapidly flows to zero while $g_2$ grows, resulting in the Mott insulator.
In the opposite limit $u \ll v$, $g_1$ is strongly relevant and hence fluctuations in the phase difference $\varphi_y -\varphi_{y+1}$ become massive, yielding the superfluid.
The two phases are divided by a separatrix that starts at $u = 2v$ for infinitesimal coupling.
Generically, higher-order terms in the RG such as renormalization of $u/v$ and generation of additional couplings will drive the system away from the $u = 2v$ condition and into one of the two phases; the separatrix is then a more complicated surface in parameter space.
When tuned to criticality, the RG flow along the separatrix terminates in the Wilson-Fisher fixed point (see Fig.~\ref{fig:boson}).

\subsection{Duality mapping}
\label{BosonDuality}

\begin{figure}[h]
\includegraphics[width=\columnwidth]{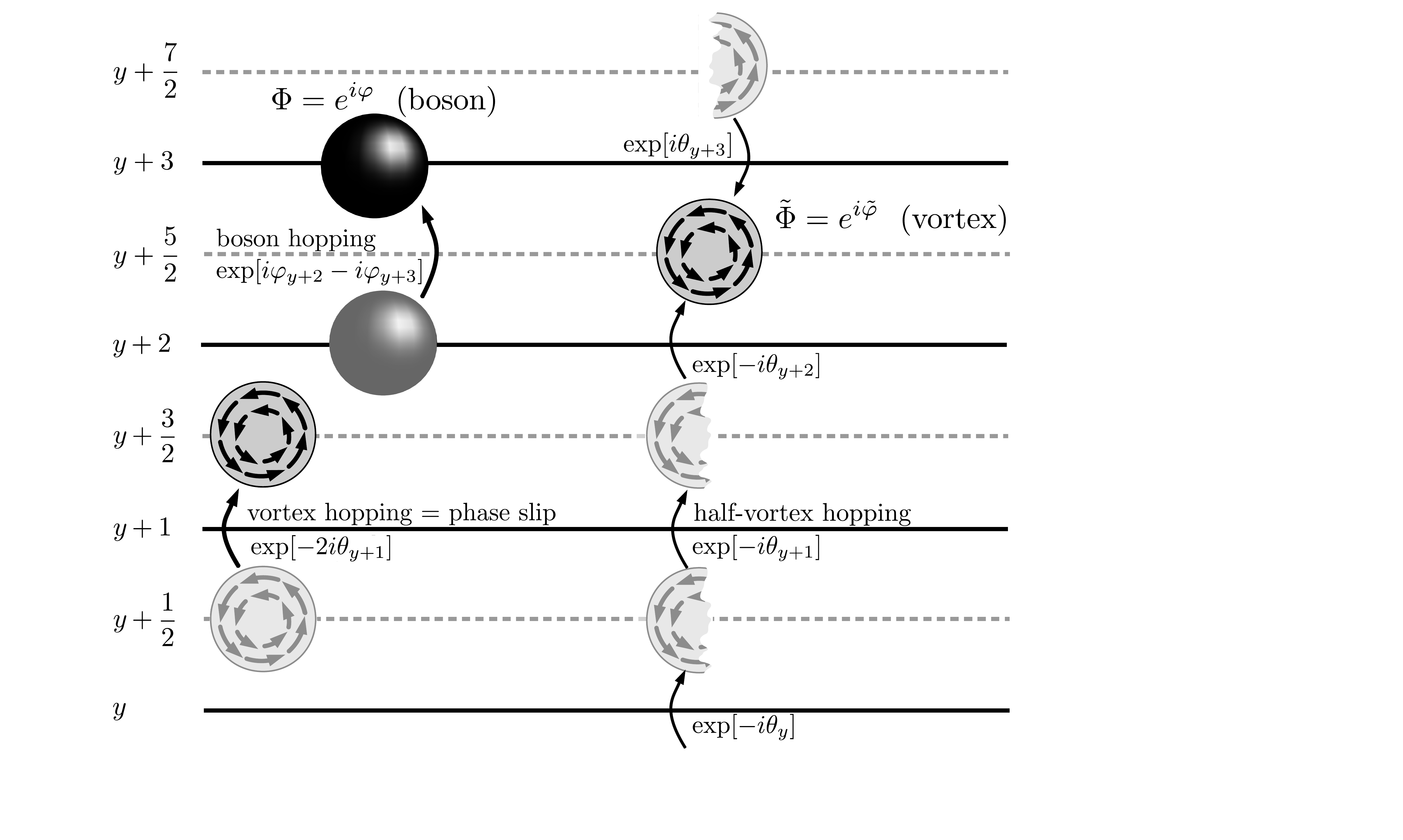}
\caption{Bosons $e^{i \varphi}$ reside on the direct lattice labeled by integers. 
Vortices $e^{i \tilde{\varphi}}$ live on the dual lattice labeled by half-integers.
Vortex hopping across wire $y$ is implemented by a phase slip $e^{-2 i \theta_y}$ on that wire.
Creating an isolated vortex requires a string of such operators emanating from infinity.
Here we define vortex creation operator by running ``half" of the string from $-\infty$ and the other half from $+\infty$.}
\label{fig:bos}
\end{figure}

To implement duality, we introduce canonically conjugate variables that are non-locally related to the original bosonic fields and can be interpreted as dual vortex degrees of freedom.
These new variables naturally live on wires labeled by half-integers (forming the dual lattice to the boson wires as shown in Fig.~\ref{fig:bos}) and are given by
\begin{subequations}
\begin{align}
&\tilde{\varphi}_{y+1/2} = \sum_{y'} \sgn\( y' - y - \frac{1}{2} \) \theta_{y'} ~, \label{tildephi} \\
&\tilde{\theta}_{y+1/2} = (\varphi_{y+1} - \varphi_y)/2 ~. \label{tildetheta}
\end{align}
\end{subequations}
Using Eq.~\eqref{eqn.canonicalcommutator}, one finds that the dual fields satisfy commutation relations identical to the original bosons,
\begin{align}
[\partial_x \tilde{\theta}_{y+1/2}(x), \tilde{\varphi}_{y'+1/2}(x')] = i \pi \delta_{y,y'} \delta(x-x') ~.
\label{eqn.vortexcommutator}
\end{align}
The inverse transformation,
\begin{subequations}
\begin{align}
&\varphi_y = -\sum_{y'} \sgn\(y' + \frac{1}{2} - y \) \tilde{\theta}_{y'+1/2} ~, \label{tildetheta2phi} \\
&\theta_y = -(\tilde{\varphi}_{y+1/2} - \tilde{\varphi}_{y-1/2})/2 ~,
\label{tildephi2theta}
\end{align}
\end{subequations}
is essentially the same as the duality mapping in Eqs.~(\ref{tildephi})-(\ref{tildetheta}) applied to the variables $\tilde{\varphi}, \tilde{\theta}$, up to an overall minus sign; schematically,
\begin{equation}
\tilde{\tilde{\varphi}} = -\varphi ~, \quad \tilde{\tilde{\theta}} = -\theta ~.
\label{DualitySquared}
\end{equation}

The operator $\tilde{\Phi}_{y+1/2} \sim e^{i \tilde{\varphi}_{y+1/2}}$ creates a phase slip of $\pi$ ($-\pi)$ on all boson wires below (above) $y+1/2$.
A $2\pi$ phase slip on any given wire may be viewed as tunneling of a $2\pi$ vortex across that wire. Consequently, $\tilde{\Phi}_{y+1/2}$ creates a $-2\pi$ vortex at $y+1/2$ by symmetrically pulling in two $\pi$ vortices, one from $y = -\infty$ and one from $y = \infty$.
The corresponding vortex density and current are
\begin{equation}
\rho_\text{vortex} = \frac{1}{\pi} \partial_x \tilde{\theta} ~, \qquad 
j_\text{vortex} = \frac{v_\text{vortex}}{\pi} \partial_x \tilde{\varphi} ~, \label{vortex_current}
\end{equation}
which directly parallel their boson counterparts,
\begin{equation}
\rho_\text{boson} = \frac{1}{\pi} \partial_x \theta ~, \qquad
j_\text{boson} = \frac{v_\text{boson}}{\pi} \partial_x \varphi ~. \label{boson_current}
\end{equation}

It is instructive to analyze symmetry transformations for the bosons and vortices.
Time-reversal symmetry acts on the bosonic variables as ${\cal T}: i \to -i,\ \varphi \to -\varphi,\ \theta \to \theta$. 
Using the definitions in Eqs.~\eqref{tildephi} and \eqref{tildetheta} one finds 
\begin{subequations}
\begin{align}
{\cal T}:&\ \ \Phi \to \Phi ~, \qquad\qquad\  &&\text{(boson time reversal)} \\
&\ \ \rho_\text{boson} \to \rho_\text{boson} ~, \nonumber \\
&\ \ j_\text{boson} \to -j_\text{boson} ~, \nonumber \\
{\cal T}:&\ \ \tilde{\Phi} \to \tilde{\Phi}^\dagger ~, \qquad\qquad &&\text{(vortex~particle-hole)} \\
&\ \ \rho_\text{vortex} \to -\rho_\text{vortex} ~, \nonumber \\
&\ \ j_\text{vortex} \to j_\text{vortex} ~. \nonumber
\end{align}
\end{subequations}
Similarly, one may define an anti-unitary particle-hole transformation of bosons as ${\cal C}: i \to -i,\ \varphi \to \varphi,\ \theta \to -\theta$.
Under this transformation one finds
\begin{subequations}
\begin{align}
{\cal C}:&\ \ \Phi \to \Phi^\dagger ~, \qquad\qquad &&\text{(boson particle-hole)} \\
&\ \ \rho_\text{boson} \to -\rho_\text{boson} ~, \nonumber \\
&\ \ j_\text{boson} \to j_\text{boson} ~, \nonumber \\
{\cal C}:&\ \ \tilde{\Phi} \to \tilde{\Phi} ~, \qquad\qquad\ &&\text{(vortex time reversal)} \\
&\ \ \rho_\text{vortex} \to \rho_\text{vortex} ~, \nonumber \\
&\ \ j_\text{vortex} \to -j_\text{vortex} ~. \nonumber
\end{align}
\end{subequations}
Both $\mathcal{T}$ and $\mathcal{C}$ act locally on bosons as well as on vortices, despite the non-local relation between the two fields.
We see above, however, that their role is effectively swapped: boson time-reversal acts as a particle-hole transformation for the vortex degrees of freedom and vice versa.

\subsection{Bosons with short-range interactions}
\label{BosonDualitySec}

The vortex variables introduced above allow us to rewrite the boson model reviewed in Sec.~\ref{sec:bosonwires}, thus obtaining a coupled-wire derivation of the familiar boson-vortex duality in $2+1$ dimensions.
We take the action to be
\begin{equation}
{\cal S} = \int_{x, \tau} \sum_y \left[ \frac{i}{\pi} \partial_x \theta_y \, \partial_\tau \varphi_y + {\cal L}^\text{boson} \right]
\end{equation}
with
\begin{align}
{\cal L}^\text{boson} =& \frac{v}{2\pi}(\partial_x \varphi_y)^2 + \frac{u}{2\pi}(\partial_x \theta_y)^2 + \frac{\tilde u}{8\pi}(\partial_x \Delta \varphi_y)^2 \\
& - g_1 \cos(\Delta \varphi_y) - g_2 \cos(2\theta_y) ~, \nonumber
\end{align}
where $\Delta \varphi_y \equiv \varphi_{y+1} - \varphi_y$.
Compared to Sec.~\ref{sec:bosonwires}, we have added the $\tilde{u}$ term, which describes specific interaction between neighboring wires; this is convenient for exposing the structure of the vortex theory but is inessential for qualitative physics.

Since the boson and vortex fields exhibit identical commutation relations, the Berry phase part of the action has an identical expression in terms of the vortex variables,
\begin{equation}
\int_{x, \tau} \sum_y \frac{i}{\pi} \partial_x \theta_y \, \partial_\tau \varphi_y = 
\int_{x, \tau} \sum_{\tilde{y}} \frac{i}{\pi} \partial_x \tilde{\theta}_{\tilde{y}} \, \partial_\tau \tilde{\varphi}_{\tilde{y}}.
\end{equation}
Here and below, $\tilde{y} = y + 1/2$ labels dual wires.
Expressing ${\cal L}^\text{boson}$ in terms of the vortex variables yields
\begin{align}
{\cal L}^\text{boson} =& \frac{v}{2\pi}(2 \Delta^{-1} \partial_x \tilde{\theta}_{\tilde{y}})^2 + \frac{u}{8\pi}(\partial_x \Delta \tilde{\varphi}_{\tilde{y}})^2 + \frac{\tilde{u}}{2\pi} (\partial_x \tilde{\theta}_{\tilde{y}})^2 \nonumber \\
& - g_1 \cos(2 \tilde{\theta}_{\tilde{y}}) - g_2 \cos(\Delta \tilde{\varphi}_{\tilde{y}})  ~. \label{eqn.bosonvortexdual1}
\end{align}
When writing the $v$ term, it was convenient to recast Eq.~\eqref{tildetheta} as $\tilde{\theta}_{y+1/2} = \frac{1}{2} \sum_{y'} \Delta_{y,y'} \varphi_{y'}$ with a matrix $\Delta_{y,y'} \equiv \delta_{y+1,y'} - \delta_{y,y'}$;
$\Delta^{-1}$ in Eq.~\eqref{eqn.bosonvortexdual1} is the inverse of this matrix and can be also read off Eq.~(\ref{tildetheta2phi}).
The $v$-term is clearly non-local in vortex variables and represents long-range interactions of vortices.

The non-local term can be replaced by
\begin{align*}
\sum_{\tilde{y}} \frac{v}{2\pi}(2 \Delta^{-1} \partial_x \tilde{\theta}_{\tilde{y}})^2
\to \sum_{\tilde{y}} \left[ -\frac{i}{\pi} \partial_x \tilde{\theta}_{\tilde{y}} \, \tilde a_{0, \tilde{y}} + \frac{(\Delta \tilde a_{0, \tilde{y}})^2}{8\pi v} \right] ~,
\end{align*}
where $\tilde a_0$ is a real-valued auxiliary field.
Performing the Gaussian integral over $\tilde a_0$ and using $\Delta^T \Delta = \Delta \Delta^T$ indeed yields precisely the first term of Eq.~\eqref{eqn.bosonvortexdual1}.
The new field $\tilde a_0$ can be viewed as the temporal component of a dynamical gauge field that mediates the long-range vortex interactions.

To bring out the gauge structure more clearly with an eye towards a $(2+1)$d continuum theory, it is convenient to additionally replace the $u$ term by
\begin{align*}
\sum_{\tilde{y}} \frac{u}{8\pi}(\partial_x \Delta \tilde{\varphi}_{\tilde{y}})^2 
\to & \sum_{\tilde{y}} \frac{u}{8\pi} [(\partial_x \tilde{\varphi}_{\tilde{y}} - \tilde a_{1, \tilde{y}})^2 + (\Delta \tilde a_{1, \tilde{y}})^2] \\
& + \sum_{\tilde{y}, \tilde{y}'} \frac{u}{8\pi} V_{\tilde{y}, \tilde{y}'} \, \partial_x (\Delta \tilde{\varphi}_{\tilde{y}}) \, \partial_x (\Delta\tilde{\varphi}_{\tilde{y}'}) ~,
\end{align*}
where in matrix notation $V \equiv \Delta^T [1 + \Delta^T \Delta]^{-1} \Delta$. Integrating over the real-valued $\tilde a_1$ (using $\Delta^T \Delta = \Delta \Delta^T$) recovers precisely the $u$ term of Eq.~\eqref{eqn.bosonvortexdual1}.
Note that $V_{\tilde{y}, \tilde{y}'}$ decays exponentially with distance $|\tilde{y} - \tilde{y}'|$.

Putting everything together, we have mapped the boson path integral onto a theory of vortices minimally coupled to a gauge field $\vect{a}$ in the specific gauge $\tilde a_2 = 0$.
Upon restoring $\tilde a_2$, the dual theory can be organized as 
\begin{align}
& {\cal L}^\text{dual} = {\cal L}^\text{vortex}_\text{wire} + {\cal L}^\text{vortex}_\text{gauge} + {\cal L}^\text{vortex}_\text{int} + {\cal L}^\text{vortex}_\text{hop\,+\,phase-slip} ~, \\
& {\cal L}^\text{vortex}_\text{wire} = \frac{u}{8\pi}(\partial_x \tilde{\varphi}_{\tilde{y}} - \tilde a_{1, \tilde{y}})^2 + \frac{\tilde{u}}{2\pi} (\partial_x \tilde{\theta}_{\tilde{y}})^2 - \frac{i}{\pi} \partial_x \tilde{\theta}_{\tilde{y}} \, \tilde a_{0, \tilde{y}} ~, \nonumber \\
& {\cal L}^\text{vortex}_{\text{gauge}} = \frac{1}{8\pi v} (\Delta \tilde a_{0, \tilde{y}} - \partial_\tau \tilde a_{2, \tilde{y}})^2 + \frac{u}{8\pi} (\Delta \tilde a_{1, \tilde{y}} - \partial_x \tilde a_{2, \tilde{y}})^2 ~, \nonumber \\
& {\cal L}^\text{vortex}_\text{int} = \frac{u}{8\pi} \sum_{\tilde{y}'} V_{\tilde{y}, \tilde{y}'} \, \partial_x (\Delta \tilde{\varphi}_{\tilde{y}} - \tilde a_{2, \tilde{y}}) \, \partial_x (\Delta \tilde{\varphi}_{\tilde{y}'} - \tilde a_{2, \tilde{y}'}) ~, \nonumber \\
& {\cal L}^\text{vortex}_\text{hop\,+\,phase-slip} = - g_1 \cos(2 \tilde{\theta}_{\tilde{y}}) - g_2 \cos(\Delta \tilde{\varphi}_{\tilde{y}} - \tilde a_{2, \tilde{y}})  ~, \nonumber
\end{align}
where ${\cal L}^\text{vortex}_\text{wire}$ is the intra-wire vortex kinetic energy, ${\cal L}^\text{vortex}_\text{gauge}$ is the Maxwell term for the dynamical gauge field, ${\cal L}^\text{vortex}_\text{int}$ encodes short-ranged vortex interaction, and ${\cal L}^\text{vortex}_\text{hop\,+\,phase-slip}$ contains both the inter-wire vortex hopping ($g_2$) and vortex phase-slip ($g_1$) terms.
Observing that the last term in ${\cal L}^\text{vortex}_\text{wire}$ naturally combines with the Berry phase term $\frac{i}{\pi} \partial_x \tilde{\theta}_{\tilde{y}} \,  \partial_\tau \tilde{\varphi}_{\tilde{y}}$, we see that the complete theory is gauge invariant.
Note that in our specific microscopic model, we do not obtain a bare Maxwell term $\sim (\partial_\tau \tilde a_1 - \partial_x \tilde a_0)^2$; however, such terms will be generated under coarse-graining, with their form dictated by gauge invariance.

We have thus established duality of the bosonic theory and the Higgs model in terms of vortices.
In particular, a Mott insulator of bosons where $\cos(2\theta)$ flows to strong coupling corresponds to a vortex condensate governed by $\cos(\Delta \tilde{\varphi})$ and vice versa.
Such correspondences between the two theories are of course well understood, and our main goal in this section was to show how the dynamical dual gauge field appears in the coupled-wire approach as a way to encode non-local interactions of vortices. 

\subsection{Bosons with Chern-Simons coupling}
\label{subsec.boson+CS}
The particle-vortex duality we have just reviewed maps bosons with short-range interactions onto vortices with long-range interactions.
More generally, when the interactions between bosons are mediated by a field with propagator $\Pi(k)$, then the interactions between vortices are mediated by another field with a propagator $\tilde{\Pi}(k) \sim \frac{1}{k^2 \Pi(k)}$.
This suggests that an intermediate-range interaction $\Pi(k) \sim k^{-1}$ may result in a self-dual model.
Such scaling is exhibited by a gauge field governed by the Chern-Simons action $\epsilon_{\mu\nu\kappa} a_\mu \partial_\nu a_\kappa$.

In the wire construction, bosons coupled to a Chern-Simons gauge field (in the $a_2 = 0$ gauge) are described by
\begin{align}
& {\cal L}^\text{boson-CS} = {\cal L}_0 + {\cal L}_\text{CS} + {\cal L}_\text{hop} + {\cal L}_\text{phase-slip} ~. \label{csbosons}
\end{align}
The first two terms read
\begin{align}
{\cal L}_0 =& \frac{v}{2\pi}(\partial_x \varphi_y - a_{1,y})^2 +  \frac{u}{2\pi} (\partial_x \theta_y)^2 \nonumber \\
& + \frac{\tilde{u} - v}{8\pi} (\partial_x \Delta \varphi_y)^2 - \frac{i}{\pi} \partial_x \theta_y \frac{a_{0, y+1/2} + a_{0, y-1/2}}{2} ~, \nonumber \\
{\cal L}_\text{CS} =& s\frac{i}{2\pi} a_{1,y} (\Delta a_0)_y ~, \nonumber
\end{align}
where $(\Delta a_0)_y = a_{0, y+1/2} - a_{0, y-1/2}$ and $ s = \pm 1$; the last two terms are once again given by Eqs.~\eqref{Lhop} and \eqref{Lumklapp}.
We will see that this model, for specific choices of parameters already anticipated in the above expressions, can realize self-duality and hence criticality exactly on the wire scale.

As discussed below, the Chern-Simons term $\mathcal{L}_\text{CS}$ attaches one flux quantum to each boson, with an orientation set by the sign $s$.
Note that the $a_0$ field here resides \emph{between} the boson wires.
This merely represents a convenient choice for enforcing a constraint on the flux---which also lives between wires (see below)---and makes subsequent manipulations particularly transparent.  In principle, one can further add a Maxwell term to the theory.  However, since the Maxwell term scales as $k^2$ while the Chern-Simons term scales as $k$, one expects it to not have a qualitative effect. Appendix~\ref{app.onwireboson} studies a generalized model featuring a Maxwell term together with a gauge field $a_0$ residing on the wires. Upon integrating out the gauge field in either model, Eq.~\eqref{csbosons} or Eq.~\eqref{app.csbosons}, we indeed recover the same (non-local) Lagrangian.

To dualize Eq.~\eqref{csbosons}, we first rewrite the coupling of the gauge field to the boson density using
\begin{align}
& \sum_y \partial_x \theta_y (a_{0, y+1/2} + a_{0, y-1/2}) \label{csbosoncnst1}\\
& = \sum_y (\Delta a_0)_y \sum_{y' \neq y} \sgn(y' - y) \partial_x \theta_{y'} ~. \nonumber
\end{align}
Integrating out the gauge field $a_0$ then yields the constraint
\begin{equation}
a_{1, y} = s\sum_{y' \neq y} \sgn(y' - y) \partial_x \theta_{y'} ~. \label{a1theta}
\end{equation}
Since we are in the $a_2 = 0$ gauge, Eq.~\eqref{a1theta} implies that the flux obeys $a_{1, y} - a_{1, y+1} = 2\pi s (\rho_y + \rho_{y+1})/2 $ with $\rho_y = \partial_x \theta_y/\pi$. That is, the flux of the gauge field between two neighboring wires is $2\pi s$ times the average boson density.
In the spirit of the formal flux-attachment approach,\cite{FradkinCS} we could equally well consider the model with dynamical fields $\varphi, \theta$ and $a_1$ fixed by Eq.~(\ref{a1theta}) as defining bosons with Chern-Simons interactions.  We will adopt this viewpoint and hereafter discard terms involving $a_0$ but retain the constraint in Eq.~\eqref{a1theta}.  

It is convenient to now organize the remaining terms in $\mathcal{L}_0 + \mathcal{L}_\text{CS}$ by their $\varphi$ and $\theta$ content:
\begin{subequations}
\begin{align}
&{\cal L}_\varphi = \frac{v}{2\pi}(\partial_x \varphi_y)^2 + \frac{\tilde{u} - v}{8\pi} [\partial_x (\varphi_{y+1} - \varphi_y)]^2 \label{lphi} \\
&\quad\ = \frac{v}{8\pi} [\partial_x (\varphi_{y+1} + \varphi_y)]^2
+ \frac{\tilde{u}}{8\pi} [\partial_x (\varphi_{y+1} - \varphi_y)]^2 ~, \nonumber \\
&{\cal L}_\theta = \frac{v}{2\pi} a_{1, y}^2 + \frac{u}{2\pi} (\partial_x \theta_y)^2 ~, \label{ltheta} \\
&{\cal L}_{\varphi,\theta} = -\frac{v}{\pi} \partial_x \varphi_y \, a_{1, y} ~. \label{lthetathetaphi}
\end{align}
\end{subequations}
In the first line we implicitly used summation over $y$ to regroup the $v$ terms.
The above writing is convenient for the duality transformation [Eqs.~\eqref{tildetheta2phi} and \eqref{tildephi2theta}] since the $\tilde{u}$ and $u$ terms interchange under this map, just like the $g_1$ (hopping) and $g_2$ (phase-slip) terms.
It is also easy to check that $a_{1, y}$ in Eq.~\eqref{a1theta} can be expressed in terms of the dual phase variables as $a_{1, y} = s\partial_x (\tilde{\varphi}_{y+1/2} + \tilde{\varphi}_{y-1/2})/2$. 
Introducing
\begin{align}
\tilde{a}_{1, y+1/2} &\equiv -s\sum_{y' \neq y} \sgn(y' - y) \partial_x \tilde{\theta}_{y'+1/2}
\label{tildea1theta} \\
&= s \frac{\partial_x (\varphi_{y+1} + \varphi_y)}{2 } ~, \nonumber
\end{align}
we can express the model entirely in terms of the dual variables as
\begin{subequations}
\begin{align}
& {\cal L}_\varphi = \frac{v}{2\pi} (\tilde{a}_{1, y+1/2})^2 + \frac{\tilde{u}}{2\pi} (\partial_x \tilde{\theta}_{y+1/2})^2 ~, \label{lphidual} \\
& {\cal L}_\theta = \frac{v}{8\pi} [\partial_x (\tilde{\varphi}_{y+1/2} + \tilde{\varphi}_{y-1/2})]^2 \nonumber\\ 
& \quad\quad\ + \frac{u}{8\pi} [\partial_x (\tilde{\varphi}_{y+1/2} - \tilde{\varphi}_{y-1/2})]^2 ~, \label{lthetadual} \\
& {\cal L}_{\varphi, \theta} = -\frac{v}{\pi} \partial_x \tilde{\varphi}_{y+1/2} \, \tilde{a}_{1, y+1/2} ~, \label{lphithetadual}
\end{align}
\end{subequations}
where in the last line we again implicitly used summation over $y$ to regroup the terms.

We see that under the duality the total Lagrangian expressed in terms of vortices coupled to a (new) Chern-Simons gauge field takes the same form as before.
Upon restoring the temporal component $\tilde a_0$ to enforce the flux-attachment constraint, the theory is described by Eq.~\eqref{csbosons} with $u \leftrightarrow \tilde{u}$, $g_1 \leftrightarrow g_2$; moreover,
because of the sign difference between Eqs.~\eqref{a1theta} and \eqref{tildea1theta}, the Chern-Simons term in the vortex theory has opposite sign compared to the original bosons.
The second vertical arrow in Eq.~\eqref{duality2} sketches this duality in the continuum language.

For the specific parameters $g_1 = g_2$ and $\tilde{u} = u$, the model is exactly self-dual in the following precise sense: The Euclidean actions ${\cal S}^\text{boson-CS}[\varphi,\theta,\vec a]$ and ${\cal S}^\text{vortex-CS}[\tilde\varphi,\tilde\theta,\vec{\tilde a}]$ for bosons and vortices satisfy 
\begin{align}
{\cal S}^\text{vortex-CS}[\tilde{\varphi}, \tilde{\theta}, \vec{\tilde{a}}] =
\left({\cal S}^\text{boson-CS}[\varphi \to \tilde{\varphi}, \theta \to -\tilde{\theta}, \vec{a} \to \vec{\tilde{a}}] \right)^* ~, \label{bosoncomplexaction}
\end{align}
i.e., the vortex action in terms of the dual fields has identical form to that of the original action in terms of the original fields, up to complex conjugation of c-numbers, and sign change of one of the conjugate fields.

We note that the complex conjugation on the r.h.s.\ does not affect the location of the critical point in terms of microscopic parameters. The sign $-\tilde\theta$ can be absorbed by a trivial re-definition of the integration variable in the path integral. This is essentially a statement that two systems related by an anti-unitary transformation have spectra and partition sums related by complex conjugation.
Moreover, the complex conjugation will be reflected in the interpretation of duality as an \textit{anti-unitary} symmetry in Sec.~\ref{subsec.symmetries}.

This self-dual point corresponds to one of three possibilities: {\it (i)} a quantum critical point describing a continuous phase transition, {\it (ii)} a first-order phase transition, or {\it (iii)} an intermediate self-dual phase.
In Sec.~\ref{sec.bosonization} we describe a second non-local mapping to Dirac fermions, which makes the fate of this model apparent. There we will see that scenario {\it (i)} occurs and that the critical theory is equivalent to a single massless Dirac fermion.   

\section{Particle-vortex duality for Dirac fermions}
\label{sec.fermions}
Next we review particle-vortex duality for Dirac fermions in $(2+1)$ dimensions \cite{Son,WangSenthil2015,MetlitskiVishwanath2015,WangSenthilReview,metlitskiduality,diracduality} As for the bosons, we will formulate the theory of a free Dirac cone as an array of quantum wires with (weak) inter-wire coupling.
We will then exploit a non-local mapping from fermions to dual fermions to establish 
{\it (i)} duality between free Dirac fermions and QED$_3$, and 
{\it (ii)} duality between Dirac fermions coupled to a Chern-Simons gauge field with coefficient $\pm 1/2$ and dual fermions coupled to a Chern-Simons gauge field with coefficient $\mp 1/2$.

\subsection{Coupled-wire formulation}
\label{WireModelDirac}

Our starting point is an array of quantum wires, which we label by integers $j$ to distinguish from the bosonic case discussed previously.
Each wire contains a single chiral fermion $\psi_j$ with alternating chirality from one wire to the next.
Such a setup can arise, e.g., from a network of $\nu = 1$ integer-quantum-Hall edge states or magnetic domain walls on a 3D-topological-insulator surface.  
We write the action as ${\cal S} = \int_{x, \tau} \sum_j \left[\psi^\dagger_j \partial_\tau \psi_j + {\cal L}_\text{chiral fermion} + {\cal L}_\text{tunnel} \right]$.
Here 
\begin{align}
{\cal L}_\text{chiral fermion} = (-1)^j v \psi^\dagger_j (-i \partial_x) \psi_j
\end{align}
encodes intra-wire kinetic energy, where the factor $(-1)^j$ accounts for the staggered chirality.
The last term,
\begin{align}
{\cal L}_\text{tunnel} = g_j (i \psi_j^\dagger \psi_{j+1} + \Hc) ~,
\label{L_tunnel}
\end{align}
with $g_j = g_1$ or $g_2$ for odd or even $j$, describes inter-wire tunneling in our model.
(This model is slightly different from the one we used in Ref.~\onlinecite{diracduality} but is more convenient here due to different choice of Klein factors naturally arising in the present setting, see below).

To expose the Dirac physics it is convenient to combine the counter-propagating modes on adjacent even and odd wires into a two-component spinor:
\begin{align}
\Psi = \begin{pmatrix} \psi_\text{even} \\ \psi_\text{odd} \end{pmatrix} ~.
\label{Psi_spinor}
\end{align}
Fourier-transforming yields an action for frequency and momentum modes ${\cal S} = \int_{k_x, \omega} \sum_{k_y} \left[-i \omega \Psi_{\vect{k}}^\dagger \Psi_{\vect{k}} + \Psi_{\vect{k}}^\dagger h_{\vect{k}} \Psi_{\vect{k}} \right]$ with
\begin{align}
h_{\vect{k}} = v k_x \sigma_z - g_+ \sin(k_y) \sigma_x + g_- \cos(k_y) \sigma_y ~,
\end{align}
where $g_\pm = g_1 \pm g_2$.
We will assume that $g_1$ and $g_2$ have the same sign, so that $g_+$ is always finite while $g_-$ can be tuned to zero.
The above Hamiltonian describes a single Dirac cone centered at $k_x = 0, k_y = 0$ with mass $g_-$.
The mass term is odd under the time-reversal-like anti-unitary transformation
\begin{subequations}
\begin{align}
&{\cal T}' \psi_j {\cal T}'^{-1} = (-1)^j \psi_{j+1} ~, \label{calT'} \\
&{\cal T}' \Psi {\cal T}'^{-1} = i \sigma_y \Psi ~.
\end{align}
\end{subequations}
The presence of $\mathcal{T}'$ symmetry thus precludes a mass, yielding a gapless Dirac cone that is protected against weak perturbations.
This symmetry can be realized microscopically at the surface of a 3D topological insulator but requires fine-tuning in strict 2D setups.

It will also be useful to consider the following anti-unitary particle-hole transformation
\begin{subequations}
\begin{align}
&{\cal C}' \psi_j {\cal C}'^{-1} = (-1)^j \psi_{j+1}^\dagger ~, \label{calC'} \\
&{\cal C}' \Psi {\cal C}'^{-1} = i \sigma_y \Psi^\dagger ~,
\end{align}
\end{subequations}
and the mass term is odd under this transformation as well.
Both ${\cal T}'$ and ${\cal C}'$ are symmetries of the above model when $g_1 = g_2$, but it is easy to consider modifications that have only one or the other present.

For the purpose of this paper, we will mainly use the following bosonized ("phase") representation of the model.
We write $\psi_j \sim e^{i \phi_j}$, where $\phi_j$ is a chiral field satisfying
\begin{align}
[\phi_j(x), \phi_{j'}(x')] =& \delta_{jj'} (-1)^j \, i \pi \, \sgn(x-x') \nonumber \\
&+ (1 - \delta_{jj'}) \, i \pi \, \sgn(j'-j) ~.
\label{phicommutator}
\end{align}
The first and second lines respectively encode proper fermion anticommutation relations within and between wires
(the specific choices are made to coincide with the "flux attachment" procedure on bosons introduced later in Sec.~\ref{sec.bosonization}).
The action becomes
\begin{align}
{\cal S} ~=~ \int_{x, \tau} \sum_j \left[ \frac{i (-1)^j}{4\pi} \partial_x \phi_j \partial_\tau \phi_j + {\cal L}_\text{chiral fermion} + {\cal L}_\text{tunnel} \right] ~,
\label{Lchiralbosonized}
\end{align} 
with
\begin{align}
&{\cal L}_\text{chiral fermions} = \frac{v}{4\pi} (\partial_x \phi_j)^2 ~, \\
&{\cal L}_\text{tunnel} = -g_j \cos(\phi_j - \phi_{j+1}) ~. \label{L_tunnel_phi}
\end{align}
In the last line we used $\exp([\phi_j, \phi_{j+1}]/2) = i$ which follows from Eq.~\eqref{phicommutator}.

\begin{figure}
\includegraphics[width=\columnwidth]{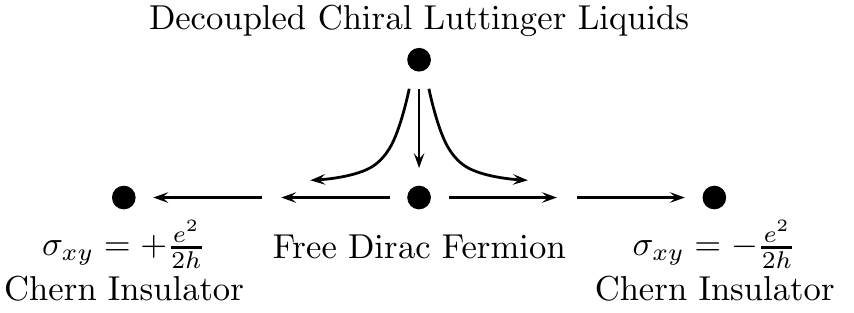}
\caption{Schematic RG flow for the array of counter-propagating chiral wires, viewed as arising on the surface of a 3D topological insulator.
Systems that preserve the time-reversal symmetry defined in Eq.~\eqref{calT'} are confined to the separatrix and flow to the free-Dirac-fermion fixed point.}
\label{fig:dirac}
\end{figure}

In the phase variables, the theory does not readily permit an exact solution due to the cosine terms.
We can nevertheless perform an RG analysis as for the Wilson-Fisher model:
Both $g_1$ and $g_2$ have scaling dimension one and grow under RG.
When $g_1 \gg g_2$, ${\cal L}_\text{tunnel}$ opens a gap by hybridizing each odd $j$ wire with its neighbor at $j+1$.
For $g_2 \gg g_1$, each odd $j$ wire hybridizes instead with its neighbor at $j-1$.
These gapped phases correspond to insulators whose Hall conductances differ by $e^2/h$ (which can be seen by introducing a boundary between the two phases at some $y$ and examining edge states).
When $g_1 = g_2$, the two competing cosines prevent each other from opening a gap, and the system is critical.
In this case, the time-reversal symmetry ${\cal T}'$ ensures that the system remains on the separatrix and flows to the fixed point of a free Dirac fermion.
Figure~\ref{fig:dirac} summarizes the flows for this model.

\subsection{Duality mapping}

Reference~\onlinecite{diracduality} leveraged the coupled-wire description reviewed above to derive an explicit duality transformation for the single Dirac fermion.
The duality proceeds by defining a new chiral field
\begin{equation}
\tilde{\phi}_j = \sum_{j' \neq j} \sgn(j-j') (-1)^{j'} \phi_{j'} ~. \label{eqn.dualfermions}
\end{equation}
Starting with the commutator of the original field, Eq.~(\ref{phicommutator}), and using properties of a matrix $D_{jj'} \equiv (1 - \delta_{jj'}) \sgn(j-j') (-1)^{j'}$ reviewed in App.~\ref{app.matrix}, it is straightforward to verify that the new field satisfies
\begin{align}
[\tilde{\phi}_j(x), \tilde{\phi}_{j'}(x')] = -[\phi_j(x), \phi_{j'}(x')] ~.
\end{align}
Dual fermions are then given by $\tilde{\psi}_j \sim e^{i \tilde{\phi}_j}$ and have opposite chiralities to the original fermions.
Important property of the dual field is
$\tilde{\phi}_{j+1} - \tilde{\phi}_j = (-1)^{j+1} (\phi_{j+1} - \phi_j)$,
so the inter-wire hopping of the original fermions is local also in the dual fermions.
The specific tunneling Hamiltonian Eq.~(\ref{L_tunnel}) expressed in phase variables in Eq.~(\ref{L_tunnel_phi}) becomes in the dual variables
\begin{align}
{\cal L}_\text{tunnel} = -g_j \cos(\tilde{\phi}_j - \tilde{\phi}_{j+1}) = g_j (-i \tilde{\psi}_j^\dagger \tilde{\psi}_{j+1} + \Hc) ~.
\label{L_tunnel_tildephi}
\end{align}

As in Eq.~\eqref{Psi_spinor}, it is useful to organize the fields on even and odd wires into a two-component spinor,
\begin{align}
\tilde{\Psi} = \begin{pmatrix} \tilde{\psi}_\text{even} \\ \tilde{\psi}_\text{odd} \end{pmatrix} ~.
\label{Psi_spinor_dual}
\end{align}
The density and current (in the $x$ direction) for Dirac fermions and dual Dirac fermions are given by
\begin{subequations}
\begin{align}
&\rho_\text{Dirac} = \Psi^\dagger \Psi ~, &&j_\text{Dirac} = v_\text{Dirac} \Psi^\dagger \sigma_z \Psi ~, \\
&\rho_\text{dual~Dirac} = \tilde{\Psi}^\dagger \tilde{\Psi} ~, &&j_\text{dual~Dirac} = v_\text{dual~Dirac} \tilde{\Psi}^\dagger \sigma_z \tilde{\Psi} ~.
\end{align}
\end{subequations}
Note that with these choices, the continuum dual Dirac fermions have opposite chirality compared to the direct Dirac fermions, i.e., $\sgn(v_\text{dual Dirac}) = -\sgn(v_\text{Dirac})$.

A proper treatment of the time-reversal and particle-hole transformations in terms of the phase fields used in the duality map requires some care and is described in App.~\ref{app:fermsymm}.
Here we can take a shortcut by considering more general inter-wire hopping allowing us to discuss ${\cal T}'$ and ${\cal C}'$ separately:
\begin{align*}
\sum_j (e^{i \alpha_j} \psi_j^\dagger \psi_{j+1} + \Hc) = \sum_j (e^{i \tilde{\alpha}_j} \tilde{\psi}_j^\dagger \tilde{\psi}_{j+1} + \Hc) ~,
\end{align*}
where easy calculation gives
$e^{i \tilde{\alpha}_\text{even}} = e^{-i \alpha_\text{even}}$,
$e^{i \tilde{\alpha}_\text{odd}} = -e^{i \alpha_\text{odd}}$.
Presence of ${\cal T'}$ would require $e^{i \alpha_\text{odd}} = -e^{-i \alpha_\text{even}}$, while presence of ${\cal C}'$ would require $e^{i \alpha_\text{odd}} = e^{i \alpha_\text{even}}$, and these conditions get swapped for the $\tilde{\alpha}$.
It is therefore natural to conclude that the action of the two symmetries also gets swapped under duality, 
${\cal T}': \tilde{\psi}_j \to (-1)^j \tilde{\psi}_j^\dagger$,
${\cal C}': \tilde{\psi}_j \to (-1)^j \tilde{\psi}_j$.  
 
Summarizing in terms of the continuum fields,
we have
\begin{subequations}
\begin{align}
&{\cal T}': \Psi \to i \sigma^y \Psi \qquad &&\text{(fermion time reversal)} \\
&{\cal T}': \tilde{\Psi} \to i \sigma^y \tilde{\Psi}^\dagger \qquad &&\text{(dual-fermion particle-hole)}.
\end{align}
\end{subequations}
and 
\begin{subequations}
\begin{align}
&{\cal C}': \Psi \to i \sigma^y \Psi^\dagger \qquad &&\text{(fermion particle-hole)} \\
&{\cal C}': \tilde{\Psi} \to i \sigma^y \tilde{\Psi} \qquad &&\text{(dual-fermion time reversal)}.
\end{align}
\end{subequations}
Both symmetries are preserved by the wire model defined in Sec.~\ref{WireModelDirac} at the special point $g_1 = g_2$.
As for the case of bosons, time-reversal and particle-hole transformations effectively swap roles for the original and dual Dirac fermions.
Moreover, both of these transformations act locally on each field despite the nonlocality of the duality mapping.

\subsection{Free Dirac fermions and QED$_3$}

Dualizing the free-Dirac-cone action in Eq.~\eqref{Lchiralbosonized} using Eq.~\eqref{eqn.dualfermions} yields a nonlocal theory (in particular because of the $v$ term).
The nonlocality reflects long-range interactions among the dual fermions, as encountered for the bosonic case in Sec.~\ref{BosonDualitySec}.
One can similarly restore locality here at the expense of introducing a gauge field $\vect{\tilde{a}}$ that mediates the interactions.  
Reference~\onlinecite{diracduality} carried out this calculation and showed that the free Dirac cone can be mapped to a coupled-wire realization of QED$_3$.
In the $\tilde{a}_2 = 0$ gauge the action is 
${\cal S}~=~\int_{x, \tau} \sum_j \left[ \frac{-i (-1)^j}{4\pi} \partial_x \tilde{\phi}_j \partial_\tau \tilde{\phi}_j + {\cal L}_\text{QED$_3$} \right]$ with
\begin{align}
&{\cal L}_\text{QED$_3$} = {\cal L}_0 + {\cal L}_\text{staggered-CS} + {\cal L}_\text{MW} + {\cal L}_\text{tunnel} ~, \label{LQED3} \\
&{\cal L}_0 = \frac{i (-1)^j}{2\pi} \partial_x \tilde{\phi}_j \, \tilde {a}_{0,j} + \frac{u}{4\pi} (\partial_x \tilde{\phi}_j - \tilde{a}_{1,j})^2 ~, \nonumber \\
&{\cal L}_\text{staggered-CS} = \frac{i (-1)^j}{8\pi}(\Delta \tilde{a}_{0,j}) (\tilde{a}_{1,j+1} + \tilde{a}_{1,j}) ~, \nonumber \\
&{\cal L}_\text{MW} = \frac{1}{16\pi} \left[\frac{1}{v} (\Delta \tilde{a}_{0,j})^2 + v (\Delta \tilde{a}_{1,j})^2 \right] ~, \nonumber
\end{align}
where $\Delta \tilde{a}_{0,j} \equiv \tilde{a}_{0,j+1} - \tilde{a}_{0,j}$, and similarly for $\Delta \tilde{a}_{1,j}$.

The ${\cal L}_\text{staggered-CS}$ contribution is required to ensure gauge-invariance of the coupled-wire formulation, but drops out in the long-wavelength limit.
We have included a ``bare velocity'' $u$ for the dual fermions into the wire model whose precise value is immaterial---it drops out once the integral over the gauge field is performed.
The dual-fermion velocity in the long-wavelength limit is instead determined by $v$ (see the supplementary material of Ref.~\onlinecite{diracduality} for a discussion of this point).
Crucially, inter-wire tunneling in the last term of Eq.~\eqref{Lchiralbosonized} takes the same form when written in terms of dual fermions, i.e., 
${\cal L}_\text{tunnel}[\phi] = {\cal L}_\text{tunnel}[\tilde{\phi}]$, 
cf.~Eq.~(\ref{L_tunnel_tildephi}). Reference~\onlinecite{diracduality} used identifications provided by this mapping to deduce some non-trivial properties of the strongly-coupled QED$_3$ theory.

\subsection{Dirac fermions with Chern-Simons coupling}
\label{subsec.ferm-CS}

As a new application of the coupled-wire duality approach, we now wish to dualize a variation of Eq.~\eqref{Lchiralbosonized} that includes a level-1/2 Chern-Simons term for the original Dirac fermions:
\begin{align}
&{\cal L}^\text{ferm-CS} = {\cal L}_0 + {\cal L}_\text{staggered-CS} + {\cal L}_\text{MW} + {\cal L}_\text{CS} + {\cal L}_\text{tunnel} ~, \label{LfermCS} \\
&{\cal L}_0 = -\frac{i (-1)^j}{2\pi} \partial_x \phi_j \, a_{0,j} + \frac{u}{4\pi}(\partial_x \phi_j - a_{1,j})^2 ~, \nonumber \\
&{\cal L}_\text{staggered-CS} = -\frac{i (-1)^j}{8\pi}(\Delta a_{0,j})(a_{1,j+1} + a_{1,j}) ~, \nonumber \\
&{\cal L}_\text{MW} = \frac{1}{8\pi} \left[\frac{1}{v} (\Delta a_{0,j})^2 + v (\Delta a_{1,j})^2 \right] ~, \nonumber \\
&{\cal L}_\text{CS} = s \frac{i}{8\pi}(\Delta a_{0,j})(a_{1,j+1} + a_{1,j}) ~, \nonumber
\end{align}
with $s = \pm 1$.
Note the similarity to the dual-fermion theory obtained in the previous subsection, the main difference being the addition of $\cal{L}_{\text{CS}}$ and the different signs in ${\cal L}_0$ and ${\cal L}_\text{staggered-CS}$ due to opposite fermion chirality. One can verify\cite{diracduality} that the above action can be obtained from a gauge-invariant model in terms of $\vect{a} = (a_0, a_1, a_2)$ in the gauge with $a_2 = 0$, and that the Chern-Simons term in the long-wavelength limit corresponds to $s \frac{i}{8\pi} \vect{a} \cdot \vect{\nabla} \times \vect{a}$.
For convenience we have also chosen a Maxwell term with a specific relation between couplings of the exhibited parts and zero coupling for the $(\partial_\tau a_1 - \partial_x a_0)^2$ part.

The Lagrangian $\cal{L}_\text{ferm-CS}$ turns out to be self-dual under the fermion duality defined in Eq.~\eqref{eqn.dualfermions}.
Upon integrating out the gauge field (see Appendix~\ref{app.csfermnew} for details) one finds
\begin{align}
{\cal L}^\text{ferm-CS} \to & \frac{v_B}{16\pi} (\partial_x \phi_j + s \partial_x \tilde{\phi}_j)^2 \label{eqn.csfermnonloc} \\
& + \frac{u_B}{16\pi} (1 + s (-1)^j) \left( \Delta \partial_x \phi_j \right)^2 + {\cal L}_\text{tunnel} \nonumber ~, 
\end{align}
with
\begin{equation}
v_B = \frac{v (2u + v)}{u + v} ~, \quad
u_B = \frac{v (u + v)}{u + 2v} ~.
\label{vBuB}
\end{equation}
Note that $[\Delta \partial_x \phi_j]^2 = [\Delta \partial_x \tilde{\phi}_j]^2$.
Recalling also Eq.~\eqref{L_tunnel_tildephi} one sees that ${\cal L}_\text{ferm-CS}$ is manifestly self-dual for arbitrary $u$ and $v$.
In the next section, we will argue that such an exactly self-dual model generically lands in a gapped phase of the fermions; thus, self-duality of the Dirac fermion system with Chern-Simons coupling does not imply criticality.
This result immediately addresses the irrelevance of adding more general Maxwell terms in the above action.

To interpret the precise statement of this duality, consider dual fermions coupled to a dual gauge field [the analogue of Eq.~\eqref{LfermCS}], i.e., 
\begin{align*}
&{\cal L}^\text{d.~ferm-CS} = {\cal L}_0 + {\cal L}_\text{staggered-CS} + {\cal L}_\text{MW} + {\cal L}_\text{CS} + {\cal L}_\text{tunnel} ~, \\
&{\cal L}_0 = \frac{i (-1)^j}{2\pi} \partial_x \tilde{\phi}_j \, \tilde {a}_{0,j} + \frac{u}{4\pi}(\partial_x \tilde{\phi}_j - \tilde{a}_{1,j})^2 ~, \nonumber \\
&{\cal L}_\text{staggered-CS} = \frac{i (-1)^j}{8\pi}(\Delta \tilde{a}_{0,j})(\tilde{a}_{1,j+1} + \tilde{a}_{1,j}) ~, \nonumber \\
&{\cal L}_\text{MW} = \frac{1}{8\pi} \left[\frac{1}{v} (\Delta \tilde{a}_{0,j})^2 + v (\Delta \tilde{a}_{1,j})^2 \right] ~, \nonumber \\
&{\cal L}_\text{CS} = -s \frac{i}{8\pi}(\Delta \tilde{a}_{0,j})(\tilde{a}_{1,j+1} + \tilde{a}_{1,j}) ~. \nonumber
\end{align*}
Note in particular the signs in ${\cal L}_0$ and ${\cal L}_\text{staggered-CS}$---which are opposite their counterparts in Eq.~\eqref{LfermCS} and reflect the different chiralities of fermions and dual fermions.
In addition, the uniform Chern-Simons term ${\cal L}_\text{CS}$ also has opposite sign. The corresponding Euclidean actions 
${\cal S}^\text{ferm-CS} = \int_{x, \tau} \sum_j \left[ \frac{i (-1)^j}{4\pi} \partial_x \phi_j \partial_\tau \phi_j + {\cal L}^\text{ferm-CS} \right]$ 
and 
${\cal S}^\text{d.~ferm-CS} = \int_{x, \tau} \sum_j \left[ \frac{-i (-1)^j}{4\pi} \partial_x \tilde{\phi}_j \partial_\tau \tilde{\phi}_j + {\cal L}^\text{d.~ferm-CS} \right]$ 
therefore obey
\begin{align}
{\cal S}^\text{d.~ferm-CS}[\tilde{\phi}, \vec{\tilde{a}}] = \left({\cal S}^\text{ferm-CS}[\phi \to \tilde{\phi}, \vec{a} \to \vec{\tilde{a}}] \right)^* ~.
\end{align}
Complex conjugation on the right side likewise suggests that fermionic duality is associated with an anti-unitary symmetry, which we will demonstrate in Sec.~\ref{subsec.fermCS2boson}.
The second vertical arrow in Eq.~\eqref{duality1} sketches this duality in the continuum language.

\section{Boson/vortex formulation of Dirac fermions }\label{sec.bosonization}

In Sec.~\ref{sec.bosons} we introduced a mapping between bosons $\Phi \sim e^{i \varphi}$ and vortices $\tilde{\Phi} \sim e^{i \tilde{\varphi}}$ as 
\begin{align}
&\tilde{\varphi}_{y+1/2} = \sum_{y'} \sgn\( y' - y - \frac{1}{2} \) \theta_{y'} ~, \label{tildephi2} \\
&\tilde{\theta}_{y+1/2} = (\varphi_{y+1} - \varphi_y)/2 ~. \nonumber
\end{align}
Section~\ref{sec.fermions} reviewed an analogous mapping between Dirac fermions $\psi \sim e^{i \phi}$ and dual Dirac fermions $\tilde{\psi} \sim e^{i \tilde{\phi}}$:
\begin{equation}
\tilde{\phi}_j = \sum_{j' \neq j} \sgn(j-j') (-1)^{j'} \phi_{j'} ~.
\end{equation}
We now relate these bosonic and fermionic fields using a coupled-wire analogue of flux attachment.  

As a starting point we combine the boson and vortex fields to form
\begin{subequations}
\begin{align}
&\phi_R(y-1/4) = \varphi_y + \tilde{\varphi}_{y-1/2} ~, \label{rightferm} \\
&\phi_L(y+1/4) = \varphi_y + \tilde{\varphi}_{y+1/2} ~. \label{leftferm}
\end{align}
\end{subequations}
Note that $\phi_R$ and $\phi_L$ naturally reside at different positions (``wires'') of the form $y-1/4$ and $y+1/4$, respectively, which lie halfway between the original boson and dual-vortex wires.
At this point we need to fix commutation of the boson phase field $\varphi$ with $\theta$ (and not just with $\partial_x \theta$), which we choose as follows:
\begin{align}
[\theta_y(x), \varphi_{y'}(x')] = \delta_{yy'} \, i \pi \, \Theta(x-x') ~,
\end{align}
where $\Theta(x-x')$ is the Heaviside step function.
The new fields then satisfy commutation relations
\begin{align*}
[\phi_{P, y-P/4}(x), \phi_{P, y'-P/4}(x')] =&~ \delta_{yy'} P\, i \pi \, \sgn(x-x') \\
&+ (1 - \delta_{yy'}) \, i \pi \, \sgn(y'-y) ~, \\
[\phi_{R, y-1/4}(x), \phi_{L, y'+1/4}(x')] =&~ i \pi \, \sgn(y' - y + 1/2) ~,
\end{align*}
where in the first equation $P = R/L = \pm 1$.
The intra-wire commutator implies that $\psi_{R/L} \sim e^{i \phi_{R/L}}$ is a right/left-moving fermion---as we have already suggested by the field labels.
Furthermore, the inter-wire commutator of the phase fields implies that such $\psi$ operators anti-commute on different wires, i.e., they are indeed fermion fields on the full array of quarter-integer wires without requiring any supplemental Klein factors.

\begin{figure}[h]
\includegraphics[width=\columnwidth]{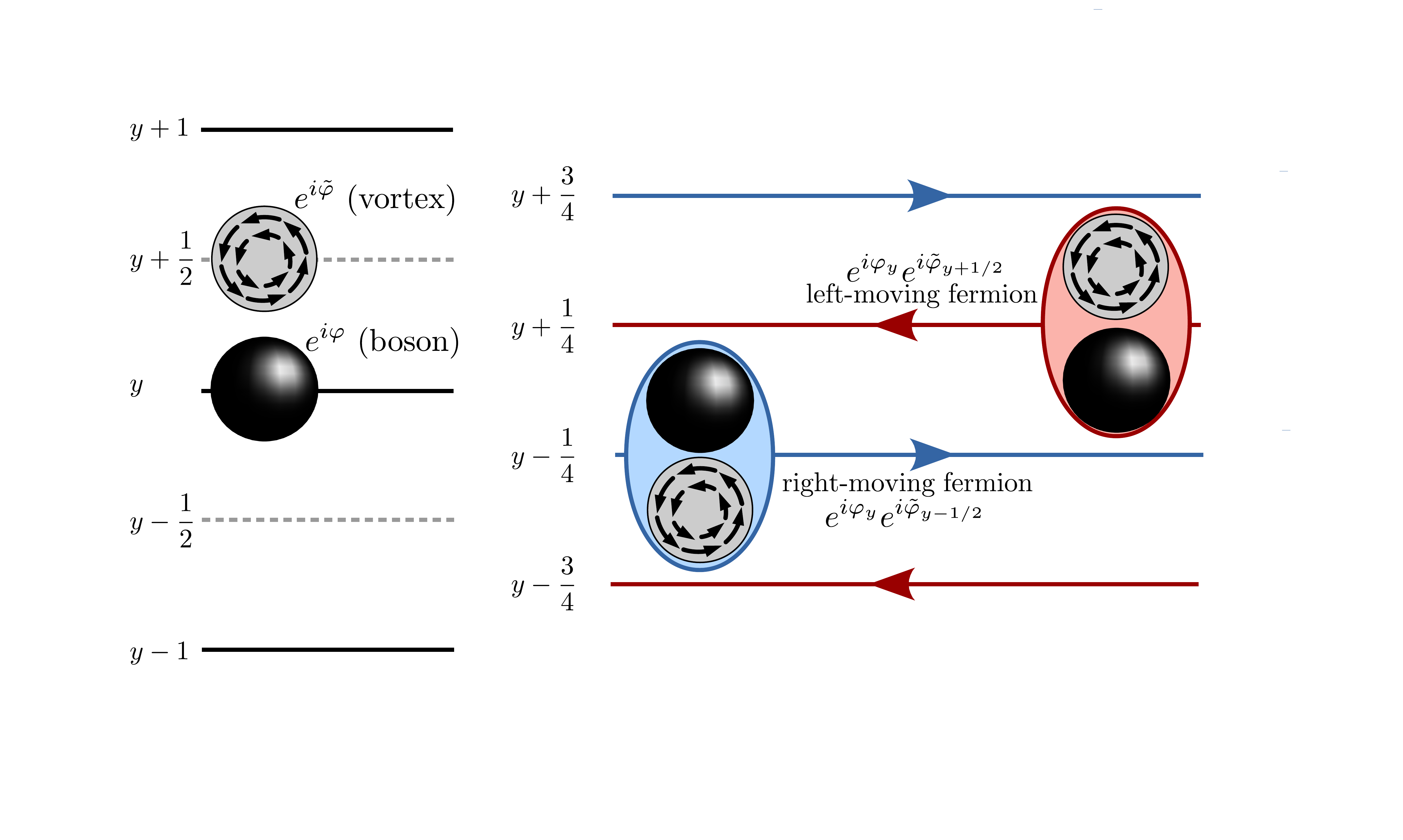}
\caption{
Composite fermions are constructed by combing a boson with a vortex.
With bosons residing on the direct lattice and vortices on the dual lattice, composite fermions are most naturally associated with new quarter-integer lattice.
The fermion chirality depends on the relative position of the boson and vortex.}
\label{fig:ferm}
\end{figure}

Using Eqs.~\eqref{rightferm} and \eqref{leftferm} we see that $\psi_{R/L}$ arise from attaching a single $2\pi$ vortex to a boson; bosons are thus transmuted to fermions a la flux attachment.
Note that when the attached vortex is below the boson we obtain a right-moving chiral fermion, while when the vortex is above the boson we obtain a left-moving fermion.
For an illustration see Fig.~\ref{fig:ferm}.
From an equivalent dual perspective, we may also view $\psi_{R/L}$ as a fermion created by attaching a ``dual flux quantum'' (i.e., a boson) to each vortex.
We remark that in the continuum implementation these two approaches appear to give different theories expressed in terms of different sets of variables that cannot be connected to each other by any simple field-theory manipulations.
Our coupled-wire approach, by contrast, reveals that they are actually identical.
We discuss this point in detail in Appendix~\ref{app:CSvsFermVort}.

Recall that in Sec.~\ref{sec.fermions} we constructed a single Dirac cone from staggered right-moving ($e^{i \phi_{j \in \text{even}}}$) and left-moving ($e^{i \phi_{j \in \text{odd}}}$) chiral fermions.
It is thus natural to identify
\begin{align}
\phi_j = \phi_P(j/2 - 1/4) ~, \nonumber
\end{align}
where $P = R$ if $j$ is even ($j = 2y$ with $y$ integer) and $P = L$ if $j$ is odd ($j = 2y + 1$).
With this identification the field commutation relations are given succinctly by Eq.~\eqref{phicommutator} (the present treatment in fact motivated this earlier specification of the fermion phase fields).

Importantly for applications, we can match the boson phase-slip/hopping terms with the fermion inter-wire tunneling terms:
\begin{align*}
&\cos(2\theta_y) = \cos(\phi_{2y} - \phi_{2y+1}) ~, \\
&\cos(\varphi_y - \varphi_{y+1}) = \cos(\phi_{2y+1} - \phi_{2y+2}) ~,
\end{align*}
see Eq.~(\ref{L_tunnel_phi}).
In particular, when the boson phase-slip and hopping terms have equal amplitudes as required for the bosonic self-dualities studied here, the fermion tunneling terms are invariant under ${\cal T}'$ and ${\cal C}'$ symmetries.

When defining the fields $\phi_{R/L}$ in Eqs.~\eqref{rightferm} and \eqref{leftferm}, we chose the same signs for the boson and vortex fields, which corresponds to binding a vortex (or flux) with a specific orientation.
One can similarly define
\begin{subequations}
\begin{align}
&\tilde{\phi}_L(y-1/4) \equiv \varphi_y - \tilde{\varphi}_{y-1/2} ~, \label{dualleftferm} \\
&\tilde{\phi}_R(y+1/4) \equiv \varphi_y - \tilde{\varphi}_{y+1/2} ~, \label{dualrightferm}
\end{align}
\end{subequations}
to reverse the orientation of the vortex (flux) attachment.
As before, $\tilde{\psi}_{R/L} \sim e^{i \tilde{\phi}_{R/L}}$ are right/left-moving fermions that obey proper anticommutation relations over the full wire array without requiring additional Klein factors.
Upon defining $\tilde{\phi}_j = \tilde{\phi}_P(j/2 - 1/4)$ with $P = L$ if $j$ is even and $P = R$ if $j$ is odd, one can verify that
\begin{equation}
\tilde{\phi}_j = \sum_{j' \neq j} \sgn(j-j') (-1)^{j'} \phi_{j'} ~,
\end{equation}
i.e., these are precisely the fields introduced in Eq.~\eqref{eqn.dualfermions} that define dual Dirac fermions via $\tilde{\psi}_j \sim e^{i \tilde{ \phi}_j}$.

\subsection{Symmetries}
\label{subsec.symmetries}

We previously discussed the action of time-reversal ${\cal T}$ and particle-hole ${\cal C}$ on bosons/vortices, as well as analogous symmetries ${\cal T}'$ and ${\cal C}'$ on fermions/dual fermions.
The explicit mapping derived above precisely relates these two sets of degrees of freedom.
Consequently, we can now further deduce the action of ${\cal T}, {\cal C}$ on fermions/dual fermions and ${\cal T}', {\cal C}'$ on bosons/vortices.
The action of these transformations on the various phase field is listed in Table~\ref{tab:symm}.
We caution that while expressions for ${\cal T}$ and ${\cal C}$ are exact, the expressions for ${\cal T}'$ and ${\cal C}'$ are only schematic and do not show pieces that come from proper treatment of the built-in Klein factors in the fermion phase fields.
Such technical details and the precise meaning of the action of ${\cal T}'$ and ${\cal C}'$ is discussed in App.~\ref{app:fermsymm}.

\begin{table}[ht]
  \centering
    \caption{Operation of the anti-unitary symmetries on the phase variables. All constant shifts are understood to be given modulo $2\pi$.}
  \begin{tabular}{|c|c|c|c|c|}
  \hline
  Symmetry & $\varphi_y$ & $\tilde{\varphi}_{y+1/2}$ & $\phi_j$ & $\tilde{\phi}_j$ \\ 
  \hline\hline
  ${\cal T}$ & $-\varphi_y$ & $\tilde{\varphi}_{y+1/2}$ & $-\tilde{\phi}_j$ & $-\phi_j$ \\
  ${\cal C}$ & $\varphi_y$ & $-\tilde{\varphi}_{y+1/2}$ & $\tilde{\phi}_j$ & $\phi_j$ \\
  ${\cal T}'$ & $-\tilde{\varphi}_{y+1/2}$ & $-\varphi_{y+1}$ & $-\phi_{j+1} + \pi j$ & $\tilde {\phi}_{j+1} - \pi j$\\
  ${\cal C}'$ &  $\tilde{\varphi}_{y+1/2}$ & $\varphi_{y+1}$ & $\phi_{j+1} - \pi j$ & $-\tilde{\phi}_{j+1} + \pi j$ \\
  \hline
  \end{tabular}
  \label{tab:symm}
\end{table}

It is instructive to also express these in terms of boson $\Phi \sim e^{i \varphi}$ and vortex $\tilde {\Phi} \sim e^{i \tilde{\varphi}}$ operators and the Dirac and dual-Dirac spinors defined in Eqs.~\eqref{Psi_spinor} and \eqref{Psi_spinor_dual}. We have
\begin{subequations}
\begin{align}
&{\cal T}: && \Phi \to \Phi \quad && \text{(boson time reversal)} ~, \\
&{\cal T}: && \tilde{\Phi} \to \tilde{\Phi}^\dagger \quad && \text{(vortex particle-hole)} ~, \\
&{\cal T}: &&\Psi \to \tilde{\Psi} \quad && \text{(fermion duality)}
\end{align}
\end{subequations}
and
\begin{subequations}
\begin{align}
&{\cal C}: \Phi \to \Phi^\dagger \quad && \text{(boson particle-hole)} ~, \\
&{\cal C}: \tilde{\Phi} \to \tilde{\Phi} \quad && \text{(vortex time reversal)} ~, \\
&{\cal C}: \Psi \to \tilde{\Psi}^\dagger \quad && \text{(fermion duality')} ~.
\end{align}
\end{subequations}
For clarity we repeated the transformations given earlier for $\Phi, \tilde{\Phi}$.
The new transformations for $\Psi$ are quite natural given the flux-attachment picture developed in the previous subsection:
Fermions and dual fermions arise from attaching opposite vorticity to the bosons.
Thus a transformation that either reverses the vorticity or conjugates the bosons must translate into fermion duality (up to a local symmetry).

Similarly, ${\cal T}', {\cal C}'$ send
\begin{subequations}
\begin{align}
&{\cal T}': \Psi \to i \sigma^y \Psi \quad && \text{(fermion time reversal)} ~, \\
&{\cal T}': \tilde{\Psi} \to i \sigma^y \tilde{\Psi}^\dagger \quad && \text{(dual-fermion particle-hole)} ~, \\
&{\cal T}': \Phi \to \tilde{\Phi} \quad && \text{(boson duality)} ~,
\end{align}
\end{subequations}
and
\begin{subequations}
\begin{align}
&{\cal C}': \Psi \to i \sigma^y \Psi^\dagger \quad && \text{(fermion particle-hole)} ~, \\
&{\cal C}': \tilde{\Psi} \to i \sigma^y \tilde{\Psi} \quad  && \text{(dual-fermion time reversal)} ~, \\
&{\cal C}': \Phi \to \tilde{\Phi}^\dagger \quad && \text{(boson duality')} ~.
\end{align}
\end{subequations}
Note that on the wire scale ${\cal T}'$ and ${\cal C}'$ also shift by one fermion wire, which corresponds precisely to moving from one boson wire to the dual vortex wire (see Fig.~\ref{fig:ferm}).

The following two statements summarize all of these cases:
\begin{itemize}
\item Each duality (boson-vortex or fermion-dual fermion) interchanges time-reversal and charge-conjugation.
\item Time-reversal and particle-hole symmetry in the bosonic theory implement (two kinds of) duality in the fermionized theory and vice versa.
\end{itemize}

We conclude our discussion of symmetries by noting that 
\begin{align}
{\cal T}{\cal C}^{-1} = {\cal T}'{\cal C}^{\prime-1} 
\end{align} 
is local in any set of variables. 
Both in terms of bosons and in terms of fermions, this is a \textit{unitary} particle-hole transformation $\Phi_y \to \Phi_y^\dagger,\  \psi_j \to \psi_j^\dagger,\ i \to i$.

\subsection{Bosons with Chern-Simons coupling revisited}
\label{subsec.boson+CS2ferm}

In Sec.~\ref{sec.bosons} we encountered a coupled-wire model, Eq.~(\ref{csbosons}), that features bosons coupled to a Chern-Simons field and is exactly self-dual for a specific parameter choice.
The discussion of symmetries in the previous section implies that this model must map to a time-reversal-symmetric theory of fermions such as a free Dirac cone or QED$_3$.
To show this correspondence, let us first use the dictionary in Eqs.~\eqref{rightferm} and \eqref{leftferm} to rewrite
\begin{eqnarray*}
&& \partial_x \varphi_y - a_{1,y} = \partial_x [\phi_R(y-1/4) + \phi_L(y+1/4)]/2 ~, \\ 
&& \theta_y = [\phi_R(y-1/4) - \phi_L(y+1/4)]/2 ~, \\
&& \varphi_{y+1} - \varphi_y = \phi_R(y+3/4) - \phi_L(y+1/4) ~,
\end{eqnarray*}
where we specialized to $s=-1$. 
In the first line $a_{1,y}$ is defined by Eq.~(\ref{a1theta}), which is the constraint obtained upon integrating out the temporal component $a_0$ of the Chern-Simons field.
Then the remaining pieces in ${\cal L}_0$ in Eq.~(\ref{csbosons}) when summed over all boson wires $y$ give
\begin{align}
\sum_j \left[ \frac{u + \tilde{u}}{8\pi} (\partial_x \phi_j)^2 
+ \frac{v - u'_j}{4\pi} \partial_x \phi_j \partial_x \phi_{j+1} \right] \nonumber
\end{align}
with $u'_j = u$ if $j$ is even and $u'_j = \tilde{u}$ if $j$ is odd; 
while $\sum_y [{\cal L}_\text{phase-slip} + {\cal L}_\text{hop}]$ in Eq.~(\ref{csbosons}) becomes
\begin{align}
-\sum_j g'_j \cos(\phi_j - \phi_{j+1})
\end{align}
with $g'_j = g_1$ if $j$ is odd and $g'_j = g_2$ if $j$ is even. 

For $u = \tilde{u} = v$ and $g_1 = g_2$, we thus obtain precisely the wire description of a single free Dirac cone with ${\cal T}'$ and ${\cal C}'$ symmetries
defined in Eqs.~(\ref{calT'}) and (\ref{calC'}). his provides an explicit realization of the duality schematically described by the first line of Eq.~\eqref{duality2}.\cite{foot}
In the present model, either symmetry actually requires only the self-duality conditions $u = \tilde{u}, g_1 = g_2$, while general $v \neq u$ adds short-range interactions to the Dirac fermion.
Phrased another way, either ${\cal T}'$ or ${\cal C}'$ interchanges $u \leftrightarrow \tilde{u}$ and $g_1 \leftrightarrow g_2$, i.e., as expected these symmetries of the fermions implement duality on the bosonic side.
We now understand that these are two independent boson dualities that just happen to act identically when applied to the terms in the present model, but we can write simple modifications of the model that maintain only one or the other self-duality (e.g., by taking the inter-wire fermion hopping terms that have only ${\cal T}'$ or ${\cal C}'$ symmetry and rewriting them in terms of the bosonic fields).
We require both self-dualities if we want the Dirac fermion to have both ${\cal T}'$ and ${\cal C}'$ (time-reversal invariant at zero chemical potential), in which case weak short-range interactions added to the Dirac cone are irrelevant and hence do not destabilize the Dirac theory for small $u - v$.
If we have only one, then the fermion can be either doped or be in magnetic field, in which case generic allowed perturbations are not easily controlled.
Fortunately, since ${\cal T}' {\cal C}^{\prime -1} = {\cal T} {\cal C}^{-1}$ is a simple unitary particle-hole transformation on the bosons, so if we have such an additional symmetry microscopically, then we can reach the ${\cal T}'$- and ${\cal C}'$-self-dual point by tuning just one parameter.

The explicit mapping of the boson-CS model to weakly interacting fermions thus guarantees that the self-dual model is described by a stable critical point, i.e., a continuous quantum phase transition occurs when the parameters of the boson-CS model are tuned across the self-dual point.
Alternatively, we could express the same model in terms of the dual-fermion variables. 
In that case one finds QED$_3$, with the character of the two anti-unitary symmetries interchanged (cf.~Sec.~\ref{sec.fermions}).

\subsection{Dirac fermions with Chern-Simons coupling revisited}
\label{subsec.fermCS2boson}
In Sec.~\ref{sec.fermions} we presented a model of Dirac fermions coupled to a level-1/2 Chern-Simons gauge field which is exactly self-dual.
It is then natural to expect that the corresponding boson/vortex model has time-reversal-like symmetry and either features short-range interactions (the Wilson-Fisher model) or long-range interactions mediated by a Maxwell photon (the Higgs model).

Based on the non-local expression of ${\cal L}^\text{ferm-CS}$ in Eq.~\eqref{eqn.csfermnonloc} one can readily obtain the corresponding model in terms of boson or vortex variables.
For $s = 1$, i.e., the positive sign of the CS term in ${\cal L}_\text{CS}$, one finds upon summation over all fermion wires $j$
\begin{align}
{\cal L}^\text{ferm-CS} &\to \sum_y \left[\frac{v_B}{2\pi}(\partial_x \varphi_y)^2 + \frac{u_B}{2\pi} (\partial_x \theta_y)^2 \right] \label{eqn.fermcs1} \\
& -\sum_y g \left[\cos(2 \theta_y) + \cos(\varphi_{y+1} - \varphi_y) \right] ~, \nonumber
\end{align}
which is the Wilson-Fisher model in terms of the boson variables (and Higgs model in the dual variables). This realizes the duality schematically described by the first line of Eq.~\eqref{duality1}.\cite{foot}
On the other hand, for the negative sign of the CS term in ${\cal L}_\text{CS}$:
\begin{align}
{\cal L}^\text{ferm-CS} &\to \sum_{\tilde{y}} \left[ \frac{v_B}{2\pi} (\partial_x \tilde{\varphi}_{\tilde{y}})^2 + \frac{u_B}{2\pi} (\partial_x \tilde{\theta}_{\tilde{y}})^2 \right]\label{eqn.fermcs2}  \\
& -\sum_{\tilde{y}} g \left[\cos(2\tilde{\theta}_{\tilde y}) + \cos(\tilde{\varphi}_{\tilde{y}+1} - \tilde{\varphi}_{\tilde{ y}}) \right] ~, \nonumber
\end{align}
i.e.,\ the Wilson-Fisher model in the vortex variables (and Higgs model in the boson variables).

Focusing on one case, e.g., $s = 1$, depending on the ratio $u_B/v_B$, the boson model ends in a phase where bosons condense ($u_B/v_B$ small) or are gapped ($u_B/v_B$ large).
From Eq.~(\ref{vBuB}) one finds (assuming $v, u > 0$)
\begin{align}
\frac{u_B}{v_B} = \frac{(u + v)^2}{2 (u + v)^2 + uv} \leq \frac{1}{2} ~,
\end{align}
which for small $g$ places the boson system in its condensed phase (the same hold for vortices in the case $s = -1$).
A transition can still be reached by tuning additional short-range interactions, e.g., by adding a $u_B$-like term (which is self-dual) directly to the fermion model.
Thus, a generic self-dual model of fermions coupled to a CS gauge field at level-1/2 is non-critical. In the low-energy limit, the Dirac fermions develop a mass whose sign changes at the critical point. 

To conclude, the self-dual fermion-(level-1/2)CS model maps to bosons with time-reversal symmetry ${\cal T}$ (and anti-unitary particle-hole symmetry ${\cal C}$), which we already anticipated from the precise formulation of the self-duality at the end of Sec.~\ref{subsec.ferm-CS}.
Similar to our discussion of the boson-CS model in the preceding subsection, our fermion-CS model here has ${\cal T}'{\cal C'}^{-1}$ unitary symmetry $(\phi_j \to -\phi_j$), which is why the self-duality can be viewed as either ${\cal T}$ or ${\cal C}$ bosonic symmetry (while in general these duality implementations are different).
Unlike bosons, even when one is thinking about only two natural phases, the self-duality of fermions does not imply criticality. 

\section{Marginal interactions and modular invariance}
\label{sec.modular}

In the previous sections, we encountered two examples where symmetry and duality are interchanged upon mapping from bosons to fermions:
{\it (i)} We saw that self-dual bosons coupled to a Chern-Simons gauge field are equivalent to a single species of time-reversal-symmetric (and hence gapless) Dirac fermions;
and {\it (ii)} self-dual fermions coupled to a level-1/2 Chern-Simons gauge field map onto time-reversal-symmetric bosons.
This suggests that a self-dual model of time-reversal-symmetric bosons (or fermions) is \emph{simultaneously} both self-dual and time-reversal-symmetric in both its bosonic and fermionic formulations.
Throughout we will assume the presence of ${\cal T}{\cal C}^{-1}$ bosonic symmetry (equivalently, ${\cal T'}{\cal C'}^{-1}$ fermionic symmetry), and so we only need to mention time-reversal separately.

Self-dual models of bosons with time-reversal symmetry are indeed known.\cite{FradkinKivelson,Geraedts2012_rangedloops}
Thus, marginally-long-range interactions mediated by a photon $\vect{a}$ described by
\begin{align}
{\cal L}_\text{marg.}[\lambda, \vect{a}] = \lambda \frac{|\vect{k} \times \vect{a}|^2}{4\pi |\vect{k}|}
\label{Lmarg}
\end{align}
correspond to similar marginally-long-range interactions for vortices mediated by a gauge field $\tilde{\vect{a}}$ that is governed by ${\cal L}_\text{marg.}[\lambda^{-1}, \tilde{\vect{a}}]$.
For $\lambda = 1$, the long-range part of the interactions between vortices is identical to the one between bosons, as required for self-duality. (For the system to be exactly self-dual and hence critical also requires tuning the short-range part of the interactions, see below and Ref.~\onlinecite{Geraedts2012_rangedloops}).
Unlike the Chern-Simons terms discussed before, such a marginally-long-range interaction does not break time-reversal symmetry.

We can also consider situation when both ${\cal L}_\text{marg.}$ and CS terms are present, which will also be the case in the dual theory:
\begin{subequations}
\begin{align}
& i \vect{j}_\text{boson} \cdot \vect{a} + \lambda_\text{boson} \frac{|\vect{k} \times \vect{a}|^2}{4\pi |\vect{k}|} - i \gamma_\text{boson} \frac{a da}{4\pi} \label{lambdagamma} \\
& \leftrightarrow i \vect{j}_\text{vortex} \cdot \tilde{\vect{a}} + \lambda_\text{vortex} \frac{|\vect{k} \times \tilde{\vect{a}}|^2}{4\pi |\vect{k}|} - i \gamma_\text{vortex} \frac{\tilde{a} d\tilde{a}}{4\pi} ~,
\end{align}
\end{subequations}
where we have abused the notation somewhat by using momentum-space and real-space expressions for different terms.
The convenience of the specific parametrization will become clear below.
The relevant functional integral for performing the duality is
\begin{align}
& \int {\cal D} \vect{a} \exp\left\{-\lambda_\text{ boson} \frac{|\vect{k} \times \vect{a}|^2}{4\pi |\vect{k}|} + i \gamma_\text{ boson} \frac{a da}{4\pi} - i \frac{a d\tilde{a}}{2\pi} \right\} \nonumber \\
& \sim \exp\left\{-\lambda_\text{vortex} \frac{|\vect{k} \times \tilde{\vect{a}}|^2}{4\pi |\vect{k}|} + i \gamma_\text{vortex} \frac{\tilde{a} d\tilde{a}}{4\pi} \right\} ~, \label{modular}
\end{align}
which gives\cite{FradkinKivelson}
\begin{subequations}
\begin{align}
& \lambda_\text{vortex} = \frac{\lambda_\text{boson}}{\lambda_\text{boson}^2 + \gamma_\text{boson}^2} ~, \label{bosonmodular1} \\
& \gamma_\text{vortex} = -\frac{\gamma_\text{boson}}{\lambda_\text{boson}^2 + \gamma_\text{boson}^2} ~. \label{bosonmodular2}
\end{align}
\end{subequations}
Writing $z = \gamma + i \lambda $ one finds the ``modular'' relationship $z_\text{vortex} = -1/z_\text{boson}$.
In particular, the unit (semi)-circle $\lambda^2 + \gamma^2 = 1$ maps back to itself, only the sign of $\gamma$ is changed, and can be called ``self-dual" line as far as these long-ranged interactions are concerned.
The points $(\lambda, \gamma) = (0, 1)$ and $(0, -1)$ correspond precisely to the boson+CS self-duality discussed in Sec.~\ref{subsec.boson+CS}.
There, we found a wire model where also all short-ranged interactions are exactly self-dual and hence the model is critical, with exact relation to a free Dirac cone with time-reversal symmetry described in Sec.~\ref{subsec.boson+CS2ferm}.
Below we will also find a larger class of exactly-self-dual (and hence critical) wire models of bosons covering the $\lambda^2 + \gamma^2 = 1$ curve and will establish precise relation to fermions with time reversal symmetry.

Before proceeding with the wire models, let us apply the formal fermionization Eq.~(\ref{duality1}) to the bosonic theory Eq.~(\ref{lambdagamma}), obtaining
\small
\begin{align}
& i \vect{j}_\text{Dirac} \cdot \vect{c} - i \frac{c dc}{8\pi} + i \frac{a dc}{2\pi} - i(1 + \gamma_\text{boson}) \frac{a da}{4\pi} + \lambda_\text{boson} \frac{|\vect{k} \times \vect{a}|^2}{4\pi |\vect{k}|} ~. \label{beforesuperduality}
\end{align}
\normalsize
Integrating over $\vect{a}$ we obtain the final version of the fermionized theory
\begin{align}
& i \vect{j}_\text{Dirac} \cdot \vect{c} + \lambda_\text{Dirac} \frac{|\vect{k} \times \vect{c}|^2}{8\pi |\vect{k}|} - i \gamma_\text{Dirac} \frac{c dc}{8\pi} ~, \nonumber
\end{align}
with
\begin{subequations}
\begin{align}
\lambda_\text{Dirac} =& \frac{2 \lambda_\text{boson}}{\lambda_\text{boson}^2 + (1 + \gamma_\text{boson})^2} ~, \label{fermionmodular1} \\
\gamma_\text{Dirac} =& \frac{\lambda_\text{boson}^2 + \gamma_\text{boson}^2 - 1}{\lambda_\text{boson}^2 + (1 + \gamma_\text{boson})^2} ~. \label{fermionmodular2}
\end{align}
\end{subequations}

In terms of $z =  \gamma + i \lambda$, this can be written as
\begin{align*}
 z_\text{Dirac} = \frac{z_\text{boson}-1}{z_\text{boson}+1}  ~.
\end{align*}
Note that we have chosen a different parametrization of the marginally-long-range interaction and CS term for Dirac fermions compared to bosons, whose convenience will become clear shortly.

In the special case $\lambda_\text{boson} = 1$ and $\gamma_\text{boson} = 0$, one finds $\lambda_\text{Dirac} = 1$ and $\gamma_\text{Dirac} = 0$.
In particular, the half-integer Chern-Simons term for the gauge field $\vect{c}$ in Eq.~\eqref{beforesuperduality} gets exactly canceled, and the resulting theory is time-reversal-symmetric both in boson and in fermion variables.
Furthermore, since time-reversal symmetry of the boson model implies self-duality of the fermion model and vice versa, the theory must also be simultaneously self-dual in either formulation.

Indeed, performing now the fermionic duality on the above theory, we find for dual Dirac fermions
\begin{align}
&i \vect{j}_\text{dual Dirac} \cdot \tilde{\vect{c}} 
+ \lambda_\text{dual Dirac} \frac{|\vect{k} \times \tilde{\vect{c}}|^2}{8\pi |\vect{k}|} - i \gamma_\text{dual Dirac} \frac{\tilde{c} d\tilde{c}}{8\pi} ~, \nonumber
\end{align}
with
\begin{subequations}
\begin{align}
\lambda_\text{dual Dirac} =& \frac{\lambda_\text{Dirac}}{\lambda_\text{Dirac}^2 + \gamma_\text{Dirac}^2} ~, \label{dualfermionmodular1} \\
\gamma_\text{dual Dirac} =& -\frac{\gamma_\text{Dirac}}{\lambda_\text{Dirac}^2 + \gamma_\text{Dirac}^2} ~. \label{dualfermionmodular2}
\end{align}
\end{subequations}
In the above parametrization, the Dirac fermions under duality thus exhibit precisely the same modular relationship as bosons, cf.~Eqs.~\eqref{bosonmodular1}-\eqref{bosonmodular2},
and $\lambda_\text{Dirac} = 1$, $\gamma_\text{Dirac} = 0$ is indeed self-dual.

\subsection{Self-dual and time-reversal-symmetric model}
\label{subsec:selfdual_and_Tsymmetric}
We have in fact already encountered an explicit wire mode which is time-reversal-symmetric and exactly self-dual in Sec.~\ref{subsec.boson+CS}.
The model ${\cal L}_\text{SD} = {\cal L}_\varphi + {\cal L}_\theta$ that includes only terms Eqs.~\eqref{lphi}-\eqref{ltheta} (but without ${\cal L}_{\theta, \varphi}$) is exactly self-dual for $\tilde{u} = u$, as can be seen from its rewriting in terms of the dual variables in Eqs.~\eqref{lphidual}-\eqref{lthetadual}.

We can also easily bring it to the manifestly self-dual form using both direct and dual variables,
\begin{align}
{\cal L}_\text{SD} = \frac{v}{2\pi} [(\partial_x \varphi_y)^2 + (\partial_x \tilde{\varphi}_{\tilde{y}})^2] + \frac{u-v}{2\pi} [(\partial_x \theta_y)^2 + (\partial_x \tilde{\theta}_{\tilde{y}})^2] ~, \label{eqn.sdboson}
\end{align}
with implicit summation over boson wires $y$ or vortex wires $\tilde{y}$.
Expressed in the fermion variables, this becomes
\begin{align}
{\cal L}_\text{SD} =& \frac{v}{8\pi} [(\partial_x \phi_j)^2 + (\partial_x \tilde{\phi}_j)^2] \label{eqn.sdfermion} \\
& + \frac{u-v}{16\pi} [(\partial_x \phi_{j+1} - \partial_x \phi_j)^2 + (\partial_x \tilde{\phi}_{j+1} - \partial_x \tilde{\phi}_j)^2] ~, \nonumber
\end{align}
with implicit summation over fermion wires $j$, where we also used $\phi_{j+1} - \phi_j = \pm (\tilde{\phi}_{j+1} - \tilde{\phi}_j)$.
The fermionic model is manifestly self-dual and invariant under ${\cal T}'$ (i.e., fermionic time-reversal).
As expected (and easy to check explicitly), the bosonic ${\cal T}$ corresponds to the fermionic self-duality while the bosonic self-duality corresponds to the fermionic ${\cal T}'$.
While we do not know how the singular gauge-field propagator Eq.~\eqref{lambdagamma} would be represented in the coupled-wire framework, it is natural to conjecture that the above boson model ${\cal L}_\text{SD}$ (supplemented by equal-amplitude phase-slip and inter-wire hopping terms) and Eq.~\eqref{lambdagamma} with $\lambda_\text{boson} = 1$ and $\gamma_\text{boson} = 0$ describe the same fixed point, which corresponds also to $\lambda_\text{Dirac} = 1$ and $\gamma_\text{Dirac} = 0$.

We note that Ref.~\onlinecite{Geraedts2012_rangedloops} studied numerically a different realization of such exactly-self-dual time-reversal-invariant bosons using a loop model on a $2+1$-dimensional lattice. The critical properties that were determined there, should then apply also to the exactly-self-dual time-reversal-invariant fermions discovered above. For example, the scaling dimension of the fermion mass term is directly related to the correlation length exponent found at the exactly-self-dual transition.

\subsection{Wire model with general modular relationship}
In the presence of the marginally-long-range interactions, the (self-dual) short-range interactions $\sim (u-v)$ in Eqs.~\eqref{eqn.sdboson},\eqref{eqn.sdfermion} can be neglected.
In addition, we now generalize the model to allow terms $(\partial_x \phi_j)^2 - (\partial_x \tilde{\phi}_j)^2$ (violating the fermionic self-duality) and $\partial_x \phi_j \partial_x \tilde{\phi}_j$ (breaking the fermionic time reversal symmetry).
In the bosonic variables, these corresponds to $\partial_x \varphi \partial_x \tilde{\varphi}$ and $(\partial_x \varphi)^2 - (\partial_x \tilde{\varphi})^2$, respectively.
A convenient parameterization of the three relevant terms is given by 
\begin{subequations}
\begin{align}
{\cal L}=& \frac{\tilde v}{4\pi} \left[ (\partial_x \tilde{\phi}_j + \gamma_\text{Dirac} \partial_x \phi_j)^2 + \lambda_\text{Dirac}^2 (\partial_x \phi_j)^2 \right]  \label{eqnwiredfermionmodular}\\ =& \frac{v}{4\pi} \left[ (\partial_x \phi_j - \gamma_\text{d.~Dirac} \partial_x \tilde{\phi}_j)^2 + \lambda_\text{d.~Dirac}^2 (\partial_x \tilde{\phi}_j)^2 \right]~, \label{eqnwirefermionmodular} 
\end{align}
\end{subequations}
with $\lambda_\text{d.~Dirac}$, $\gamma_\text{d.~Dirac}$ given in Eqs.~\eqref{dualfermionmodular1}-\eqref{dualfermionmodular2} and $\tilde v = v (\lambda_\text{d.~Dirac}^2 + \gamma_\text{d.~Dirac}^2)$. 
Equation~\eqref{eqnwirefermionmodular} with $\lambda_\text{Dirac}=\lambda_\text{d.~Dirac} = 0$ is precisely what one obtains upon integrating out a Chern-Simons gauge field with a generic coefficient $-\gamma_\text{Dirac} = \gamma_\text{d.~Dirac}$, see Appendix~\ref{app.csfermnew}. [The opposite sign in front of $\gamma_\text{d.~Dirac}$ compared to $\gamma_\text{Dirac}$ in Eq.~(\ref{eqnwirefermionmodular}) arises since the fermions $e^{i \phi_j}$ and dual fermions $e^{i \tilde{\phi}_j}$ on the same wire $j$ have opposite chiralities.] 

In terms of the boson/vortex variables, the Lagrangian Eq.~\eqref{eqnwirefermionmodular} becomes
\begin{subequations}
\begin{align}
{\cal L}= \frac{\tilde v'}{2\pi} [& (\partial_x \tilde{\varphi}_{y+1/2})^2 + (\gamma_\text{boson}^2 + \lambda_\text{boson}^2) (\partial_x \varphi_y)^2 \nonumber \\
& - \gamma_\text{boson} (\partial_x \tilde{\varphi}_{y+1/2})(\partial_x \varphi_{y+1} + \partial_x \varphi_y) ] \\ = \frac{v'}{2\pi} [& (\partial_x \varphi_y)^2 + 
(\gamma_\text{vortex}^2 + \lambda_\text{vortex}^2) (\partial_x \tilde{\varphi}_{y+1/2})^2 \nonumber \label{eqnwirefermionbosonmodular}\\
& + \gamma_\text{vortex} (\partial_x \varphi_y)(\partial_x \tilde{\varphi}_{y+1/2} + \partial_x \tilde{\varphi}_{y-1/2}) ]  ~,\end{align}
\end{subequations}
where
\begin{subequations}
\begin{align}
&v' = v \frac{\lambda_\text{Dirac}^2 + (1 + \gamma_\text{Dirac})^2}{\lambda_\text{Dirac}^2 + \gamma_\text{Dirac}^2} ~, \\
& \lambda_\text{boson} = \frac{2 \lambda_\text{Dirac}}{\lambda_\text{Dirac}^2 + (1 - \gamma_\text{Dirac})^2} ~, \\
&\gamma_\text{boson} = \frac{1 - \lambda_\text{Dirac}^2 - \gamma_\text{Dirac}^2}{\lambda_\text{Dirac}^2 + (1 - \gamma_\text{Dirac})^2} ~.
\end{align}
\end{subequations}
The last two equations give exactly the inverse transformation to Eqs.~(\ref{fermionmodular1})-(\ref{fermionmodular2}).
In the vortex variables, $\lambda_\text{vortex}$ and $\gamma_\text{vortex}$ are given by Eqs.~\eqref{bosonmodular1}-\eqref{bosonmodular2}, while
\begin{align}
\tilde v' = v' [\lambda_\text{vortex}^2 + \gamma_\text{vortex}^2] ~,
\end{align}
Note that in the above Lagrangian, we implicitly used summation over $j$ for fermions/dual fermions and summation over $y$ for bosons/vortices.

Equation~\eqref{eqnwirefermionbosonmodular} with $\lambda_\text{boson} = 0$ is precisely what one obtains upon integrating out a Chern-Simons gauge field with a generic coefficient, see Appendix~\ref{app.onwireboson}.
Given the similarity in the behavior of the couplings $\lambda, \gamma$ under dualities and fermionization/bosonization to those for the continuum theories with both marginally-long-range and CS interactions, we conjecture that the above Lagrangians provide precise wire realizations of these field theories.

\section{Gapped phases and parent Hamiltonians}
\label{GappedPhases}

While in this paper we are primarily concerned with gapless field theories, our approach is well suited for studying interesting gapped phases in this context. A handy property of the explicit wire-model dualities presented here is that they readily generate parent Hamiltonians for a wide range of topologically ordered phases.
Numerous wire-models for such phases have been constructed in recent years (see, e.g., References~ \onlinecite{KaneWires,TeoKaneChains,LuWires,AshvinSenthil,QuantumWiresParafendleyons,Mong,Seroussi,Neupert,Sagi,STO}), 
mostly on a case-by-case basis. 
The duality transformations described here provide a straightforward algorithm to construct parent Hamiltonians for any phase that has a simple ``Hartree-Fock'' description in at least one of its composite-fermion/boson formulations:
Wire models for conventional phases (such as superfluids and integer quantum Hall states) can be readily written down. Equations~\eqref{tildephi}-\eqref{tildetheta} (boson duality), \eqref{eqn.dualfermions} (fermion duality) and \eqref{rightferm}-\eqref{leftferm} (boson-fermion mapping) then instantly yield the corresponding Hamiltonian in the variables of choice, which may describe a more exotic, topologically ordered phase.
A concrete example is the ``T-Pfaffian'' topological order whose parent Hamiltonian is related to the one of a Fu-Kane superconductor\cite{FuKane} by the fermionic duality mapping of Eq.~\eqref{eqn.dualfermions}.
This mapping was carried out explicitly in Ref.~\onlinecite{diracduality}.

It is useful to characterize gapped phases according to the response of the matter field only to a minimally coupled probing field.
Some natural possibilities are given in Table~\ref{tab.phases}.

\begin{center}
\begin{table}[h]
\begin{tabular}{ |c| c| }
\hline
Fermions & Bosons \\ \hline \hline
  $\sigma_{xy} = +\frac{e^2}{2h}$ & Superfluid  \\
  $\sigma_{xy} = -\frac{e^2}{2h}$ & Mott insulator  \\
   ``Mott'' insulator &  $\sigma_{xy} = +\frac{e^2}{h}$  \\
  Superfluid & $\sigma_{xy} = -\frac{e^2}{h}$ \\
  \hline
\end{tabular}
\caption{
Gapped fermion phases and their bosonic duals can be characterized by their bulk electromagnetic response (to the respective total gauge fields seen by the fermions or bosons). 
Gapped fermions with a quantized Hall response $\sigma_{xy}^\text{fermion} = \pm e^2/(2h)$ correspond to boson superfluids and Mott insulators (with $\sigma_{xy}^\text{boson} = 0$).
Conversely, bosons with a quantized Hall response $\sigma_{xy}^\text{boson} = \pm e^2/h$ correspond to fermion superfluids and Mott insulators (with $\sigma_{xy}^\text{fermion} = 0$).
``Mott insulator'' of Dirac fermions here refers to any phase that is dual to a (gauged) superfluid of fermions.
This includes, e.g., the T-Pfaffian (dual to a Fu-Kane superconductor\cite{FuKane}) which exhibits  vanishing $\sigma_{xy}^\text{fermion} = 0$ and conserves particle number. 
Note, however, that this relationship between response properties of gapped phases does not presume the presence of time-reversal or particle-hole symmetries.
}
\label{tab.phases}
\end{table}
\end{center}

We note that, depending on symmetries, there may be multiple distinct phases corresponding to each entry in this table.
For example, a superfluid of fermions can be of the Fu-Kane\cite{FuKane} type which is compatible with ${\cal T}'$, or feature general odd-angular-momentum pairing which breaks ${\cal T}'$.
Each of these possibilities corresponds to a specific Mott-insulating phase of dual fermions as well as a specific $\nu = -1$ quantum-Hall state of bosons or vortices.
Such bosonic quantum Hall states are characterized by identical charge-carrying chiral edge modes, but distinct neutral modes (a well known example being the bosonic Moore-Read state).
Among this class of bosonic quantum Hall states, there is thus a subset that is compatible with self-duality (${\cal T}'$ in its fermionic formulation).
Since this is a gapped phase, its properties (i.e., its quasiparticle content) should reflect this compatibility regardless of whether this self-duality is microscopically present.
(This is analogous to the PH-Pfaffian which, as a phase of matter, need not be PH-symmetric\cite{zucker}). 
It might be interesting to explore more generally how the requirement of self-duality in this sense constrains possible phases and their properties.

\section{Extensions to $N = 2$ Dirac fermions}
\label{sec.furtherapps}

Before concluding, we will now discuss some additional applications of the duality mappings described above.
We will sketch how some useful relations for systems with multiple fermion species can be inferred by applying the various mappings separately to each fermion species.

\subsection{Self-dual QED$_3$ with two fermion flavors}
Consider two flavors of Dirac fermions coupled to a single propagating gauge field $\vect{a}$, schematically described by
\begin{align}
i \left(\vect{j}_\text{Dirac, 1} + \vect{j}_\text{Dirac, 2} \right) \cdot \vect{a} ~. \label{N2qed}
\end{align}
We will take the model to have the fermionic anti-unitary symmetries
\begin{subequations}
\begin{align}
& {\cal T}'_{N=2}: \qquad
\Psi_\alpha \to i \sigma^y \Psi_\alpha ~, \quad \alpha = 1,2 ~, \\
& {\cal C}'_{N=2}: \qquad\Psi_\alpha \to i \sigma^y \Psi_\alpha^\dagger ~, \quad \alpha = 1,2 ~,
\end{align}
\end{subequations}
i.e., the same time-reversal and particle-hole transformations as in Sec.~\ref{sec.fermions} but now acting on each flavor $\alpha = 1,2$.
We will also consider a unitary $\mathbb{Z}_2$ symmetry ${\cal R}$ that interchanges the two fermion flavors,
\begin{align}
{\cal R}: \qquad \Psi_1 \leftrightarrow \Psi_2 ~.
\end{align}
Applying the Dirac-QED$_3$ duality of Eq.~\eqref{duality2} separately to each species yields
\begin{align*}
i \vect{j}_\text{dual Dirac, 1} \cdot \tilde{\vect{a}}_1 + i \vect{j}_\text{dual Dirac, 2} \cdot \tilde{\vect{a}}_2 + i\frac{a d(\tilde{a}_1 + \tilde{a}_2)}{4\pi} ~,
\end{align*}
whereupon integrating out $\vect{a}$ yields
\begin{align}
i (\vect{j}_\text{dual Dirac, 1} - \vect{j}_\text{dual Dirac, 2}) \cdot \tilde{\vect{a}} ~. \label{dualN2QED3}
\end{align}
At this point, redefining particle and hole for one of the two flavors, e.g., $\vect{j}_\text{dual Dirac, 2} \to - \vect{j}_\text{dual Dirac, 2}$ in the path integral, returns the dual action to the same form as in Eq.~\eqref{N2qed}.  
In this sense $N = 2$ QED$_3$ is self-dual as discussed in Ref.~\onlinecite{XuYouSPT}.  We will make this statement rigorous below. 

An explicit self-dual wire model of $N=2$ QED$_3$ (with flavors again labeled by $\alpha = 1, 2$) can be readily written down as
${\cal S}~=~\int_{x, \tau} \sum_j \left[ \sum_\alpha \frac{i (-1)^j}{4\pi} \partial_x \phi_{j,\alpha} \partial_\tau \phi_{j,\alpha} + {\cal L}_\text{QED$_3$}^{N=2} \right]$ with
\begin{align}
&{\cal L}_\text{QED$_3$}^{N=2} = {\cal L}_0 + {\cal L}_\text{staggered-CS} + {\cal L}_\text{MW} + {\cal L}_\text{tunnel} ~, \\
&{\cal L}_0 = \sum_\alpha \frac{-i (-1)^j}{2\pi} \partial_x \phi_{j,\alpha} \, a_{0,j} + \frac{v}{8\pi} (\partial_x \phi_{j,1} - \partial_x \phi_{j,2})^2 ~, \nonumber \\
&{\cal L}_\text{staggered-CS} = 2 \frac{-i (-1)^j}{8\pi} (\Delta a_{0,j}) (a_{1,j+1} + a_{1,j}) ~, \nonumber \\
&{\cal L}_\text{MW} = \frac{2}{16\pi} \left[ \frac{1}{\tilde{v}} (\Delta a_{0,j})^2 + \tilde{v} (\Delta a_{1,j})^2 \right] ~. \nonumber
\end{align}
The structure parallels that for $N=1$ QED$_3$ in Eq.~(\ref{LQED3}), except that here we start with opposite fermion chiralities.\footnote{We use the opposite chirality here since Eq.~\eqref{LQED3} describes dual fermions while the $N = 2$ QED$_3$ theory is meant to describe ``original'' fermions.  (Recall that the fermion chirality flips under duality.)} The first term in ${\cal L}_0$ couples the two fermion flavors to the gauge field with the same charge, while the $v$ term introduces an energy cost for the gauge-neutral combination of fermion fields whose role will become clear below.
The extra factor of $2$ in ${\cal L}_\text{staggered-CS}$ is required for gauge invariance with two flavors; we have also added an extra factor of $2$ in ${\cal L}_\text{MW}$ and parametrized the Maxwell term by $\tilde v$---both for later convenience.  
Finally, ${\cal L}_\text{tunnel}$ contains the same tunneling terms for each species as before and will not be written out explicitly.

Integrating out the gauge field as in the coupled-wire derivation of the Dirac-QED$_3$ duality\cite{diracduality} yields
\begin{align*}
{\cal L}^{N=2}_{\text{QED}_3} = \frac{v}{8\pi}(\partial_x \phi_{j,1} - \partial_x \phi_{j,2})^2 + \frac{\tilde{v}}{8\pi}(\partial_x \tilde{\phi}_{j,1} + \partial_x \tilde{\phi}_{j,2})^2 ~,
\end{align*}
where $\tilde \phi_{j,\alpha}$ are dual-fermion variables with opposite chirality relative to $\phi_{j,\alpha}$. By reversing the treatment that produced the $\tilde{v}$ term in the last equation, but instead for the $v$ term, we can rewrite this theory entirely in terms of dual fermions coupled to a new gauge field $\tilde a$.  The key difference is that $\tilde{\phi}_1$ and $\tilde{\phi}_2$ will carry opposite gauge charges with respect to $\tilde a$, in agreement with Eq.~(\ref{dualN2QED3}).

When $\tilde{v} = v$, the model is exactly self-dual in the following precise sense: The coupled-wire action is explicitly invariant under $\phi_1 \leftrightarrow -\tilde{\phi}_1$, $\phi_2 \leftrightarrow \tilde{\phi}_2$ together with overall complex conjugation (due to the opposite chiralities for dual and original fermions).
In terms of continuum Dirac fields, this duality corresponds to the anti-unitary operation 
\begin{align}
  {\cal S}_{N=2}: \qquad &\Psi_1 \to \tilde{\Psi}_1 ~, \quad \Psi_2 \to \tilde{\Psi}_2^\dagger ~. \label{SN2a}
\end{align}
As in previous sections, we expect that such a self-duality condition corresponds to a local symmetry in an equivalent bosonic formulation where fermions $\Psi_\alpha$ are traded for bosons $\Phi_\alpha \sim e^{i \varphi_\alpha}$.   Using a straightforward extension of the dictionary from Sec.~\ref{subsec.symmetries}, Eq.~\eqref{SN2a} indeed yields the local anti-unitary transformation
\begin{align}
{\cal S}_{N=2}: \qquad &\Phi_1 \to \Phi_1 ~, \quad \Phi_2 \to \Phi_2^\dagger ~. \label{SN2}
\end{align}

We can further rewrite the self-dual wire model ${\cal L}^{N=2}_{\text{QED}_3}$ in terms of bosonic variables as
\begin{align}
{\cal L}^{N=2}_{\text{QED}_3} = & \frac{v}{2\pi} \sum_\alpha \left[ (\partial_x \varphi_{y, \alpha})^2 + (\partial_x \tilde {\varphi}_{y+1/2, \alpha})^2 \right] \label{N2QED3boson} \\
&- \frac{v}{2\pi} \sum_\alpha \partial_x \varphi_{y, \alpha} \left(\partial_x \tilde\varphi_{y+1/2, -\alpha} + \partial_x \tilde\varphi_{y-1/2, -\alpha} \right) ~. \nonumber
\end{align}
Here $\tilde \varphi_\alpha$ are dual vortex fields, and in the last line $-\alpha$ denotes the ``opposite'' flavor relative to $\alpha$.
We readily see that Eq.~\eqref{N2QED3boson} satisfies ${\cal S}_{N=2}$, which ``time-reverses'' the first species of bosons and ``particle-hole-conjugates'' the second.
(Explicitly, we have ${\cal S}_{N=2}: \varphi_1 \to -\varphi_1, \varphi_2 \to \varphi_2$ 
and hence $\tilde{\varphi}_1 \to \tilde{\varphi}_1, \tilde{\varphi}_2 \to -\tilde{\varphi}_2$.) 
Moreover, both the original fermion theory and boson reformulation are invariant under the flavor interchange ${\cal R}: \Phi_1 \leftrightarrow \Phi_2$.

It is instructive to observe that ${\cal L}^{N=2}_{\text{QED}_3}$ is \emph{not} invariant under time-reversal or particle-hole conjugation of both species, corresponding to the transformations
\begin{subequations}
\begin{align*}
&{\cal T}_{N=2}: \qquad \Phi_\alpha \to \Phi_\alpha ~, \quad \alpha = 1, 2 ~, \quad \text{(not present)} ~, \\
&{\cal C}_{N=2}: \qquad \Phi_\alpha \to \Phi_\alpha^\dagger ~, \quad \alpha = 1, 2 ~, \quad \text{(not present)} ~.
\end{align*}
\end{subequations}
The absence of these symmetries reflects the fact that $N=2$ QED$_3$ is self-dual only in the above sense where duality is followed by particle-hole conjugation of one dual fermion flavor.
(Of course, bosonic models satisfying ${\cal T}_{N=2}$ and/or ${\cal C}_{N=2}$ are also possible but are not of interest here.)

Let us develop more understanding of the bosonic reformulation of ${\cal L}_\text{QED$_3$}^{N=2}$. The non-local action for the bosonic variables that couples $\varphi_\alpha$ and $\tilde{\varphi}_{-\alpha}$ can alternatively be viewed as a theory of bosons with a mutual Chern-Simons term.
This can be inferred from the formal bosonization of the schematic continuum theory in Eq.~(\ref{N2qed}) to
\begin{align}
&i \vect{j}_\text{boson, 1} \cdot \vect{c}_1 + i \vect{j}_\text{boson, 2} \cdot \vect{c}_2 - i \frac{c_1 dc_2}{2\pi} ~,
\label{boson2pi}
\end{align}
which describes two particles with mutual statistics $2\pi$.
Under the bosonic duality sketched in the left side of Eq.~\eqref{duality1}, the theory of vortices also has a mutual Chern-Simons term but with opposite sign:
\begin{align}
&i \vect{j}_\text{vortex, 1} \cdot \tilde{\vect{c}}_1 + i \vect{j}_\text{vortex, 2} \cdot \tilde{\vect{c}}_2 + i \frac{\tilde{c}_1 d\tilde{c}_2}{2\pi} ~.
\end{align}
Given the exhibited long-wavelength structure, such a bosonic theory can be self-dual in the sense that upon additional complex conjugation (and also changing the signs of either currents or gauge fields),
the path integral in terms of the dual fields has identical structure to that in terms of the original fields.  
Self-duality for the bosonic theory can be readily related to the local fermionic symmetries ${\cal T}'_{N=2}$ and ${\cal C}'_{N=2}$, both of which implement boson-vortex duality in a slightly different manner: the former operates as $\varphi_{y, \alpha} \to -\tilde{\varphi}_{y+1/2, \alpha}$ while the latter sends $\varphi_{y, \alpha} \to \tilde{\varphi}_{y+1/2, \alpha}$.
Since both ${\cal T}'_{N=2}$ and ${\cal C}'_{N=2}$ are present in the original fermion formulation of the wire model ${\cal L}^{N=2}_{\text{QED}_3}$, the bosonic version in Eq.~(\ref{N2QED3boson}) is exactly self-dual in either sense.\\
\hiddensubsubsection{Connection to self-dual EP-NCCP1 model}

It is interesting to consider one more form of the model obtained by dualizing only one of the boson species, yielding
\begin{align}
& i \left( \vect{j}_\text{boson, 1} + \vect{j}_\text{vortex, 2} \right) \cdot \vect{c} 
\equiv i \left( \vect{l}_1 + \vect{l}_2 \right) \cdot \vect{c} ~,  \label{nccp1}
\end{align}
where for later convenience we introduced $\vect{l}_1 \equiv \vect{j}_\text{boson, 1}$ and $\vect{l}_2 \equiv \vect{j}_\text{vortex, 2}$.
This model (with implicit Maxwell term on $\vect{c}$) is otherwise known as the easy-plane non-compact CP$^1$ (EP-NCCP1) model\cite{shortlight,deccp_science} and was first mapped to QED$_3$ in Ref.~\onlinecite{senthilmatthewduality}. (This connection was recently revisited in greater detail in Ref.~\onlinecite{dccdual}.)

We first verify that this reformulation has the required properties of the EP-NCCP1 model in terms of $\vect{l}_1$ and $\vect{l}_2$.
The EP-NCCP1 model as defined in Refs.~\onlinecite{shortlight,deccp_science} has a unitary symmetry 
$\vect{l}_{1/2} \to -\vect{l}_{1/2}, \vect{c} \to -\vect{c}$ 
and an anti-unitary symmetry 
$\vect{l}_{1/2} \to \vect{l}_{1/2}, \vect{c} \to -\vect{c}$.
The first is enforced here by ${\cal T}'_{N=2} ({\cal C}')^{-1}$ ($\Phi_\alpha \to \Phi_\alpha^\dagger, \tilde{\Phi}_\alpha \to \tilde{\Phi}_\alpha^\dagger$ specialized to operators $\Phi_1$ and $\tilde{\Phi}_2$ which annihilate $\vect{l}_1$ and $\vect{l}_2$ particles), while the second symmetry is ${\cal S}_{N=2}$ ($\Phi_1 \to \Phi_1, \tilde{\Phi}_2 \to \tilde{\Phi}_2$).
Next, of main interest is the EP-NCCP1 model with species interchange symmetry $\vect{l}_1 \leftrightarrow \vect{l}_2$.
This property is in fact present here as well and is non-trivially related to the above-mentioned exact self-duality of the bosonic reformulation Eq.~(\ref{boson2pi}) as realized by individual ${\cal T}'_{N=2}$ or ${\cal C}'$.
Specifically, combining ${\cal T}'_{N=2}$ with ${\cal R}$ and $S_{N=2}$ gives the unitary symmetry interchanging $\vect{l}_1$ and $ \vect{l}_2$:
\begin{align}
{\cal S}_{N=2} {\cal R} {\cal T}'_{N=2}: \quad \Phi_1 \to \tilde{\Phi}_2 ~, \quad \tilde{\Phi}_2 \to \Phi_1 ~.
\end{align}

The EP-NCCP1 model enjoys the possibility of self-duality:\cite{shortlight,deccp_science}
Dualizing both fields gives a model of two species coupled to a new gauge field with opposite charges:
\begin{align}
i (\tilde{\vect{l}}_1 - \tilde{\vect{l}}_2) \cdot \tilde{\vect{c}} = i (\vect{j}_\text{vortex, 1} + \vect{j}_\text{boson, 2}) \cdot \tilde{\vect{c}} ~,  \label{nccp1dual}
\end{align}
where on the l.h.s., tildes mark the dual currents.
Upon an additional particle-hole transformation on $\tilde{\vect{l}}_2$, the corresponding path integral has identical long-wavelength structure as Eq.~(\ref{nccp1}) and the model can be self-dual in this sense.
The expression on the r.h.s.~is obtained by recalling that $\tilde{\vect{j}}_\text{boson, 1} = \vect{j}_\text{vortex, 1}$ and $\tilde{\vect{j}}_\text{vortex, 2} = -\vect{j}_\text{boson, 2}$ which follows from Eq.~(\ref{DualitySquared}), so using these variables we have effectively performed a particle-hole transformation on $\tilde{\vect{l}}_2$.
Comparing the content of Eqs.~(\ref{nccp1}) and (\ref{nccp1dual}), we can infer that the present EP-NCCP1 model is in fact exactly self-dual due to the symmetry ${\cal R}$ in the original $N=2$ QED$_3$ model.
More formally, ${\cal R}$ acts on the relevant operators as
\begin{align*}
{\cal R}: \quad \Phi_1 \to \Phi_2 = (\tilde{\tilde{\Phi}}_2)^\dagger ~, \quad \tilde{\Phi}_2 \to \tilde{\Phi}_1 ~,
\end{align*}
where we used Eq.~(\ref{DualitySquared}).
We can view this as implementing $\vect{l}_1 \to -\tilde{\vect{l}}_2$, $\vect{l}_2 \to \tilde{\vect{l}}_1$, which is equivalent to the above-mentioned self-duality of the EP-NCCP1 (different definitions of the self-duality are possible but become equivalent in the presence of the $\vect{l}_1$ and $\vect{l}_2$ interchange symmetry).

To summarize, the original fermionic symmetries ${\cal T}'_{N=2}$, ${\cal C}'_{N=2}$, and ${\cal R}$, together with the requirement of the self-duality ${\cal S}_{N=2}$ lead to the EP-NCCP1 model with its own species-interchange symmetry and tuned exactly to self-duality.
Such a theory was studied in Ref.~\onlinecite{Geraedts2012_pi} and was found to reside, over some parameter range, at a phase boundary between phases where $\vect{l}_{1,2}$ are both gapped and where they are both condensed.
In the specific model in that study, the phase boundary turned out to be first-order.
It is an interesting open problem whether such a phase boundary can be second-order, which would then correspond to critical self-dual $N=2$ QED$_3$.

\subsection{Dualities for generalized two-flavor models}
We can consider a more general class of $N = 2$ models with both marginally-long-range and Chern-Simons interactions of the form
\begin{align}
&i \vect{j}_\text{Dirac, 1} \cdot(\vect{a}_c + \vect{a}_n) + i \vect{j}_\text{Dirac, 2} \cdot (\vect{a}_c - \vect{a}_n) \label{eqn.twospeciesgeneralfermions} \\
&- i \gamma_c \frac{a_c da_c}{4\pi} + \lambda_c {\cal L}[a_c] - i \gamma_n \frac{a_n da_n}{4\pi} + \lambda_n {\cal L}[a_n] ~, \nonumber
\end{align}
where as in Sec.~\ref{sec.modular}
\begin{align}
{\cal L}[a] = \frac{|\vect{k} \times \vect{a}|^2}{4\pi |\vect{k}|} ~, \nonumber
\end{align}
and it is convenient to introduce ``charge'' $\vect{a}_c$ and ``neutral'' $\vect{a}_n$ ``flavors'' of the gauge fields. In general, there is no anti-unitary ${\cal T}'_{N=2}$ or ${\cal C}'_{N=2}$ symmetry, but we will assume that the unitary particle-hole symmetry ${\cal T}'_{N=2} ({\cal C}'_{N=2})^{-1}$ still holds.
In addition, we focus on cases with the species-interchange symmetry ${\cal R}$.
For $\lambda_c = \lambda_n$ and $\gamma_c = \gamma_n$, the two species of fermions decouple and are described by the $N = 1$ case discussed before.

Performing duality in the general case yields
\begin{align}
&i \vect{j}_\text{dual Dirac, 1} \cdot (\tilde{\vect{a}}_c + \tilde{\vect{a}}_n) + i \vect{j}_\text{dual Dirac, 2} \cdot (\tilde{\vect{a}}_c - \tilde{\vect{a}}_n) \\
&- i \gamma_{c, \text{~dual}} \, \frac{\tilde{a}_c d\tilde{a}_c}{4\pi} + \lambda_{c,~\text{dual}} \, {\cal L}[\tilde{a}_c] \nonumber \\
&- i \gamma_{n,~\text{dual}} \, \frac{\tilde{a}_n d\tilde{a}_n}{4\pi} + \lambda_{n,~\text{dual}} \, {\cal L}[\tilde{a}_n] ~; \nonumber
\end{align}
for the chosen conventions on the couplings and in terms of $z_{c/n} \equiv \gamma_{c/n} + i \lambda_{c/n}$, we obtain the modular relationship 
\begin{align*}
z_\text{dual} = -\frac{1}{z}
\end{align*} 
for both ``c'' and ``n'' flavors of the gauge field. A coupled-wire representation of this model and its duality is given in Appendix~\ref{app.csfermnew}. For the rest of this section, we will use schematic continuum expressions rather than explicit wire models to emphasize the long-wavelength structure. The duality/symmetry relations can often be deduced through examining the action for the dynamical gauge fields in each formulation. Using the two-flavor model of Appendix~\ref{app.csfermnew} and the techniques developed throughout this paper, it is straightforward to translate the following discussion into concrete wire models (including operator forms of the duality/symmetry relations).

The above yields three families of potential exact self-dualities: 
\begin{enumerate}[(i)]
\item $z_c = -z_{c, \text{~dual}}^*$ and $z_n = -z_{n, \text{~dual}}^*$ (i.e., $|z_{c/n}|^2 = 1$); 
\item $z_c = z_{n, \text{~dual}}$ and $z_n = z_{c, \text{~dual}}$ (i.e., $z_c z_n = -1$);
\item $z_c = -z_{n, \text{~dual}}^*$ and $z_n = -z_{c, \text{~dual}}^*$ (i.e., $z_c z_n^* = 1$).
\end{enumerate}
[Note that (i) includes the special case $z_c = z_{c, \text{~dual}}$ and $z_n = z_{n, \text{~dual}}$.]
The previously discussed case of self-dual $N = 2$ QED$_3$ formally corresponds to $\gamma_c = \gamma_n = 0$ and $\lambda_c = 1/\lambda_n \to 0$.
It is a special point that is part of both families (ii) and (iii).

In terms of the bosonic variables, the general model becomes
\begin{align}
&i \vect{j}_\text{boson, 1} \cdot (\vect{c}_c + \vect{c}_n) + i \vect{j}_\text{boson, 2} \cdot (\vect{c}_c - \vect{c}_n) \label{eqn.twospeciesgeneralbosons} \\
&- i \gamma_{c, \text{~boson}} \, \frac{c_c d c_c}{2\pi} + 2 \lambda_{c, \text{~boson}} \, {\cal L}[c_c] \nonumber \\
&- i \gamma_{n, \text{~boson}} \, \frac{c_n d c_n}{2\pi} + 2 \lambda_{n, \text{~boson}} \, {\cal L}[c_n] ~; \nonumber
\end{align}
with these conventions we find
\begin{align}
z_\text{boson} = \frac{1 + z_\text{Dirac}}{1 - z_\text{Dirac}}
\end{align}
for both ``c'' and ``n'' flavors.
Furthermore, under the bosonic duality we again have
\begin{align*}
z_\text{vortex} = -1/z_\text{boson}
\end{align*}
for both ``c'' and ``n'' flavors.
Similarly to the fermionic model, we can consider exact bosonic self-dualities of the type (i), (ii), and (iii).

We can now discuss interplay of symmetries and dualities in this general model. We first note that the fermionic self-duality of type (ii) implies bosonic self-duality of type (ii) and vice versa, and by itself is apparently not related to symmetry in either formulation.
In contrast, fermionic (bosonic) self-duality of type (iii) corresponds to a bosonic (fermionic) anti-unitary symmetry.

Let us consider case with fermionic self-duality (iii).
This implies $z_{n, \text{~boson}} = -z_{c, \text{~boson}}^*$, i.e., $\lambda_{n, \text{~boson}} = \lambda_{c, \text{~boson}} \equiv \lambda_\text{boson}$ and $\gamma_{n, \text{~boson}} = -\gamma_{c, \text{~boson}} \equiv -\gamma_\text{boson}$.
The bosonic formulation can be rewritten as
\begin{align*}
&i \vect{j}_\text{boson, 1} \cdot \vect{c}_1 + i \vect{j}_\text{boson, 2} \cdot \vect{c}_2 \\
&+ \lambda_\text{boson} \, \left( {\cal L}[c_1] + {\cal L}[c_2] \right) - i \gamma_\text{boson} \, \frac{c_1 d c_2}{2\pi} ~.
\end{align*}
Boson models of this type were studied numerically in Ref.~\onlinecite{Geraedts2012_SL2Z_longranged_loops}.
Specifically, integrating out the gauge fields $c_1$ and $c_2$ yields intra-species marginally-long-range interactions parameterized by $g_1 = g_2 = \lambda_\text{boson}/(\lambda_\text{boson}^2 + \gamma_\text{boson}^2)$ and inter-species statistical interaction parameterized by $\eta = \gamma_\text{boson}/(\lambda_\text{boson}^2 + \gamma_\text{boson}^2)$ in the notation of Ref.~\onlinecite{Geraedts2012_SL2Z_longranged_loops}, cf.~Eqs.~(1)-(3) there.
The fermionic self-duality (iii) is related to an anti-unitary symmetry of the bosonic model that acts as
$\vect{j}_\text{boson, 1}  \to \vect{j}_\text{boson, 1}$, 
$\vect{j}_\text{boson, 2}  \to -\vect{j}_\text{boson, 2}$,
$\vect{c}_1 \to -\vect{c}_1$, $\vect{c}_2 \to \vect{c}_2$.
This is identical to ${\cal S}_{N=2}$ in Eq.~(\ref{SN2}) discussed in the context of self-dual $N=2$ QED$_3$; it is precisely this symmetry that enabled a sign-free reformulation and Monte Carlo study of the model in Ref.~\onlinecite{Geraedts2012_SL2Z_longranged_loops}.

If we now in addition require bosonic self-duality condition of the type (ii), i.e., $z_{c, \text{~boson}} z_{n, \text{~nboson}} = -1$, we obtain $|z_{c, \text{~boson}}|^2 = 1$, which is equivalent to $g^2 + \eta^2 = 1$ studied in that reference and established to represent a phase transition line in the specific model for $|\eta| < 1/2$, cf.~Fig.~2 there.

Returning to the fermionic representation, we find $\lambda_{c/n} = 2\lambda_\text{boson}/[\lambda_\text{boson}^2 + (1 \pm \gamma_\text{boson})^2]$ and $\gamma_{c/n} = 0$.
These fermions have no Chern-Simons interactions but only marginally-long-range interactions with $\lambda_c$ and $\lambda_n$ related by the condition $\lambda_c \lambda_n = 1$ [this could be established more easily by noting that we now have both fermionic self-dualities (iii) and (ii) present].
Thus, we found an interesting fermionic representation of the phase transitions studied in Ref.~\onlinecite{Geraedts2012_SL2Z_longranged_loops}.

We conclude by considering simultaneous bosonic self-dualities of type (ii) and (iii).
In this case $\gamma_{c/n, \text{~boson}} = 0$ and $\lambda_{c, \text{~boson}} \lambda_{n, \text{~boson}} = 1$.
This is a special line in a more general two-parameter space with independent $\lambda_{c, \text{~boson}}$ and $\lambda_{n, \text{~boson}}$, which can be viewed as a two-species generalization of the single-species model with such interactions [this case also arises when one considers the fermionic self-duality of type (i)]. It includes the case $\lambda_c = \lambda_n = 1$ where the two species decouple and are known to be critical at self-duality.\cite{Geraedts2012_rangedloops} We expect that criticality persists over a finite range of $\lambda_c = 1/\lambda_n \neq 1$ where the two species are coupled, i.e., that over some range the special line represents a phase transition where both species go from insulating to condensed state.

\section{Conclusions and Outlook}
\label{sec.conclusions}
We have demonstrated a family of duality and statistical-transmutation mappings between bosonic and fermionc theories within a coupled-wire framework.
This technique allowed us to implement these mappings as exact, non-local transformations that leave the quantum partition function invariant.
In particular, our transformations show explicitly how local symmetry operators on one kind of variable become duality transformations on a different kind.  
A particularly interesting application of these mappings is the special point described in Sec.~\ref{sec.modular} with marginally-long-range interactions,
where both bosons and fermions have respective time-reversal symmetry and consequently both are self-dual.
The bosonic model is amenable to numerical studies\cite{Geraedts2012_rangedloops} that may be tested against field-theoretic treatments of the fermionic theory.  This last feature is shared also by the two-species models described in Sec.~\ref{sec.furtherapps}, which exhibit an even richer interplay of symmetries and dualities in various reformulations. 

While we focused primarily on relativistic theories with both particle-hole and time-reversal symmetries, these properties are not all required for the symmetry-duality relationship.
The external ``probing'' vector potential, which we mostly suppressed for clarity, can be carried through without any assumptions on it smallness. [This property has been crucial for constructing explicit (wire) models that realize particle-hole symmetric composite-Fermi liquids of fermions as well as bosons.\cite{diracduality,nu1bosons}] One can therefore dope Dirac fermions to obtain a Fermi surface, thus breaking ${\cal C}'$. The resulting non-relativistic model with only ${\cal T}'$ symmetry is still self-dual in its boson/vortex formulation.
It could be interesting to explore such models, e.g., in the context of quantum critical points with dynamical exponent $z \neq 1$.

The coupled-wire formulation provides a direct connection between the duality of the quantum Ising chain and that of the 2D boson theory via the rough correspondence $\sigma^z_i \leftrightarrow \Phi_y$ and $\sigma^x_i \leftrightarrow e^{2 i \theta_y}$.
In particular, the duality relations take a very similar form, i.e.,
\begin{align*}
&\sigma^z_i \sigma^z_{i+1} = \tau^x_{i+1/2} & \leftrightarrow & &\Phi_y^\dagger \Phi_{y+1} = e^{2 i \tilde{\theta}_{y+1/2}} ~, \\
&\tau^z_{i-1/2} \tau^z_{i+1/2} = \sigma^x_i & \leftrightarrow & &\tilde{\Phi}_{y-1/2}^\dagger \tilde{\Phi}_{y+1/2} = e^{-2 i \theta_y} ~.
\end{align*}
It is worth noting some further parallels and distinctions with the symmetry-duality relation in the  quantum Ising chain.
There we saw two natural dualities corresponding to unitary $T$ and anti-unitary $T'$ symmetries of the Majorana-fermion reformulation.
These symmetries are in general independent, but become related when the Ising model has an additional anti-unitary time-reversal symmetry $K$.
For 2D bosons, we similarly have two dualities realized by in-general independent anti-unitary ${\cal T}'$ and ${\cal C}'$ fermionic symmetries that become related when the bosonic model has additional unitary particle-hole symmetry ${\cal T} {\cal C}^{-1}$.
The 1D self-dual critical theory is more restrictive, however, in that its critical properties persist even when we break one of $T$ or $T'$; by contrast, the 2D theory changes its form qualitatively if we break ${\cal T}'$ or ${\cal C}'$.

These tantalizing correspondences with the Ising chain raise an interesting prospect that, more generally, dualities of bosons in $d+1$ dimensions can be inferred from dualities of quantum-Ising-type spin models in $d$ dimensions.
Given the wealth of spin models for which analogues of Ising duality are known, this could be a fruitful avenue to discover new mappings for bosons or fermions in two or three dimensions.

We conclude by pointing out two possible generalization of the techniques developed here. The first is the dual formulation (bosonization) of a $(2+1)$-dimensional Majorana cone, which was introduced in Refs.~\onlinecite{maxcenkemajorana,Aharony2017}.
Here, an extension of our approach (which keeps symmetries manifest at all stages) could help understand how time-reversal symmetry of the Majorana fermions is implemented on the dual variables.
A second interesting direction could be a generalization to non-Abelian symmetries. 
Ref.~\onlinecite{dccdual} analyzed the relationship between symmetry and duality for a model with SU(2) symmetry in the context of exotic quantum critical points (see also Ref.~\onlinecite{Benini2017}).
We hope that combining our approach with non-Abelian bosonization techniques could lead to such generalized duality mappings.

\textit{Note added:} 
We have recently learned that W.-H.~Hsiao and D.~T.~Son [to appear] have also studied $(2+1)$d Dirac fermions with marginally-long-range interactions at self-duality (similar to our Sec.~\ref{sec.modular}) focusing in particular on transport properties.  We thank them for sharing their manuscript with us.

\section{Acknowledgements}
We gratefully acknowledge Chong Wang, Matthew Fisher, Yin-Chen He, and D.~T.~Son for valuable discussions. This work was supported by NSF through grants  DMR-1341822 (JA) and DMR-1619696 (OIM).
We also acknowledge support by the Institute for Quantum Information and Matter, an NSF Physics Frontiers Center, with support of the Gordon and Betty Moore Foundation.

\appendix
\renewcommand{\subsection}{\hiddensubsection}\renewcommand{\subsubsection}{\hiddensubsubsection}

\section{Translation symmetry in the Majorana chain}
\label{app.ising}

We obtain the action of the unitary transformation $T$, Eq.~(\ref{TIsing}), on the spin variables as follows.
Let us write out explicitly $T: \Gamma(r-1/4) \to \Gamma(r+1/4)$ and $\Gamma(r+1/4) \to \Gamma(r+3/4)$:
\begin{align}
T:\quad
& \left(\prod_{r'=0}^{r-1} \sigma_{r'}^x \right) \sigma_r^z ~&\to~ -\left(\prod_{r'=0}^{r-1} \sigma_{r'}^x \right) \sigma_r^y ~, \label{Tdetailed1} \\
& -\left(\prod_{r'=0}^{r-1} \sigma_{r'}^x \right) \sigma_r^y ~&\to~ \left(\prod_{r'=0}^r \sigma_{r'}^x \right) \sigma_{r+1}^z ~, \label{Tdetailed2}
\end{align}
with carefully specified string operators on the chain starting at position $r'=0$.
Multiplying the two equations, we find:
\begin{align}
T:\quad \sigma_r^x \to \sigma_r^z \sigma_{r+1}^z = \tau_{r+1/2}^x ~.
\end{align}
This gives the transformation of the string $T: \prod_{r'=0}^{r-1} \sigma_{r'}^x \to \sigma_0^z \sigma_r^z$, and then from equations Eqs.~(\ref{Tdetailed1}) and (\ref{Tdetailed2}) we deduce
\begin{eqnarray}
T:\quad
& \sigma_r^z ~&\to~ i \sigma_0^z \left(\prod_{r'=0}^r \sigma_{r'}^x \right) = i \sigma_0^z \tau_{r+1/2}^z ~, \\
& \sigma_r^y ~&\to~ \sigma_0^z \left(\prod_{r'=0}^r \sigma_{r'}^x \right) \sigma_r^z \sigma_{r+1}^z = i \sigma_0^z \tau_{r+1/2}^y ~,
\end{eqnarray}
where in the last line we defined $\tau_{r+1/2}^y = -i \tau_{r+1/2}^z \tau_{r+1/2}^x$.

Similar reasoning for the anti-unitary $T'$, Eq.~(\ref{T'Ising}), gives:
\begin{eqnarray}
T': \quad 
& \sigma_r^x ~&\to~ \tau_{r+1/2}^x ~, \\
& \sigma_r^z ~&\to~ i \sigma_0^z \tau_{r+1/2}^z ~, \\
& \sigma_r^y ~&\to~ -i \sigma_0^z \tau_{r+1/2}^y ~, \\
& i ~&\to -i ~.
\end{eqnarray}
In the main text, we quoted action of $T$ and $T'$ on $\sigma_r^{x,z}$.
Both $T$ and $T'$ implement Ising duality, but clearly these are different transformations, and the meaning of self-duality is different in the two cases.

\begin{figure}[ht]
\includegraphics[width=.8\columnwidth]{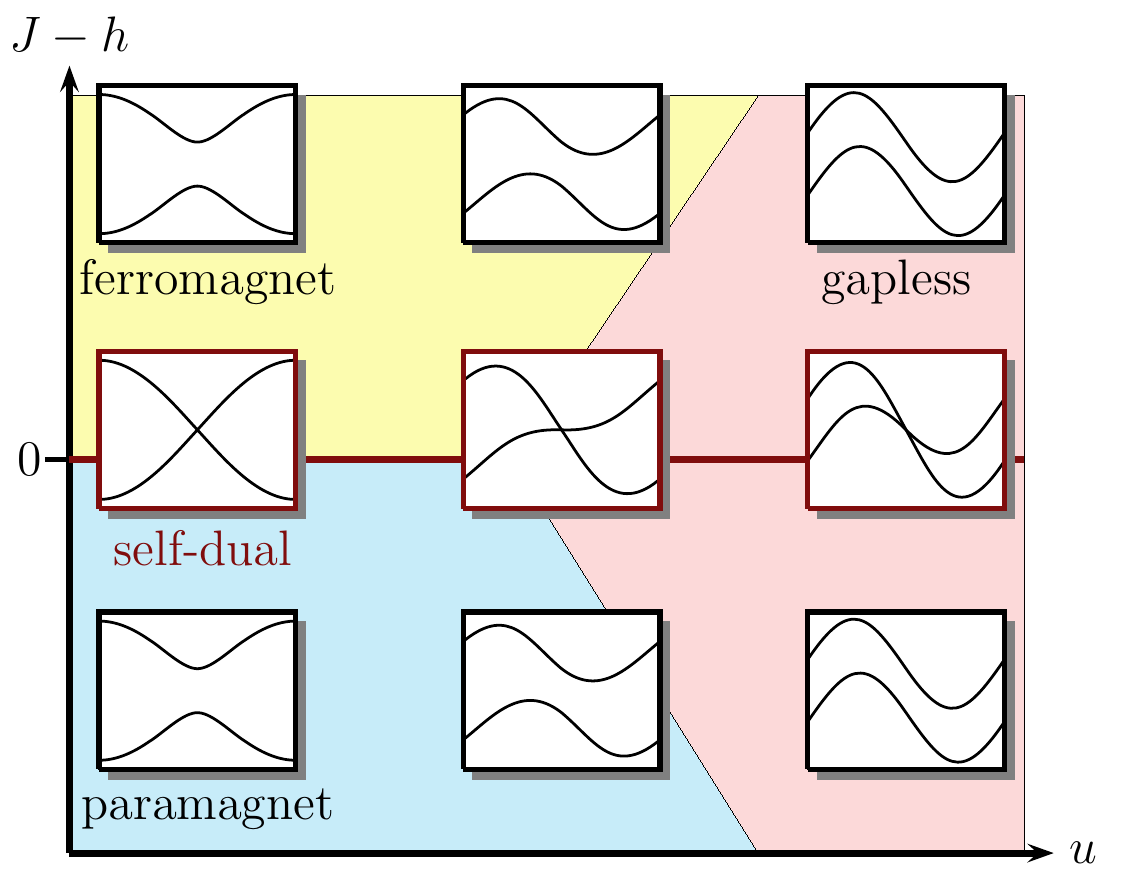}
\includegraphics[width=.8\columnwidth]{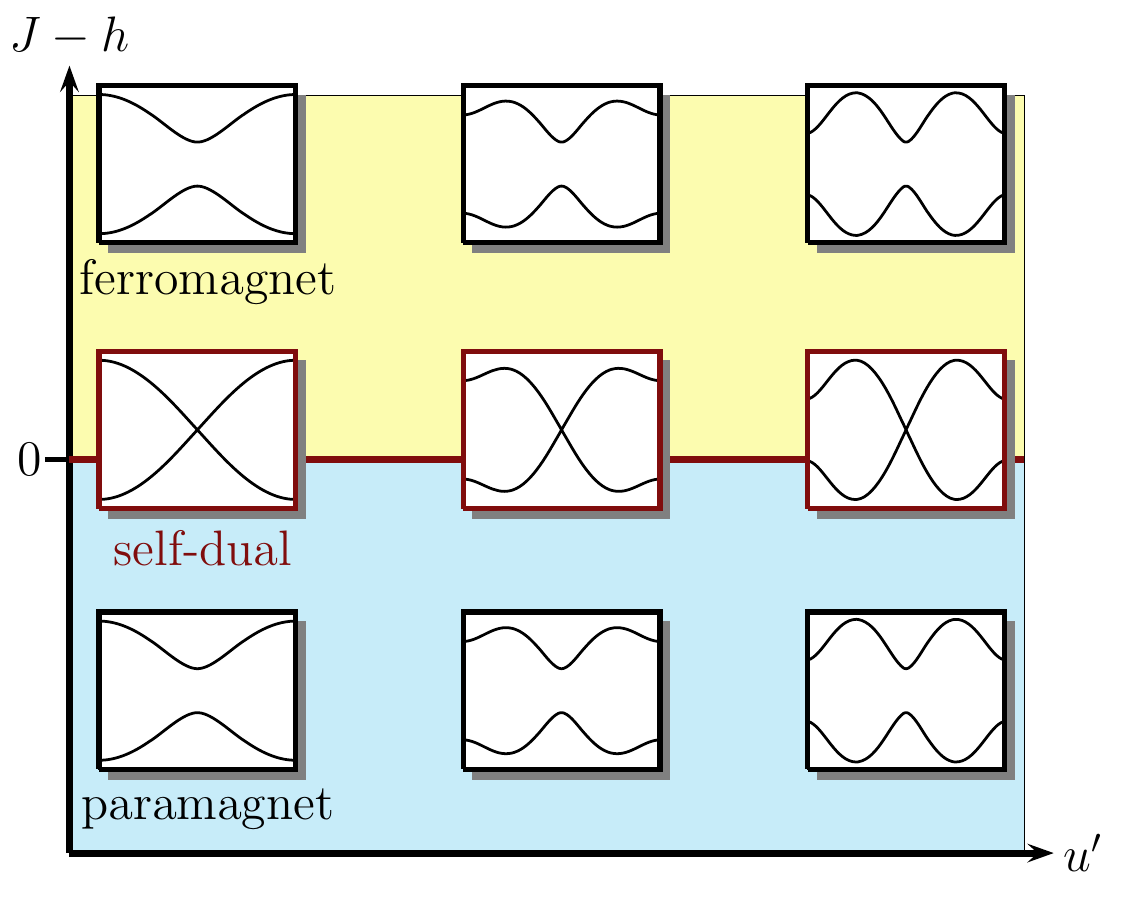}
\caption{Energy spectra of the perturbed Ising models $H_0  + \delta H$ and $H_0 + \delta H'$. The former preserves $T$ and is gapless along the self-dual line $J = h$. 
For small $u$ this line separates gapped paramagnetic and ferromagnetic phases. In contrast, for large $u$ the system enters an extended gapless phase with central charge $1$, and self-duality is not required for the gaplessness.
The model $H_0 + \delta H'$ preserves $T'$ and is gapless only at self-duality.}
\label{fig:isingspectra}
\end{figure}

To better separate aspects associated with $T$ and $T'$, we can consider situations where the Ising time reversal $K$ defined after Eq.~(\ref{T'Ising}) is broken while $T$ or $T'$ is preserved (both cannot be preserved in this case).
It is easy to write down deformations of the Ising model $H_0$ [whose various representations are written in Eqs.~(\ref{Hsigma}), (\ref{Htau}), and (\ref{Hgamma})]
that achieve this:
\begin{eqnarray*}
\delta H &=& u \sum_r \( \sigma_r^z \sigma_{r+1}^y - \sigma_r^y \sigma_{r+1}^z \) = u \sum_j i \gamma_j \gamma_{j+2} \\
&=& u \sum_r \( -\tau_{r+1/2}^y \tau_{r+3/2}^z + \tau_{r-1/2}^z \tau_{r+1/2}^y \) ~, \\
\delta H' &=& u' \sum_r \( \sigma_r^z \sigma_{r+1}^y + \sigma_r^y \sigma_{r+1}^z \) = u' \sum_j (-1)^j i \gamma_j \gamma_{j+2} \\
&=& u' \sum_r \( -\tau_{r+1/2}^y \tau_{r+3/2}^z - \tau_{r-1/2}^z \tau_{r+1/2}^y \) ~.
\end{eqnarray*}
Here each perturbation $\delta H$ or $\delta H'$ is also expressed in terms of the Majorana variables and in terms of the dual spin variables.
The corresponding Majorana energy spectra are plotted in Fig.~\ref{fig:isingspectra}
In the Majorana language, $\delta H$ is invariant under the unitary $T$, while its expression in terms of the dual spins $\tau$, upon using the translational invariance of the spin chain, has exactly the same form as in terms of the original spins $\sigma$.
On the other hand, $\delta H'$ in the Majorana language is invariant under the anti-unitary $T'$, while its expression in terms of the $\tau$ variables obtains exactly the same form as in terms of the $\sigma$ variables upon additional action of the complex conjugation $\mathcal{K}$.
In both cases, the expressions in terms of $\tau$ variables can be formally obtained from the expressions in terms of $\sigma$ variables by applying $T$ and $T'$ acting on $\sigma_r^{x,y,z}$ as specified above, remembering that $\sigma_0^z$ anticommutes with $\tau_{r+1/2}^{y,z}$.

In the simplest case with only two competing phases, such a self-duality condition (with or without $\mathcal{K}$) can guarantee that the model sits at the critical point separating the two gapped phases. 
This is the situation for small $u$ in the model $H_0 + \delta H$ and for any $u'$ in the model $H_0 + \delta H'$, see respectively top and bottom panels in Fig.~\ref{fig:isingspectra}.
On the other hand, for large $u$ in the former model, the system is in a critical phase for any $J$ and $h$; along the self-dual line $J = h$ the system has some additional properties but the self-duality itself is not required for criticality.

It is instructive to take an alternative point of view on the interpretation of duality as a symmetry. One could \emph{define} dual spins $\tau '$ through the action of the symmetries $T$ or $T'$ on $\sigma$, e.g.,
\begin{align}
\tau^{\mu\prime}_{r+1/2} \equiv T \sigma_r^\mu T^{-1} ~, \qquad \mu = x, y, z ~. \label{isingprimed}
\end{align}
(For the anti-unitary symmetry $T'$ one should instead use $\tau_{r+1/2}^{y\prime} = T' \sigma^y_r T^{-1\prime}$.)
Then, by construction, spins $\sigma$ and dual spins $\tau'$ are exactly related by the fermionic symmetry $T$.
These alternative dual variables differ from the conventional choice, $\tau$, only by a boundary term, e.g., $\tau^{x \prime} = \tau^x$ and $\tau^{z \prime} = i \sigma_0^z \tau^z$.
In particular, any local term in the Hamiltonian that preserves the global $\mathbb{Z}_2$ symmetry is identical whether expressed in terms of $\tau$ or $\tau'$.

\section{Duality of Xu-Moore and quantum-compass models}
\label{app.2dising}

Consider the Xu-Moore model, given by the $J_z = J_x = 0$ limit of Eq.~\eqref{Ising2D}, i.e.,
\begin{align}
 H_\text{XM} =&-h\sum_{\bf r} \sigma^x_{\bf r} -K \sum_{\bf r} \sigma^z_{\bf r}\sigma^z_{\bf r + \hat{x}}\sigma^z_{\bf r + \hat{y}}\sigma^z_{\bf r + \hat{x} + \hat{y}}~.
\end{align}  
In the main text we used the duality transformation 
\begin{subequations}
\begin{align}
 &\tau^z_{{\bf r + \hat{x}}/2} = \prod_{{\bf r'} < {\bf r +  \hat{x}}/2} \sigma^x_{\bf r'} ~, \\
&\tau^x_{{\bf r + \hat{x}}/2} = \sigma^z_{\bf r}\sigma^z_{\bf r + \hat{x}} ~,
\end{align}
\end{subequations}
to map it onto the quantum compass model 
\begin{align}
&H_\text{QC} = -h \sum_{\bf r} \tau^z_{{\bf r - \hat{x}}/2} \tau^z_{{\bf r + \hat{x}}/2}
-K \sum_{\bf r} \tau^x_{{\bf r + \hat{x}}/2} \tau^x_{{\bf r + \hat{x}}/2 + {\bf \hat{y}}} ~.
\end{align}  
\normalsize
(A closely related mapping between these models was performed in Ref.~\onlinecite{NussinovFradkin}.) We can now subject $H_\text{QC}$ to a second duality transformation 
\begin{subequations}
\begin{align}
 &\mu^x_{{\bf r + \hat{x}}/2+{\bf \hat{y}}/2} = \prod_{{\bf r'} < {\bf r  + \hat{x}}/2+{\bf \hat{y}}/2} 
 \tau^z_{{\bf r'+\hat{x}}/2}~,\\
 &\mu^z_{{\bf r + \hat{x}}/2+{\bf \hat{y}}/2}=
\tau^x _{{\bf r + \hat{x}}/2}
 \tau^x_{{\bf r + \hat{x}}/2+{\bf \hat{y}}}~.
\end{align}
\end{subequations}
Here, the string in the first line begins at the bottom left and runs upward through each column until the termination (in typewriter fashion, but vertical). This results in 
\begin{align}
 &H_\text{XM}' = -K\sum_{\bf r} \mu^z_{{\bf r + \hat{x}}/2+{\bf \hat{y}}/2}\\ 
 &-h\sum_{\bf r} \mu^x_{{\bf r + \hat{x}}/2+{\bf \hat{y}}/2}
 \mu^x_{{\bf r + \hat{x}}/2-{\bf \hat{y}}/2}
 \mu^x_{{\bf r - \hat{x}}/2+{\bf \hat{y}}/2}
 \mu^x_{{\bf r - \hat{x}}/2-{\bf \hat{y}}/2}~,\nonumber
\end{align}
which for $h = K$ yields the self-duality first observed in Ref.~\onlinecite{XuMoore}. The mapping between $H_\text{XM}$ and $H_\text{XM}'$  is schematically given by $\sigma \rightarrow \tau \rightarrow \mu$ and the quantum compass model lies ``half-way'' between the two. At $h = K$ it is symmetric under a 90 degree lattice rotation followed by a discrete 90 degree spin rotation about $\tau^y$ axis. 

In contrast, the model discussed in the main text is self-dual under a single mapping $\sigma \rightarrow \tau$. In this case, self-duality corresponds to a symmetry of variables that again lie ``half-way'' between $\tau$ and $\sigma$. These are the Majorana fermions $\gamma \sim \tau^z \sigma^z$ introduced in the main text.

\section{Dirac-fermion symmetries and Klein factors}
\label{app:fermsymm}
In this appendix we discuss some subtleties related to ``Klein factors"---pieces that ensure fermion anticommutation on different wires---and their transformation under symmetries.

\subsubsection{Transformation of fermion variables under ${\cal T}'$ and ${\cal C}'$}
In Sec.~\ref{sec.fermions} we defined the anti-unitary time-reversal and particle-hole symmetries for fermions as
\begin{align*}
&{\cal T}' \psi_j {{\cal T}'}^{-1} = (-1)^j \psi_{j+1} ~, \\
&{\cal C}' \psi_j {{\cal C}'}^{-1} = (-1)^j \psi_{j+1}^\dagger ~,
\end{align*}
and expressed the fermion as $\psi_j \sim e^{i \phi_j}$ with
\begin{align}
[\phi_j(x), \phi_{j'}(x')] = \ & \delta_{jj'} (-1)^j \, i \pi \, \text{sgn}(x-x') \nonumber \\
& + (1 - \delta_{jj'}) \, i \pi \, \text{sgn}(j'-j) ~, \label{fermcommapp}
\end{align}
where the first line on the right-hand-side describes chiral fermions with alternating wire chirality, and the second line ensures anticommutation of the fermion fields on different wires.
It is tempting---but incorrect---to infer that the above transformations act on the phase fields as
\begin{subequations}
\begin{align}
&{\cal T}' \phi_j {{\cal T}'}^{-1} = -\phi_{j+1} + \pi j ~, &&(\text{incorrect}) ~, \label{eqn.tfermphase} \\
&{\cal C}' \phi_j {{\cal C}'}^{-1} = \phi_{j+1} - \pi j ~, &&(\text{incorrect}) ~. \label{eqn.cfermphase}
\end{align}
\end{subequations}
To see the problem, consider applying either transformation on both sides of Eq.~(\ref{fermcommapp}).
While the first line on the right-hand-side would transform properly, the second line would obtain the wrong sign, i.e., the commutation relations would not be preserved; this indicates that such transformations of the phase variables do not exist.

To correct Eqs.~\eqref{eqn.tfermphase} and \eqref{eqn.cfermphase}, we introduce operators
\begin{align}
\xi_j \equiv 2 \pi \sum_{j'<j} N_{j'} = \sum_{j'<j} (-1)^{j'} \int_x \partial_x \phi_{j'} ~\label{xidirect},
\end{align}
where $N_{j'}$ is the fermion number operator on wire $j'$.
We note that $\kappa_j = e^{i \xi_j/2}$ is similar to a Jordan-Wigner string between wires and could be one possible representation of a Klein factor in some settings. For us $\xi_j$ will play a slightly different role, helping to maintain commutation relations of the phase fields that have Klein factors built into them.
It is easy to check that $\xi_j$ satisfies the commutation relations
\begin{align*}
&[\xi_j, \xi_{j'}] = 0 ~, \\
&[\xi_j, \phi_{j'}(x')] = 
\begin{cases} 
2 i \pi ~, & j > j'~, \\ 0 ~, & j \leq j'~, 
\end{cases} \\
&[\xi_j, \phi_{j'}(x')] + [\phi_j(x), \xi_{j'}] = 2 i \pi (1 - \delta_{jj'}) \text{sgn}(j-j') ~.
\end{align*}
We can now readily provide a faithful implementation of the two symmetries:
\begin{subequations}
\begin{align}
&{\cal T}' \phi_j {{\cal T}'}^{-1} = -\phi_{j+1} - \xi_{j+1} + \pi j ~, \label{eqn.tfermphasecorrect} \\
&{\cal C}' \phi_j {{\cal C}'}^{-1} = \phi_{j+1} + \xi_{j+1} - \pi j ~. \label{eqn.cfermphasecorrect}
\end{align}
\end{subequations}
(The signs in front of $\pi j$ in each line are not essential as the phases are defined only modulo $2\pi$. This particular choice makes treatments of ${\cal C}'$ and ${\cal T}'$ essentially identical via ${\cal C}' \phi_j {{\cal C}'}^{-1} = -{\cal T}' \phi_j {{\cal T}'}^{-1}$.
Of course, these are different transformations on the physical fermion fields $\psi_j$ and can be independent symmetries.)
Since $\phi_{j+1}$ commutes with $\xi_{j+1}$ and $e^{i \xi_{j+1}} = 1$ (verified by acting on any state in the fermion Fock space), it follows that $\psi_j = e^{i \phi_j}$ indeed transforms under ${\cal T}'$ and ${\cal C}'$ as stated at the beginning of this appendix.

We remark that care is needed when we use phase variables and encounter $\xi_j$ operators under a cosine.
While it may be tempting to drop them, we cannot do so if there are also parts that do not commute with $\xi_j$.
As an example, consider ${\cal T}'$- and ${\cal C}'$-invariant inter-wire hopping expressed in the phase variables as
\begin{align}
i \psi_j^\dagger \psi_{j+1} + \Hc = -2 \cos(\phi_j - \phi_{j+1}) ~.
\label{fermhopphivar}
\end{align}
Taking ${\cal T}'$ for concreteness (similar analysis holds for ${\cal C}'$), the expression under the cosine transforms as
\begin{align*}
\phi_j - \phi_{j+1} \to -\phi_{j+1} + \phi_{j+2} - \pi + 2\pi N_{j+1} ~.
\end{align*}
When taking cosine of the last expression, it is important to remember that $\phi_{j+1}$ and $N_{j+1}$ do not commute.
Thus, when separating out $e^{\pm i 2\pi N_{j+1}}$ from the other terms (only after this separation we can safely replace $e^{\pm i 2\pi N_{j+1}} = 1$), an extra minus sign arises via $e^{\pi [\phi_{j+1}, N_{j+1}]} = -1$.
Hence under ${\cal T}'$,
\begin{align*}
\cos(\phi_j - \phi_{j+1}) \to \cos(\phi_{j+1} - \phi_{j+2}) ~,
\end{align*}
which ensures that the right-hand-side of Eq.~(\ref{fermhopphivar}) transforms in the same way as expected from the left-hand-side.

We finally turn to the transformation properties of dual fermions.
To analyze a given wire-model, it is sufficient to understand how \textit{differences} of $\tilde{\phi}_j$ transform.
Recalling that
\begin{align*}
\tilde{\phi}_{j+1} - \tilde{\phi}_j = (-1)^{j+1} (\phi_{j+1} - \phi_j) ~,
\end{align*}
we can directly read off their transformations from Eqs.~\eqref{eqn.tfermphasecorrect} and
\eqref{eqn.cfermphasecorrect}.

For completeness we also provide expressions for the transformation of individual phase variables $\tilde{\phi}_j$.
Using Eq.~\eqref{eqn.dualfermions} we find
\begin{subequations}
\begin{align}
&{\cal T}' \tilde{\phi}_j {{\cal T}'}^{-1} = \tilde{\phi}_{j+1} + \tilde{\xi}_{j+1} - \pi j ~, \label{eqn.tdfermphasecorrect} \\
&{\cal C}' \tilde{\phi}_j {{\cal C}'}^{-1} = -\tilde{\phi}_{j+1} - \tilde{\xi}_{j+1} + \pi j ~, \label{eqn.cdfermphasecorrect}
\end{align}
\end{subequations}
with
\begin{align}
\tilde{\xi}_j = \sum_{j' \neq j} \sgn(j - j') (-1)^{j'} \xi_{j'} ~.
\end{align}
As before, the signs of the $\pi j$ terms are not essential, and we used this fact to simplify the equations.
In terms of the dual-fermion numbers $\tilde {N}_j \equiv \frac{1}{2\pi}(-1)^{j+1} \int_x \partial_x \tilde{\phi}_j$ we find
\begin{align}
\tilde{\xi}_j &= -\pi \sum_{j''} \left[|j - j''| + \text{sgn}(j - j'' + 0^+) \right] \tilde{N}_{j''} ~.
\label{xidual}\end{align}
We emphasize that the apparent difference between Eqs.~\eqref{xidirect} and \eqref{xidual} has no impact on physical operators.
Their given form merely correspond to a particular choice of fermionic Klein factors. 
\subsubsection{Transformation of bosonic variables under ${\cal T}'$ and ${\cal C}'$}
We want to translate the action of ${\cal T}'$ and ${\cal C}'$ from the fermionic to the bosonic variables.
Recall the definition of the fermionic variables in the main text,
\begin{subequations}
\begin{align}
&\phi_{2y} \equiv \phi_R(y-1/4) \equiv \varphi_y + \tilde{\varphi}_{y-1/2} \label{app.fermimonize.even}~, \\
&\phi_{2y+1} \equiv \phi_L(y+1/4) \equiv \varphi_y + \tilde{\varphi}_{y+1/2} \label{app.fermimonize.odd}~.
\end{align}
\end{subequations}
Inverting these expressions to obtain boson $\theta_y$ and vortex $\tilde{\theta}_{y+1/2}$ variables yields
\begin{align*}
&2\theta_y = \tilde{\varphi}_{y-1/2} - \tilde{\varphi}_{y+1/2} = \phi_{2y} - \phi_{2y+1} ~, \\
&2\tilde{\theta}_{y+1/2} = \varphi_{y+1} - \varphi_y = \phi_{2y+2} - \phi_{2y+1} ~.
\end{align*}
We now define new fields
\begin{subequations}
\begin{align}
\tilde{\varphi}_{y+1/2}' \equiv & -{\cal T}' \varphi_y {{\cal T}}'^{-1} = {\cal C}' \varphi_y {{\cal C}}'^{-1} ~, \label{tildevarphiprime} \\
\tilde{\theta}_{y+1/2}' \equiv & {\cal T}' \theta_y {{\cal T}}'^{-1} = -{\cal C}' \theta_y {{\cal C}}'^{-1} ~. \label{tildethetaprime} 
\end{align}
\end{subequations}
The key advantage of these new variables over $\tilde \varphi$ and $\tilde \theta$ is their simple transformation property under ${\cal T}'$ and ${\cal C}'$.  [Analogous expressions for the Ising model were introduced in Eq.~\eqref{isingprimed}.] The primed and unprimed variables are related through
\begin{align*}
&\tilde{\varphi}_{y+1/2}' - \tilde{\varphi}_{y+3/2}' = \tilde{\varphi}_{y+1/2} - \tilde{\varphi}_{y+3/2} + \pi - 2\pi N_{2y+2} ~, \\
&2\tilde{\theta}_{y+1/2}' = 2\tilde{\theta}_{y+1/2} - \pi + 2\pi N_{2y+1} ~.
\end{align*}
Using $e^{\pi [\phi_j, N_j]} = -1$, it is easy to see that 
\small
\begin{align}
&\exp[i (\tilde{\varphi}_{y+1/2}' - \tilde{\varphi}_{y+3/2}')] = \exp[i (\tilde{\varphi}_{y+1/2} - \tilde{\varphi}_{y+3/2})] ~, \label{exp_tildevarphiprime} \\
&\exp[i 2\tilde{\theta}_{y+1/2}'] = \exp[i 2\tilde{\theta}_{y+1/2}] ~. \label{exp_tildethetaprime}
\end{align}
\normalsize
Since ${\cal T}'$ is anti-unitary, by construction $\tilde{\varphi}', \tilde{\theta}'$ have the same commutation relations as the original $\varphi, \theta$, and by Eqs.~(\ref{eqn.canonicalcommutator}) and (\ref{eqn.vortexcommutator})
\begin{align*}
&[\partial_x \tilde{\theta}_{y+1/2}'(x), \tilde{ \varphi}_{y'+1/2}'(x')] = [\partial_x \tilde{\theta}_{y+1/2}(x), \tilde{\varphi}_{y'+1/2}(x')] ~.
\end{align*}
We trivially have
$\partial_x \tilde{\theta}_{y+1/2}' = \partial_x \tilde{\theta}_{y+1/2}$,
and also expect
$\partial_x \tilde{\varphi}_{y+1/2}' = \partial_x \tilde{\varphi}_{y+1/2}$ (see below).
This together with Eqs.~(\ref{exp_tildevarphiprime})-(\ref{exp_tildethetaprime}) cover all terms that can appear in the Hamiltonian, and consequently $H[\tilde{\varphi}', \tilde{\theta}'] = H[\tilde{\varphi}, \tilde{\theta}]$.
This provides a precise interpretation of the fermionic time reversal ${\cal T}'$ as boson-vortex duality.

Let us also consider the transformation of the individual phase variables $\varphi$.
For this we first need to solve Eqs.~\eqref{app.fermimonize.even},\eqref{app.fermimonize.odd} for $\varphi$ and $\tilde{\varphi}$.
A convenient choice is
\begin{align*}
&\varphi_y = \frac{1}{2} \sum_j \sgn(2y + 1/2 - j) \, (-1)^j \, \phi_j ~,\\
&\tilde{\varphi}_{y+1/2} = -\frac{1}{2} \sum_j \sgn(2y + 3/2 - j) \, (-1)^j \, \phi_j ~.
\end{align*}
[Note that the solution is not unique, since we can add a $j$-independent operator $\beta(x)$ in the first line and subtract it in the second line.
Since ${\cal T}'$ and ${\cal C}'$ also translate by one wire, it is natural to require ${\cal T}' \beta(x) {{\cal T}'}^{-1} = \beta(x) + \text{const}$, ${\cal C}' \beta(x) {{\cal C}'}^{-1} = -\beta(x) + \text{const}$ and the results below are then insensitive to the above choice.]
We find
\small
\begin{align*}
\tilde{\varphi}_{y+1/2}' &= \tilde{\varphi}_{y+1/2} + \frac{1}{2} \sum_j \sgn(2y + 1/2 - j) \, (-1)^j \, (\xi_{j+1} - \pi j) \\
 &= \tilde{\varphi}_{y+1/2} 
+ \sum_{j \in \text{even}} \sgn(2y + 1 - j) (\pi N_j - \pi/2) ~.
\end{align*}
\normalsize
The main point for us is that the second part is independent of the $x$ coordinate, and hence as discussed above we can equivalently use primed-tilde or unprimed-tilde variables in the Hamiltonian. In particular, the primed-tilde field $\tilde{\Phi}_{y+1/2}' \equiv e^{i \tilde{\varphi}_{y+1/2}'}$ satisfies by construction
\begin{align}
& {\cal T}' \Phi_y {{\cal T}'}^{-1} = \tilde{\Phi}_{y+1/2}' ~, \\
& {\cal C}' \Phi_y {{\cal C}'}^{-1} = \tilde{\Phi}_{y+1/2}'^\dagger ~,
\end{align}
while the originally defined vortex field $\tilde{\Phi}_{y+1/2}$ picks up a sign $\exp[i \pi \sum_{j \in \text{even}} N_j] = \pm 1$.
While this sign is state-dependent, it does not affect the Hamiltonian.
(Note the similarity with the discussion of alternative dual Ising variables in the last paragraph of App.~\ref{app.ising}.) When describing the action of the fermionic symmetries ${\cal T}'$ and ${\cal C}'$ on bosonic variables in Sec.~\ref{subsec.symmetries} we therefore tacitly use $\tilde{\Phi}'$ rather than $\tilde{\Phi}$; in all other parts the distinction between the two is immaterial.

\section{Continuum treatments of Chern-Simons fermions and fermionized vortices}
\label{app:CSvsFermVort}

We briefly review apparent difference between Chern-Simons fermions and fermionized vortices in continuum. Below, we will then explain how the more microscopic wire treatment allows to reconcile these differences.
Consider bosons with short-range interactions coupled to a static external (probe) field $\vect{A}$.
Attaching $2\pi$ or $-2\pi$ flux to convert bosons to Chern-Simons fermions gives, schematically:
\begin{align}
i \vect{j}_\text{bos} \cdot \vect{A}
\to ~&~ i \vect{j}_\text{CSferm} \cdot (\vect{\alpha} + \vect{A}) \mp \frac{i}{4\pi} \alpha d\alpha \label{JCSferm} \\
& = i \vect{j}_\text{CSferm} \cdot \vect{c} \mp \frac{i}{4\pi} (c - A) d(c - A) ~,
\end{align}
where $\vect{\alpha}$ is a dynamical Chern-Simons gauge field and $\alpha d\beta$ is a short-hand for $\vect{\alpha} \cdot (\vect{\nabla} \times \vect{\beta})$.
Here and below, we also implicitly understand having some short-range interactions among original bosons.
These interactions ultimately determine which phase one is in (e.g., gapped or condensed bosons or some composites), but are not important for keeping track of qualitative aspects.
As indicated in the last line, we can alternatively use $\vect{c} = \vect{\alpha} + \vect{A}$ as a new dynamical gauge field; this, however, does not change the fact that there is a Chern-Simons term governing the gauge field dynamics.

Consider now $\pm 2\pi$ flux attachment to vortices, starting with the dual description in terms of (bosonic) vortices coupled to a dynamical gauge field $\tilde{\vect{a}}$:
\begin{align*}
& i \vect{j}_\text{bos} \cdot \vect{A} \to i \vect{j}_\text{vort} \cdot \tilde{\vect{a}} + \frac{i}{2\pi} A d\tilde{a} \\
& \to i \vect{j}_\text{vortferm} \cdot (\vect{\beta} + \tilde{\vect{a}}) \mp \frac{i}{4\pi} \beta d\beta + \frac{i}{2\pi} A d\tilde{a} \\
& \quad = i \vect{j}_\text{vortferm} \cdot \tilde{\vect{c}} \mp \frac{i}{4\pi} (\tilde{c} - \tilde{a}) d(\tilde{c} - \tilde{a}) + \frac{i}{2\pi} A d\tilde{a} \\
& \to i \vect{j}_\text{vortferm} \cdot \tilde{\vect{c}} + \frac{i}{2\pi} A d\tilde{c} \pm \frac{i}{4\pi} A dA ~.
\end{align*}
In the last line, we integrated out the dynamical field $\tilde{\vect{a}}$ keeping only leading pieces, assuming short-range interactions of the original bosons.
[Note that generically we would generate also higher-order CS-like term $i (\vect{\nabla} \times \tilde{\vect{c}}) \cdot (\vect{\nabla} \times \vect{\nabla} \times \tilde{\vect{c}})$ with a non-universal coefficient, see discussion in Ref.~\onlinecite{Aliceafermvort2015}].
This is the dual vortex description which has no CS term on the dynamical gauge field $\tilde{\vect{c}}$ that $\vect{j}_\text{vortferm}$ couples to, in contrast to the theory in terms of $\vect{j}_\text{CSferm}$.
While keeping $\vect{j}_\text{CSferm}$ and $\vect{j}_\text{vortferm}$ intact, there is no way to relate these two theories.
On the other hand, in the wire construction in the main text, we claim that attaching $-2\pi$ flux to the original bosons gives identical fields to attaching $2\pi$ flux to the dual bosons. 
The resolution is that they are indeed identical fields when right- and left-movers are resolved separately, as we now demonstrate.

There are four natural choices depending whether we consider fermions or dual fermions and whether we group them around boson wires or around vortex wires.
We consider each choice in turn.

\subsubsection{Grouping fermions $\psi_R(y-1/4)$ and $\psi_L(y+1/4)$: $2\pi$ flux attachment on boson $\Phi_y$}
To connect the definition of the lattice variables to the familiar continuum treatment, we begin by defining ``Chern-Simons fermion'' phase and density fields
\begin{align}
\varphi^\text{fCS}_y &\equiv [\phi_R(y-1/4) + \phi_L(y+1/4)]/2 \\
&= \varphi_y + \sum_{y' \neq y} \sgn(y'-y) \, \theta_{y'} ~, \\
\theta^\text{fCS}_y &\equiv [\phi_R(y-1/4) - \phi_L(y+1/4)]/2 = \theta_y ~.
\end{align}
(These definitions are equivalent to $\phi_{R/L} = \varphi^\text{fCS} \pm \theta^\text{fCS}$ familiar in descriptions of 1D electrons, so these are indeed non-chiral fermion fields).
Note that the density operator of these Chern-Simons fermions is the same as the one for the microscopic bosons, $\partial_x \theta^\text{fCS}_y/\pi = \partial_x \theta_y/\pi$.

Consider the boson intra-wire terms (keeping also the Berry phase term for completeness)
\begin{align}
{\cal L} =& \sum_y \frac{i}{\pi} \, \partial_x \theta_y (\partial_\tau \varphi_y - A_{0,y}) \\
& + \sum_y \left[ \frac{v}{2\pi} (\partial_x \varphi_y - A_{1,y})^2 + \frac{u}{2\pi}(\partial_x \theta_y)^2 \right] ~.
\end{align}
This can be written as
\begin{align*}
{\cal L} =& \sum_y \frac{i}{\pi} \, \partial_x \theta^\text{fCS}_y (\partial_\tau \varphi^\text{fCS}_y - A_{0,y}) \\
& + \sum_y \left[ \frac{v}{2\pi} (\partial_x \varphi^\text{fCS}_y - a_{1,y} - A_{1,y})^2 + \frac{u}{2\pi} (\partial_x \theta^\text{fCS}_y)^2 \right] ~,
\end{align*}
where we used 
$\int_{\tau, x} \sum_y \sum_{y' \neq y} \partial_x \theta_y \, \sgn(y'-y) \, \partial_\tau \theta_{y'} = 0$
and introduced 
$a_{1,y} \equiv \sum_{y' \neq y} \sgn(y'-y) \, \partial_x \theta_{y'}$.
This satisfies $a_{1,y} - a_{1,y+1} = \partial_x \theta_y + \partial_x \theta_{y+1} = 2\pi (\rho_y + \rho_{y+1})/2$, which is naturally interpreted as attaching $2\pi$ flux to the bosons. 
We implement the definition of $a_1$ via Lagrange multipliers $a_0$ as
\begin{align}
\delta {\cal L} = & \sum_y i \, \frac{a_{0,y+1/2} - a_{0,y-1/2}}{2\pi} a_{1,y} \\
& - \sum_y i \, \frac{a_{0,y+1/2} - a_{0,y-1/2}}{2\pi} \sum_{y' \neq y} \sgn(y'-y) \, \partial_x \theta_{y'} \nonumber \\
= & \sum_y i \, \frac{a_{1,y} (\Delta a_0)_y - a_{0,y+1/2} (\Delta a_1)_{y+1/2}}{4\pi} \nonumber \\
& - \sum_y i \, \frac{\partial_x \theta_y}{\pi} \, \frac{a_{0,y+1/2} + a_{0,y-1/2}}{2} ~.
\end{align}
Putting everything together one obtains
\begin{align}
{\cal L} =& \sum_y \frac{i}{\pi} \, \partial_x \theta^\text{fCS}_y \( \partial_\tau \varphi^\text{fCS}_y - \frac{a_{0,y+1/2} + a_{0,y-1/2}}{2} - A_{0,y} \) \nonumber \\
& + \sum_y \left[ \frac{v}{2\pi} (\partial_x \varphi^\text{fCS}_y - a_{1,y} - A_{1,y})^2 + \frac{u}{2\pi} (\partial_x \theta^\text{fCS}_y)^2 \right] \nonumber \\
& + \sum_y i \, \frac{a_{1,y} (\Delta a_0)_y - a_{0,y+1/2} (\Delta a_1)_{y+1/2}}{4\pi} ~.
\end{align}
This is the wire model of a system of non-chiral fermions that are minimally coupled both to the external electromagnetic field $\vect{A}$ and to a dynamical gauge field $\vect{a}$ with a Chern-Simons term (expressed in the gauge $a_2 = 0$).
The structure is similar to performing flux attachment in the continuum, cf.~Eq.~(\ref{JCSferm}). Note that slight care is needed when comparing the sign of the CS term in our wire model with the continuum writing in Eq.~(\ref{JCSferm}).
In the wire model our conventions dictate that the density couples to $a_0$ as $-i \rho a_0$, which is opposite to the convention used in Eq.~(\ref{JCSferm}).
Thus, the above wire model corresponds to attaching $2\pi$ flux to the boson, even though the sign of the CS term is opposite to what we called $2\pi$ flux attachment in Eq.~(\ref{JCSferm}).

\subsubsection{Grouping dual fermions $\tilde{\psi}_L(y-1/4)$ and $\tilde{\psi}_R(y+1/4)$: $-2\pi$ flux attachment on boson $\Phi_y$}
Let us now consider using dual chiral fermions, Eq.~(\ref{dualleftferm})-(\ref{dualrightferm}), grouped around boson wires $y$ to define new non-chiral fermions
\begin{align}
\varphi^\text{fCS-}_y &\equiv [\tilde{\phi}_R(y+1/4) + \tilde{\phi}_L(y-1/4)]/2 \\
&= \varphi_y - \sum_{y' \neq y} \sgn(y'-y) \, \theta_{y'} ~, \\
\theta^\text{fCS-}_y &\equiv [\tilde{\phi}_R(y+1/4) - \tilde{\phi}_L(y-1/4)]/2 = \theta_y ~.
\end{align}
Comparing with the variables $\varphi^\text{fCS}_y, \theta^\text{fCS}_y$ and manipulations leading to their interpretation as $2\pi$ flux attachment to bosons, we immediately see that $\varphi^\text{fCS-}_y, \theta^\text{fCS-}_y$ correspond to exactly opposite flux attachment on the original bosons, which we indicated by the minus sign in the label ``fCS-''.
Which composite fermion variables to use, ``fCS'' or ``fCS-'', of course depends on the problem at hand.
For example, if we have bosons in an external magnetic field, in typical fractional quantum Hall applications we would strive to have the average CS flux cancel the external field.
Importantly, we note here that there is no local transformation between the ``fCS'' and ``fCS-'' variables.

\subsubsection{Grouping dual fermions $\tilde{\psi}_R(y+1/4)$ and $\tilde{\psi}_L(y+3/4)$: $2\pi$ flux attachment on vortex $\tilde{\Phi}_{y+1/2}$}
Consider now grouping $\tilde{\phi}_R(y+1/4)$ and $\tilde{\phi}_L(y+3/4)$ which gives non-chiral fields residing on half-integer (vortex) wires:
\begin{align}
\varphi^\text{fv}_{y+1/2} &\equiv [\tilde{\phi}_R(y+1/4) + \tilde{\phi}_L(y+3/4)]/2 \\
&= -\tilde{\varphi}_{y+1/2} - \sum_{y' \neq y} \sgn(y'-y) \, \tilde{\theta}_{y'+1/2} ~, \\
\theta^\text{fv}_{y+1/2} &\equiv [\tilde{\phi}_R(y+1/4) - \tilde{\phi}_L(y+3/4)]/2 = -\tilde{\theta}_{y+1/2} ~.
\end{align}
Thus, $-\varphi^\text{fv}, -\theta^\text{fv}$ are related to $\tilde{\varphi}, \tilde{\theta}$ in exactly the same way as $\varphi^\text{fCS}, \theta^\text{fCS}$ are related to $\varphi, \theta$.
A moment's thought shows that we can then interpret $\varphi^\text{fv}, \theta^\text{fv}$ as describing $2\pi$ ``dual flux'' attachment to vortices.
[An alternative argument is to note that we can view $-\tilde{\phi}_{R/L}$ as obtained from the vortex variables $\tilde{\varphi}, \tilde{\theta}$ by applying identical procedure (with the same ``orientation'') as $\phi_{R/L}$ from $\varphi, \theta$:
The procedure Eq.~(\ref{rightferm}) applied to the vortex fields $\tilde{\varphi}, \tilde{\theta}$ would give at wire $y+1/4$ a composite $\tilde{\varphi}_{y+1/2} + \tilde{\tilde{\varphi}}_y$, but since $\tilde{\tilde{\varphi}} = -\varphi$, this gives exactly $-\tilde{\phi}_R(y+1/4)$, and similarly for the left-moving field at $y+3/4$.]

\subsubsection{Grouping fermions $\psi_L(y+1/4)$ and $\psi_R(y+3/4)$: $-2\pi$ flux attachment on vortex $\tilde{\Phi}_{y+1/2}$}
Finally, this grouping gives
\begin{align}
\varphi^\text{fv-}_{y+1/2} &\equiv [\phi_R(y+3/4) + \phi_L(y+1/4)]/2 \\
&= \tilde{\varphi}_{y+1/2} - \sum_{y' \neq y} \sgn(y'-y) \, \tilde{\theta}_{y'+1/2} ~, \\
\theta^\text{fv-}_{y+1/2} &\equiv [\phi_R(y+3/4) - \phi_L(y+1/4)]/2 = \tilde{\theta}_{y+1/2} ~.
\end{align}
This is naturally interpreted as $-2\pi$ flux attachment on vortices, which we marked as ``fv-''.

\subsubsection{Equivalence between Chern-Simons fermions and fermionized vortices and resolution of the CS term puzzle}
Note that the ``fCS'' and ``fv-'' fields are obtained by different local groupings of the same chiral fermion fields (and similalry ``fCS-'' and ``fv'' are different local groupings of the dual chiral fermions).
Focusing on the first pair, this means that there is a local relation between the ``fCS'' and ``fv-'' fields, which is easy to find explicitly:
\begin{align}
\varphi^\text{fv-}_{y+1/2} = & \left( \varphi^\text{fCS}_{y+1} + \theta^\text{fCS}_{y+1} + \varphi^\text{fCS}_y - \theta^\text{fCS}_y \right)/2 ~, \\
\theta^\text{fv-}_{y+1/2} = & \left( \varphi^\text{fCS}_{y+1} + \theta^\text{fCS}_{y+1} - \varphi^\text{fCS}_y + \theta^\text{fCS}_y \right)/2 ~.
\end{align}
In this sense, at the microscopic wire level, Chern-Simons fermions ``fCS'' and fermionized vortices ``fv-" are essentially the same objects.
This appears to pose a puzzle relating to the continuum treatment in App.~\ref{app:CSvsFermVort}, where the ``fCS'' fermions are coupled to a gauge field with the Chern-Simons term, while the ``fv-'' fermions are coupled to a gauge field with no Chern-Simons term, and there is no simple way to relate these two formulations via continuum manipulations.
The resolution is that microscopic densities and currents of the Chern-Simons fermions and fermionic vortices are different.
As a consequence, we will see that the theory of the "fCS" fermions coupled to the dynamical gauge field $\vect{a}$ with the Chern-Simons term can be exactly translated to the ``fv-'' fermions coupled to a new gauge field with no Chern-Simons term.

For simplicity, let us consider a setup where both $a_0$ and $a_1$ reside on the same wires and the Chern-Simons term is written as $i (a_{1,y+1} + a_{1,y})(a_{0,y+1} - a_{0,y})/(4\pi)$, cf.~App.~\ref{app.onwireboson}.
(Treatment where $a_0$ resides between boson wires as in the earlier presentation of the ``fCS'' fermions is more tedious but leads to the same qualitative conclusion.)
The ``fCS'' fermion coupling to the gauge field $\vect{a}$ can be rewritten in terms of the ``fv-'' fermion coupling to a new gauge field $\vect{\tilde{a}}$ (residing on the vortex wires) as follows: 
\begin{align}
&-\sum_y \left( \frac{i}{\pi} \partial_x \theta^\text{fCS}_y \, a_{0,y} + \frac{v}{\pi} \partial_x \varphi^\text{fCS}_y \, a_{1,y} \right) \\
&= -\sum_y \left( \frac{i}{\pi} \partial_x \theta^\text{fv-}_{y+1/2} \, \tilde{a}_{0,y+1/2} + \frac{v}{\pi} \partial_x \varphi^\text{fv-}_{y+1/2} \, \tilde{a}_{1,y+1/2} \right) \nonumber
\end{align}
with
\begin{align*}
&a_{0,y} = \frac{1}{2} (S \tilde{a}_0)_y + \frac{i v}{2} (\Delta \tilde{a}_1)_y ~, \\
&a_{1,y} = \frac{1}{2} (S \tilde{a}_1)_y - \frac{i}{2 v} (\Delta \tilde{a}_0)_y ~,
\end{align*}
where we introduced short-hand notation $(S \tilde{a}_0)_y = \tilde{a}_{0,y+1/2} + \tilde{a}_{0,y-1/2}$, $(\Delta \tilde{a}_0)_y = \tilde{a}_{0,y+1/2} - \tilde{a}_{0,y-1/2}$, and similarly for $(S \tilde{a}_1)$ and $(\Delta \tilde{a}_1)$.
We can now plug these expressions into terms in the Lagrangian that are quadratic in the gauge field $\vect{a}$, which consist of the ``diamagnetic'' and Chern-Simons pieces:
\begin{align*}
\sum_y \left[ \frac{v}{2\pi} a_{1,y}^2 + \frac{i}{4\pi} (a_{1,y+1} + a_{1,y})(a_{0,y+1} - a_{0,y}) \right] ~.
\end{align*}
It is easy to see that the diamagnetic piece gives a cross term $-i/(4\pi) (S\tilde{a}_1) (\Delta \tilde{a}_0)$ which cancels the leading long-wavelength cross term from the Chern-Simons piece.
When all contributions are written out microscopically, we find a diamagnetic piece for the $\tilde{a}_1$ field as well as Maxwell-like pieces for the $\tilde{a}_1$ and $\tilde{a}_0$ fields. The leading cross term,
\begin{align*}
 \frac{i}{16\pi}\sum_y  \left[(\Delta \tilde{a}_0)_{y+1} +(\Delta \tilde{a}_0)_y \right]\left[   (\Delta \tilde{a}_1)_{y+1} -  (\Delta \tilde{a}_1)_y\right] ~,
\end{align*}
effectively contains three derivatives in $y$.

To conclude, the regrouping of the chiral constituents at the wire level indeed allows us to connect the Chern-Simons fermion theory with CS term to the fermionized vortex theory with no CS term.
The key point is that this regrouping mixes densities $\rho = \partial_x \theta/\pi$ and currents $j = v \partial_x \phi/\pi$ when going between the ``fCS'' and ``fv-'' fermions:
At long wavelengths,
\begin{align}
\rho^\text{fv-} &= \rho^\text{fCS} + \frac{1}{2v} \partial_y j^\text{fCS} ~, \\
j^\text{fv-} &= j^\text{fCS} + \frac{v}{2} \partial_y \rho^\text{fCS} ~,
\end{align}
which corresponds to long-wavelength version 
$a_0 = \tilde{a}_0 + i (v/2) \partial_y \tilde{a}_1$,
$a_1 = \tilde{a}_1 - i/(2v) \partial_y \tilde{a}_0$
of the above transformation between the gauge fields.
Such possibility is lost when one is working with continuum complex fermion fields for the Chern-Simons fermions and the fermionized vortices, and this resolves the above-mentioned puzzle when we said that microscopically they are essentially the same objects.

\section{Alternate model for bosons with Chern-Simons coupling}
\label{app.onwireboson}

In this section we demonstrate that the same non-local boson action,
${\cal L}_\varphi + {\cal L}_\theta + {\cal L}_{\varphi, \theta}$ in Eqs.~(\ref{lphidual})-(\ref{lthetadual}),
can be obtained when starting with a gauge field whose temporal component lives on the wires.
Specifically, we consider the following model
\begin{align}
{\cal L} = {\cal L}_0 + {\cal L}_\text{CS} + {\cal L}_\text{MW} \label{app.csbosons}
\end{align}
with
\begin{align*}
{\cal L}_0 =& \frac{v}{2\pi} (\partial_x \varphi - a_1)^2 + \frac{u}{2\pi} (\partial_x \theta)^2 \\
& -\frac{i}{\pi} \partial_x \theta \, a_0 + \frac{\lambda}{8\pi} (\Delta \partial_x \varphi)^2 ~, \\
{\cal L}_\text{CS} =& \frac{i}{4\pi n} (S a_1) (\Delta a_0) ~, \\
{\cal L}_\text{MW} =& \frac{1}{8\pi} \left[\frac{\alpha}{v} (\Delta a_0)^2 + \beta v (\Delta a_1)^2 \right] ~.
\end{align*}
Here, $a_{0,y}$ and $a_{1,y}$ reside on the same wires as the boson fields; $(\Delta a_0)_y \equiv a_{0,y+1} - a_{0,y}$ and similarly for $(\Delta a_1)_y$; and $(S a_1)_y \equiv a_{1,y+1} + a_{1,y}$.
We also used schematic vector notation with implicit indices running over the wire labels $y$.
Matrix representation of operators $\Delta$ and $S$ and some useful identities are reviewed in App.~\ref{app.matrix}.
We note that the above model can be turned into a gauge-invariant Lagrangian in terms of $(a_0, a_1, a_2)$ written in the gauge $a_2 = 0$; we will maintain this gauge throughout.

Integrating out $a_0$ yields
\begin{align*}
{\cal L} =& \frac{v}{2\pi} (\partial_x \varphi - a_1)^2 + \frac{u}{2\pi} (\partial_x \theta)^2 + \frac{\lambda}{8\pi} (\Delta \partial_x \varphi)^2 \\ 
&+ \frac{2v}{\pi \alpha} \left( \partial_x \Delta^{-1,T} \theta \right)^2 - \frac{v}{\pi \alpha n} \left( S a_1 \right) \left( \partial_x \Delta^{-1,T} \theta \right) \\
&+ \frac{\beta v}{8\pi} (\Delta a_1)^2 + \frac{v}{8\pi \alpha n^2} \left( S a_1 \right)^2 ~.
\end{align*}
Using Eq.~(\ref{STSpDTD}), the quadratic terms in $a_1$ combine to
\begin{align*}
\frac{v}{2\pi} (a_1) \left[ 1 + \frac{1}{\alpha n^2} + \frac{\beta - \frac{1}{\alpha n^2}}{4} \Delta^T \Delta \right] (a_1) \equiv \frac{v}{2\pi} (a_1) M (a_1) ~.
\end{align*}
We will need the inverse of the matrix $M$, and for reasons that will become clear momentarily, we write it as
\begin{align*}
& M^{-1} = \frac{\alpha n^2}{1 + \alpha n^2} + \Delta^T W \Delta ~, \\
& W = \frac{(1 - \beta \alpha n^2) \alpha n^2}{4 (1 + \alpha n^2)^2} \left[ 1 + \frac{\beta \alpha n^2 - 1}{4 (1 + \alpha n^2)} \Delta^T \Delta \right]^{-1} ~,
\end{align*}
which can be checked by simple algebra, also remembering that $\Delta$ and $\Delta^T$ commute.
It is easy to see that for sufficiently small $|\beta \alpha n^2 - 1|$, matrix elements $W_{y,y'}$ decay exponentially with $|y-y'|$.
Now integrating out $a_1$ we find
\begin{align*}
{\cal L} =& \frac{v}{2\pi} (\partial_x \varphi)^2 + \frac{u}{2\pi} (\partial_x \theta)^2 + \frac{\lambda}{8\pi} (\Delta \partial_x \varphi)^2 \\ 
&+ \frac{2v}{\pi \alpha} \left( \partial_x \Delta^{-1,T} \theta \right)^2 \\
&- \frac{v}{2\pi} \frac{\alpha n^2}{1 + \alpha n^2} \left( \partial_x \varphi + \frac{1}{\alpha n} \partial_x S^T \Delta^{-1,T} \theta \right)^2 \\
&- \frac{v}{2\pi} \left( \partial_x \Delta \varphi - \frac{1}{\alpha n} \partial_x S \theta \right) W \left( \partial_x \Delta \varphi - \frac{1}{\alpha n} \partial_x S \theta \right) ~,
                                                                                                                                                                              \end{align*}
where we used $\Delta S^T = -S \Delta^T$ to explicitly show that the combinations of fields multiplying $W$ are local.

Defining $v_B = \frac{v}{1 + \alpha n^2}$ we finally get
\begin{align*}
{\cal L} =& \frac{v_B}{2\pi} (\partial_x \varphi)^2 + \frac{u + v_B/\alpha}{2\pi} (\partial_x \theta)^2 + \frac{\lambda}{8\pi} (\Delta \partial_x \varphi)^2 \\
&+ \frac{v_B n^2}{2\pi} \left( 2 \partial_x \Delta^{-1,T} \theta \right)^2 \\
&- \frac{v_B n}{\pi} (\partial_x \varphi) \left( \partial_x S^T \Delta^{-1,T} \theta \right) \\
&- \frac{v}{2\pi} \left( \partial_x \Delta \varphi - \frac{1}{\alpha n} \partial_x S \theta \right) W \left( \partial_x \Delta \varphi - \frac{1}{\alpha n} \partial_x S \theta \right) ~,
\end{align*}
where we again used Eq.~(\ref{STSpDTD}) for terms quadratic in $\theta$.
The last term describes exponentially decaying inter-wire interactions, which are expected to be present in generic models but do not effect any universal properties.
In the special case $\beta = 1/(\alpha n^2)$, we have $W = 0$ and such interactions are absent.

Focusing on the second and third lines in the above equation, it is straightforward to check that $2 (\Delta^{-1,T} \theta)_y =  \tilde{\varphi}_{y+1/2}$ and $2 (S^T \Delta^{-1,T} \theta)_y = \tilde{\varphi}_{y+1/2} + \tilde{\varphi}_{y-1/2}$.
We can then see that for $W = 0$ the above Lagrangian coincides with the model described by Eqs.~(\ref{lphi})-(\ref{lthetathetaphi}) for $n = 1$ and an appropriate choice of the parameters $u, v, \lambda, \alpha$:
Namely, $v_B$ here corresponds to $v$ in Eqs.~(\ref{lphi})-(\ref{lthetathetaphi}), $u + v_B/\alpha$ corresponds to $u - v$, and $\lambda$ corresponds to $\tilde{u} - v$.
All discussions at the end of Sec.~\ref{subsec.boson+CS} now apply.
Since $\Delta \varphi = 2 \tilde{\theta}$, the Lagrangian is clearly self-dual when $u + v_B/\alpha = \lambda$.

We conclude by re-iterating that the above special choice of the coupling $\beta$ in the Maxwell terms leads to a model that is identical to the one in Eqs.~(\ref{lphi})-(\ref{lthetathetaphi}). 
The latter was obtained in Sec.~\ref{subsec.boson+CS} from the theory that had $a_0$ residing between wires and had only Chern-Simons term.
Thus, special finite Maxwell term for $a_0$ on the wires corresponds to zero Maxwell term for $a_0$ between wires.

As remarked earlier, generic Maxwell terms will not change the universal properties of the critical point, while our choices allow us to find exact parameters for the self-duality on the microscopic wire scale and hence criticality.

\section{Integrating out the Chern-Simons field in a fermion wire model}

\label{app.csfermnew}

Here we provide details of the treatment of the model in Sec.~\ref{subsec.ferm-CS}.
We adopt the matrix notation of App.~\ref{app.matrix} to write the action ${\cal L}^\text{ferm-CS}$ concisely as
\begin{align}
&{\cal L}^\text{ferm-CS} = {\cal L}_0 + {\cal L}_\text{staggered-CS} + {\cal L}_\text{MW} + {\cal L}_\text{CS} ~, \\
&{\cal L}_0 = - \frac{i P \partial_x \phi \, a_0}{2\pi} + \frac{u}{4\pi}(\partial_x \phi - a_1)^2 ~, \nonumber \\
&{\cal L}_\text{staggered-CS} = -i \frac{P}{8\pi}(\Delta a_0)(S a_1) ~, \nonumber \\
&{\cal L}_\text{MW} = \frac{\beta}{16\pi} \left[\frac{1}{v} (\Delta a_0)^2 + v (\Delta a_1)^2 \right] ~, \nonumber \\
&{\cal L}_\text{CS} = \frac{i}{8\pi f_F}(\Delta a_0)(S a_1) ~. \nonumber
 \end{align}
Here the matrix indices refer to the fermionic wire labels $j$ rather than the bosonic labels $y$, but all matrix identities in App.~\ref{app.matrix} hold (and we only use fermionic wires in this section).
In the above model, we also allowed for a more general coefficient of the CS term and a more general Maxwell term parameterized by parameters $\beta$ and $v$.

The coupling between $\partial_x \phi$ and $a_0$ can be written as
\begin{align}
-\frac{i a_0 P \partial_x \phi}{2\pi} = \frac{i (\Delta a_0) (D + P) \partial_x \phi}{4\pi} ~,
\end{align}
where we used matrix identity Eq.~\eqref{DeltaDplusPeq2P}.
Integrating out $a_0$ then yields
\begin{align*}
& {\cal L}^\text{ferm-CS} \to
 \frac{u}{4\pi}(\partial_x \phi - a_1)^2 + \frac{\beta v}{16\pi} (\Delta a_1)^2 \\
& + \frac{v}{16\pi \beta} a_1 
\left[4 + \frac{4}{f_F^2} - (1 + \frac{1}{f_F^2}) \Delta^T \Delta + \frac{2}{f_F} \Delta^T P \Delta \right] a_1 \\
& + \frac{v}{4\pi \beta} \partial_x \phi (D^T D + 1) \partial_x \phi + \frac{v}{2\pi \beta} a_1 \left(\frac{1}{f_F} D - 1 \right) \partial_x \phi ~.
\end{align*}
In the intermediate steps, we used the matrix identities Eqs.~\eqref{STSpDTD},\eqref{DPeqPDT},\eqref{STDplusPeq2D},\eqref{DeltaPDelta}.
We can now integrate out $a_1$, which requires inverting a matrix of the form $A + B \Delta^T \Delta + C \Delta^T P \Delta$.
This inversion can in principle be carried out by using 
$(\Delta^T \Delta)_{j,j'} = 2\delta_{j,j'} - \delta_{j,j'+1} - \delta_{j,j'-1}$, 
$(\Delta^T P \Delta)_{j,j'} = (-1)^j (\delta_{j,j'+1} - \delta_{j,j'-1})$,
and using Fourier transform in the wire label.
One finds two bands of eigenvalues parameterized by the momentum $Q$ in the $y$ direction.
Assuming $A \gg B > |C|$, one finds that the smallest eigenvalues are near zero momentum and are given by
\begin{align}
E(Q) = A + \frac{B^2 - C^2}{4 B} Q^2 ~.
\end{align}
It follows that under these conditions
\begin{align}
[A + B \Delta^T \Delta + C \Delta^T P \Delta]^{-1} = \frac{1}{A} + \mathcal{O}(\Delta^2) ~.
\end{align}
Crucially, only the leading term $\sim A^{-1}$ enters in the universal properties, while the derivative terms correspond to exponentially decaying interactions.
Still, in the special case $C = \pm B$ the inverse can be obtained analytically by virtue of the identity Eq.~(\ref{1pPDTD1pP}):
\begin{align*}
\left[ A + B \Delta^T (1 \pm P) \Delta \right]^{-1} = \frac{1}{A} - \frac{B}{A(A + 4B)} \Delta^T (1 \pm P) \Delta ~.
 \end{align*}
The above condition is satisfied for special $\beta = 1 + 1/|f_F|$, which gives $C = \sgn(f_F) B$.
We will assume this $\beta$ below.
 
Integrating out $a_1$ using the above formula, one finds after lengthy but straightforward algebra
\begin{align}
{\cal L}^\text{CS-ferm} \to \ \ \ &  \frac{v_B}{16\pi}(\partial_x \phi + f_F \partial_x \tilde{\phi})^2 \label{eqn.finalcsfermions} \\
& + \frac{u_B}{16\pi} \left[1 + \sgn(f_F) P \right] \left(\Delta \partial_x \phi \right)^2 ~, \nonumber
\end{align}
with the dual $\tilde{\phi} = D \phi$ and
\begin{align*}
v_B &= 4v \, \frac{|f_F|(u + v) + u}{(|f_F| + 1) \bigg[f_F^2 (u + v) + |f_F| u + v \bigg]} ~, \\
u_B &= v \, \frac{2f_F^2 (u + v)^2}{ \bigg[|f_F|(u + v) + v \bigg] \bigg[f_F^2 (u + v) + |f_F| u + v \bigg]} ~.
\end{align*}

\subsection{Generalization to two fermion species}
We now sketch the treatment of multiple fermions species, i.e., the model of Eq.~\eqref{eqn.twospeciesgeneralfermions}.
We can write a coupled-wire model realization as
\begin{align}
&{\cal L}^\text{ferm-CS}_{N=2} = {\cal L}_0 + {\cal L}_\text{staggered-CS} + {\cal L}_\text{MW} + {\cal L}_\text{CS} ~, \\
&{\cal L}_0 = -\frac{i P \partial_x \phi_1 \, (a_{0,c} + a_{0,n})}{2\pi} - \frac{i P \partial_x \phi_2 \, (a_{0,c} - a_{0,n})}{2\pi} \nonumber \\
& \quad\ \ \ + \frac{u}{4\pi} \left[ (\partial_x \phi_1 - a_{1,c} - a_{1,n})^2 + (\partial_x \phi_2 - a_{1,c} + a_{1,n})^2 \right] ~, \nonumber \\
&{\cal L}_\text{staggered-CS} = -i \frac{P}{4\pi} (\Delta a_{0,c})(S a_{1,c}) - i \frac{P}{4\pi} (\Delta a_{0,n})(S a_{1,n}) ~, \nonumber \\
&{\cal L}_\text{MW} = \frac{\beta_c}{8\pi} \left[\frac{1}{v_c} (\Delta a_{0,c})^2 + v_c (\Delta a_{1,c})^2 \right] \nonumber \\
&\quad\ \ \ + \frac{\beta_n}{8\pi} \left[\frac{1}{v_n} (\Delta a_{0,n})^2 + v_n (\Delta a_{1,n})^2 \right] ~, \nonumber \\
&{\cal L}_\text{CS} = \frac{i}{4\pi f_c} (\Delta a_{0,c})(S a_{1,c}) + \frac{i}{4\pi f_n} (\Delta a_{0,n})(S a_{1,n}) ~. \nonumber
\end{align}
To model marginally-long-range interactions, $\lambda_{c/n} \neq 0$, we can simply include additional terms $\sim (\partial_x \tilde{\phi}_{c/n})^2$, similar to the single-flavor case described in Sec.~\ref{sec.modular}. 

We can integrate out the gauge fields and obtain the model in terms of $\phi_{1/2}$ and $\tilde{\phi}_{1/2}$ fields as follows.
We rescale the gauge fields $\vect{a}_{c/n} \to \vect{a}_{c/n} / \sqrt{2}$ and define $\phi_{c/n} = (\phi_1 \pm \phi_2) / \sqrt{2}$ to get 
${\cal L}^\text{ferm-CS}_{N=2} = \sum_{i = c, n} \left[ {\cal L}_{0,i} + {\cal L}_{\text{staggered-CS}, i} + {\cal L}_{\text{MW}, i} + {\cal L}_{\text{CS}, i} \right]$ with
\begin{align}
&{\cal L}_{0,i} = -\frac{i P \partial_x \phi_i \, a_{0,i}}{2\pi} +  \frac{u}{4\pi} (\partial_x \phi_i - a_{1,i})^2 ~, \nonumber \\
&{\cal L}_{\text{staggered-CS}, i} =
-i \frac{P}{8\pi} (\Delta a_{0,i})(S a_{1,i}) ~, \nonumber \\
&{\cal L}_{\text{MW}, i} = \frac{\beta_i}{16\pi} \left[\frac{1}{v_i} (\Delta a_{0,i})^2 + v_i (\Delta a_{1,i})^2 \right] ~, \nonumber \\
&{\cal L}_{\text{CS}, i} = \frac{i}{8\pi f_i} (\Delta a_{0,i})(S a_{1,i}) ~. \nonumber
\end{align}
This coincides with two copies of the model described at the start of this Appendix, and hence integrating out the $\vect{a}_i$ gauge field for special $\beta_i = 1 + 1/|f_i|$ yields Eq.~\eqref{eqn.finalcsfermions} separately for $\phi_c$ and $\phi_n$.
Note that $\phi_c$ and $\phi_n$ terms remain coupled in the presence of inter-wire tunneling terms.
The (non-local) linear transformation between fermionic and bosonic phase fields can be performed directly in the ``spin'' and ``charge'' variables, leading to coupled-wire model with two boson species corresponding to the schematic model in Eq.~\eqref{eqn.twospeciesgeneralbosons}. One can then establish precise relations among various self-dualities and symmetries similar to other coupled-wire examples.

\section{Matrix notation and identities for coupled wires}
\label{app.matrix}

The analysis of coupled-wire models, especially those with non-local terms, is greatly facilitated by adopting the matrix notation introduced in Ref.~\onlinecite{diracduality}.
In the main text, we already introduced the lattice derivative $\Delta$ and its inverse $\Delta^{-1}$ as
\begin{align}
&\Delta_{y,y'} = \delta_{y+1,y'} - \delta_{y,y'} ~, \\
&\Delta^{-1}_{y,y'} = \frac{1}{2} \text{sgn}(y - y' - 0^+) ~, \\
&\sum_{y''} \Delta_{y,y''} \Delta^{-1}_{y'',y'} = \delta_{y,y'} ~.
\end{align}
We further define
\begin{align}
S_{y,y'} = \delta_{y+1,y'} + \delta_{y,y'} ~,
\end{align}
which commutes with the derivative, $S \Delta = \Delta S$, and further satisfies
\begin{align}
&S^T \Delta = -\Delta^T S ~, \quad 
S \Delta^T = -\Delta S^T ~, \label{SDT} 
\\
&S^T S + \Delta^T \Delta = 4 ~. \label{STSpDTD}
\end{align}
It is also easy to check that $\Delta^T \Delta = \Delta \Delta^T$ and $S^T S = S S^T$.

We note that these relations hold both for bosonic wires (labeled by integers $y$) and for fermionic wires (labeled by integers $j$). In the present work we use matrix notation exclusively within either bosonic or fermionic formulations, so there is no need to distinguish between sets of matrices used in either case.  For fermionic wire models, it is further useful to define 
\begin{align}
&P_{j,j'} = (-1)^j \delta_{j,j'} ~, \\
&D_{j,j'} = (1 - \delta_{j,j'}) \text{sgn}(j-j') (-1)^{j'} ~.
\end{align}
These matrices satisfy a number of useful relations, such as
\begin{align}
&D^2 = P^2 = 1 ~, \\
&D P D^T = -P ~, \label{DPeqPDT} \\
&\Delta D = -P \Delta ~, \\
&P S P = -\Delta ~, \\
&\Delta^T (D + P) = -2 P ~, \label{DeltaDplusPeq2P} \\
&S^T (D + P) = 2 D ~, \label{STDplusPeq2D} \\
&\Delta^T P \Delta= -S^T P S ~ \label{DeltaPDelta}, \\
&(1 \pm P) \Delta^T \Delta (1 \pm P) = 4 (1 \pm P) \label{1pPDTD1pP} ~.
\end{align}

 \bibliography{duality}

\begin{thebibliography}{66}
\bibitem [{foo()}]{foot}%
  \BibitemOpen
  \href@noop {} {\bibinfo  {journal} {When relating the wire models to the
  schematic continuum expressions, note that the coupling between gauge and
  matter fields in the former is $- i \rho a_0$ while it is $i \rho a_0$ in the
  latter. It follows that ${\cal S}_\text{wire}$ formally corresponds to ${\cal
  S}_\text{schematic}^* $ and the signs of the Chern-Simons coefficients are
  opposite between the two formulations. This overall sign leaves the
  interpretation of the fermions/dual fermions as bosons with attached fluxes
  of certain sign unaffected}\ }\BibitemShut {NoStop}%
\end{thebibliography}
\end{document}